\documentclass[aps, twocolumn, prb,preprintnumbers,superscriptaddress]{revtex4-2}

\usepackage{mathtools} 

\usepackage{times}
\usepackage{float}
\usepackage{multirow}
\usepackage[dvipsnames]{xcolor}
\usepackage{amsmath}
\usepackage{amsthm}
\usepackage{amssymb}
\usepackage{amsbsy}
\usepackage{nccmath}
\usepackage{xfrac}
\usepackage{enumitem}
\usepackage{wasysym}
\usepackage[english]{babel}
\usepackage[T1]{fontenc}
\usepackage[utf8]{inputenc}
\usepackage{graphicx}
\usepackage[colorlinks,bookmarks=false,citecolor=blue,linkcolor=blue,urlcolor=blue]{hyperref}
\usepackage[normalem]{ulem}
\usepackage{pstricks}
\usepackage{rotating}			       
\usepackage{tabularx,hhline}	
\usepackage[caption=false]{subfig}	
\usepackage{appendix}
\usepackage{bigdelim}
\usepackage{tikz-feynman,contour}
\newcolumntype{P}[1]{>{\centering\arraybackslash}p{#1}}

\usepackage{hyperref}
\usepackage{bbold}
\usepackage{bm}
\hypersetup{linktocpage,colorlinks=true,citecolor=red}

\usepackage[final]{changes}
\newcommand{\stkout}[1]{\ifmmode\text{\sout{\ensuremath{#1}}}\else\sout{#1}\fi}
\setdeletedmarkup{\stkout{#1}}
\definecolor{deeppink}{rgb}{1.0, 0.08, 0.58}
\newcommand{\kim}{\textcolor{deeppink}}
  
\definecolor{deeppurple}{rgb}{0.4, 0.25, 1}
\DeclareMathOperator{\sgn}{sgn}	        
\DeclareMathOperator{\csch}{csch}       

\newcommand{\nn}{\nonumber \\}

\def \be {\begin{equation}}
	\def \ee {\end{equation}}
\def \bee{\begin{equation*}}
	\def \eee{\end{equation*}}

\def \dd  {{\rm d}}
\def \ddp  {{\partial}}

\begin{document}

\let\subsectionautorefname\sectionautorefname
\let\subsubsectionautorefname\sectionautorefname
\def\chapterautorefname~#1\null{Chapter~#1\null}
\def\sectionautorefname~#1\null{Section~#1\null}
\def\appendixautorefname~#1\null{Appendix~#1\null}
\def\figureautorefname~#1\null{Fig.~#1\null}
\def\tableautorefname~#1\null{Table~#1\null}
\def\equationautorefname~#1\null{Eq.~(#1)\null}

\newcommand{\Autoref}[1]{%
	\begingroup%
	\def\chapterautorefname~##1\null{Chapter~(##1)\null}%
	\def\sectionautorefname~##1\null{Section~##1\null}%
	\def\appendixautorefname~##1\null{\hyperref[~##1]{Appendix~}\null}%
	\def\subsectionautorefname~##1\null{Sub--Section~##1\null}%
	\def\figureautorefname~##1\null{Fig.~##1\null}%
	\def\tableautorefname~##1\null{Table~(##1)\null}%
	\def\equationautorefname~##1\null{Eq.~(##1)\null}%
	\autoref{#1}%
	\endgroup%
}
\newcommand{\aref}[1]{\hyperref[#1]{Appendix~\ref*{#1}}}
\title{Universality of $T$--linear and $B$--linear Planckian scattering rate in high--$T_c$ cuprate superconductors}

\author{K. Remund}
\affiliation{Physics Division, National Center for Theoretical Sciences, Taipei 10617, Taiwan}
\affiliation{Department of Electrophysics, National Yang Ming Chiao Tung University, Hsinchu 30010, Taiwan}
\author{K. V. Nguyen}
\affiliation{Department of Electrophysics, National Yang Ming Chiao Tung University, Hsinchu 30010, Taiwan}
\author{P.-H. Chou}
\affiliation{Physics Division, National Center for Theoretical Sciences, Taipei 10617, Taiwan}
\affiliation{Department of Electrophysics, National Yang Ming Chiao Tung University, Hsinchu 30010, Taiwan}
\author{P. Giraldo-Gallo}
\affiliation{Department of Physics, Universidad de Los Andes, Bogotá 111711, Colombia}
\author{J. A. Galvis}
\affiliation{School of Sciences and Engineering, Universidad del Rosario, Bogotá, 111711, Colombia}
\author{G. S. Boebinger}
\affiliation{National High Magnetic Field Laboratory, Florida State University, Tallahassee, FL 32310, USA}
\affiliation{Department of Physics, Florida State University, Tallahassee, FL 32310, USA}
\author{C.-H. Chung}
\affiliation{Physics Division, National Center for Theoretical Sciences, Taipei 10617, Taiwan}
\affiliation{Department of Electrophysics, National Yang Ming Chiao Tung University, Hsinchu 30010, Taiwan}
\affiliation{Center for Theoretical and Computational Physics (CTCP), National Yang Ming Chiao Tung University, Hsinchu 30010, Taiwan}

\date{\today}

\begin{abstract}
One of the long-standing puzzles in strongly correlated materials is the microscopic origin of the quantum critical Planckian strange metal phase with universal linear-in-temperature scattering rate from which unconventional superconductivity directly emerges by lowering temperatures.
Recently, the linear-in-temperature and linear-in-field resistivity have been simultaneously observed in
high-$T_c$ cuprate superconductors, manifested by the universal field-to-temperature ($B/T$) scaling in magneto--resistivity. To date, there has been a lack of coherent and unified understanding of these co-existing linear behaviors and their possible link to quantum criticality.
In this work, we establish the universality in $T$--linear and $B$-linear Planckian behaviors in underdoped LSCO near optimal doping. Experimentally, we observe the $B$--linear Planckian  scattering rate and its relation to its $T$--linear counterpart.
Theoretically, we propose a spin-based common microscopic mechanism
based on Kondo-like charge fluctuations near local quantum criticality of heavy-fermion formulated $t$--$J$ model subject to a Zeeman term. Similar to frequency-to-temperature ($\omega/T$)--scaling near quantum criticality, we find the magnetic field here effectively introduces a Zeeman energy, reminiscent of an external energy ($\hbar \omega$) in the quantum critical regime, leading to $B/T$--scaling. Our analytically predicted universal $B/T$--scaling in isotropic scattering rate and the relation between the $T$--linear and $B$--linear Planckian coefficients, unifies these two phenomena over an extended doping range, pointing toward a unified quantum-critical origin of Planckian transport in cuprates.
\end{abstract}

\maketitle

\section{Introduction}\label{sec:intro}
A major mystery in condensed matter systems is the microscopic origin of the strange metal with perfect linear-in-temperature resistivity extending to highest temperature without saturation, from which unconventional superconductivity directly emerges by lowering temperatures.
In ordinary metals, however, electrical resistivity due to scattering among long-lived quasi-particles is bounded by the Mott-Ioffe-Regel (MIR) limit, where the mean-free path $l$ cannot be shorter than the Fermi wavelength $\lambda_F$, $l \ge \lambda_F$.
The $T$-linear resistivity in cuprates, which exceeds the MIR limit, is often referred to “Planckian dissipation”. This suggests a universal Planckian scattering rate in which transport relaxation rate $1/\tau$ is limited only by the thermal energy $k_BT$ ($k_B$ is Boltzmann constant) divided by the universal constant $\hbar$ ($\hbar=h/2\pi$ with $h$ being Planck’s constant): $1/\tau = \alpha_T k_BT/\hbar$ with $\alpha_T \sim 1$ being a coupling-independent universal constant Planckian coefficient, instead of limited by the quasi-particle interaction strength.
Experiments on overdoped cuprates have shown that this Planckian coefficient remains constant across different samples \cite{Taillefer-planckian-2019,Taillefer-Organic-PRB,George-NatComm-SM,Taillefer-annuphys-2019,michon-nature-2019-QC-cuprate,taillefer-NatLett-changeofcarrier-YBCO-2016,taillefer-PRB-FS-transformation-LSCO} and doping levels \cite{Ramshaw-MottPlanckian-arxiv-2024}, indicating a universal state of matter. Meanwhile, frequency-to-temperature ($\hbar\omega/k_BT$) scaling from optical conductivity measurement \cite{George-NatComm-SM} suggest that it is a quantum critical state.

\begin{figure*}[ht!]
\begin{center}
 		\begin{minipage}[b]{1\textwidth}
 			{
 				\includegraphics[width=0.3\linewidth]{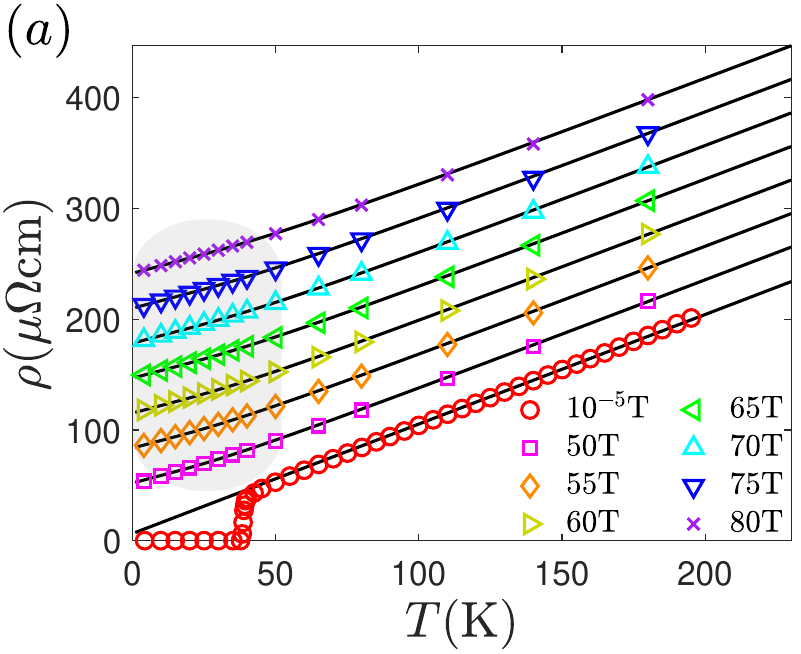}
			}
            {
 				\includegraphics[width=0.365\linewidth]{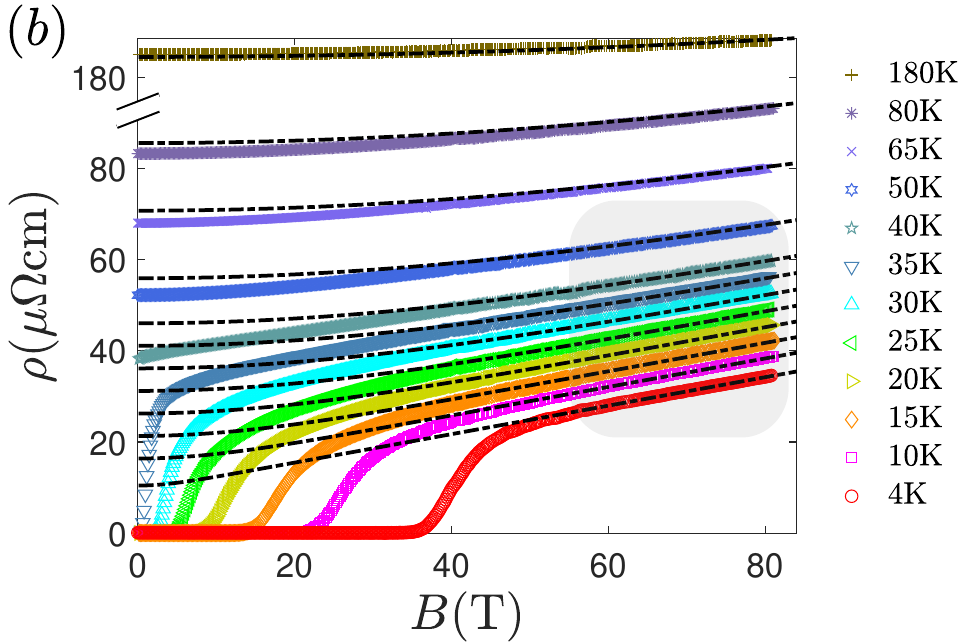}
			}
             {
 				\includegraphics[width=0.3\linewidth]{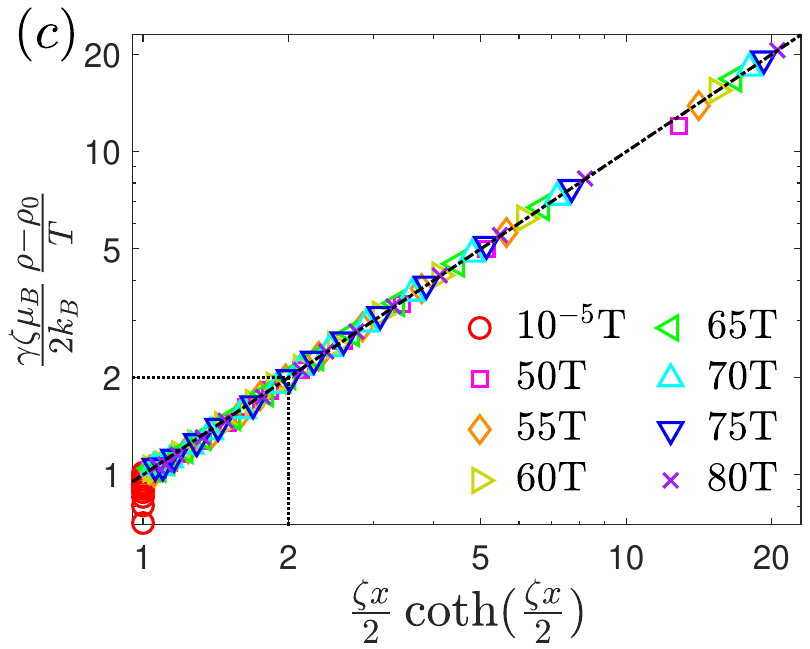}
			}
  	\end{minipage}
 	\end{center}
    \caption{ Magneto--resistivity data for LSCO at doping $p=0.19$, in which $T$--linear and $B$--linear coexist over the range of temperatures ($T<$ 50K) and magnetic fields ($B>$ 60T) indicated by the shaded regions. (a) Magneto--resistivity versus temperature at different magnetic fields. Colored symbols represent experimental data and black solid lines correspond to the theoretical prediction given in \protect \autoref{eq:rho-tt}, with $c_0=7.2 \mu\rm{\Omega cm}$, $c_1=0.654\mu\rm{\Omega cm K^{-1}}$, $\gamma=3.2
    (\mu\rm{\Omega cm})^{-1}\rm{T}$ , $\zeta=3.06$.
    For visibility purposes, the data points and curves have been shifted by $30 \mu\rm{\Omega cm}$ in the y-axis successively for each value of magnetic field.  
    (b) Magneto--resistivity data versus magnetic field at different temperatures.
    Colored symbols represent experimental data and black dotted lines correspond to the theoretical prediction given in \protect \autoref{eq:rho-tt} with same fitting parameters values as in (a). 
    (c) Scaling of the field dependent part of the magneto--resistivity at different magnetic field values in function of $\zeta x/2 \coth(\zeta x/2 )$, with $x=\mu_B B/(k_B T)$. This is obtained by subtracting to the total magneto--resistivity the constant and temperature-only dependent ($\rho_0(T)$)  and dividing it by temperature using the same parameters as in (a).
    Data is the same as in ref. \cite{Beobinger-2018-Science-LSCO} for LSCO.
}\label{fig:rho}
\end{figure*}

To reveal the Planckian normal state of high-$T_c$ cuprates and its link to putative quantum criticality inside the superconducting dome, magneto--transport measurements have been performed on various cuprates in high magnetic field to suppress superconductivity.
The mysterious universal quantum-critical-like field-to-temperature ($B/T$) scaling in magneto--resistance in strange metal state has been observed in various cuprates over an extended doping range \cite{hussey-incoherent-nature-2021,Ayres-NatComm-HT}, indicating this is a quantum critical “phase”.
Interestingly, both linear-in-temperature and linear-in-field resistivity have been simultaneously observed therein, manifested more clearly by
$k_BT \gg \mu_BB$ and $\mu_BB \gg k_BT$ limits of universal $B/T$-scaling in magneto--resistivity, respectively,
suggesting a strong correlation between these two strange metal behaviors \cite{Ayres-NatComm-HT}. This suggests that both phenomena may share a common microscopic origin. To date, there has not been a unified and coherent understanding of these linear behaviors that co-exist in the regime of high fields ($B>$ 60T) and low temperatures ($T<$ 50K) (gray shaded areas in \Autoref{fig:rho}(a) and (b)). This puts challenges on our understanding of the magneto--transport in metals. 

Magneto--resistance in ordinary metals originates from
the cyclotron motion of electrons around the Fermi surface,
determined by the product of cyclotron frequency,
$\omega_c = eB/m^*$ (where $m^*$ is the quasiparticle mass),
and the quasiparticle relaxation time $\tau_B$
\cite{Abrikosov-book-metals,Pippard-book-metals};
consequently, the Kohler's rule with quadratic-field magneto--resistivity $\rho(B)$ is expected:
$\Delta\rho(B)/\rho(0,T) = f(B/\rho(0,T)) \propto (B/T)^2$
with $\Delta\rho(B) = \rho(B) - \rho(B=0)$
\cite{Ayres-NatComm-HT}.
The $B/T$-linear scaling in magneto--resistivity at high fields observed in cuprates clearly violates the Kohler's rule,
indicating that the field-dependent transport relaxation rate is set directly by the magnetic field through $1/\tau_B(B)=\alpha_B \mu_B B/\hbar$.
Whether there exists a universal B-linear Planckian scattering rate
$1/\tau_B(B) =\alpha_B\mu_B B/\hbar$  with $\alpha_B\sim 1$ in the same strange metal state of cuprates, similar to that in the corresponding $T$-linear counterpart is therefore an issue of great fundamental significance.
If it does exist, whether a universality between these two Planckian metal behaviours and a possible common origin is an outstanding open problem.

Previous attempts to address this problem 
based on an orbitally driven origin
include an approach based on Boltzmann transport of conventional electrons in the presence of an anisotropic Fermi surface \cite{Taillefer-Boltzmann-2022}, and the Sachdev-Ye-Kitaev-like models with extreme external disorder \cite{Sachdev_PRX_2018,Kim-PNAS-2024}.
However, it is still an open problem whether the former (semi-classical) approach--significantly relying on specific parametrization--is able to capture the generic and robust feature of the data: an extended regime of $B$--linear magneto--resistivity with universal quantum critical $B/T$--scaling \cite{hussey-incoherent-nature-2021,Ayres-NatComm-HT}.
We note that recent experiments have demonstrated that the strange-metal behavior is independent of external disorder \cite{Alloul_fphy_2024}, which might pose challenges to approaches based upon SYK-like models.

In this work, we establish an unprecedented match between the experimental $B$-linear Planckian scattering rate and its relation to the $T$-linear counterpart near optimal doping of underdoped LSCO. We propose a spin-based common microscopic mechanism for this observation based on Kondo-like charge fluctuations near local quantum criticality via the recently developed heavy-fermion approach to the $t$--$J$ model \cite{YYC-ROPP-2025}. Within our approach, we assume the field-dependent scattering rate $\tau_B$ arises from the Zeeman splitting of fermions coupled to the isotropic component of the Fermi surface, a distinctly different approach than previous attempts based on electron orbital motion \cite{Taillefer-Boltzmann-2022,Sachdev_PRX_2018,Kim-PNAS-2024}. We find the isotropic electron scattering rate of our model near criticality simultaneously leads to both $T$--linear and $B$--linear resistivity. The Zeeman energy here plays a similar role as an external energy $\hbar\omega$ near criticality, giving rise to universal quantum critical $B/T$-scaling in transport, similar to $\omega/T$-scaling observed in AC transport at zero field. Our analytically predicted universal $B/T$--scaling in scattering rate excellently unifies the $T$--linear and $B$--linear Planckian behaviors over an extended doping range, pointing toward a unified quantum-critical origin of Planckian transport in cuprates. 

\section{Theoretical model: a unifying framework for $T$-linear and $B$-linear Planckian behaviors}\label{sec:theory}

The $T$-linear and $B$-linear resistivity have been simultaneously observed and systematically studied in the strange metal phase of LSCO over an extended doping range \cite{Beobinger-2018-Science-LSCO}. 
Inspired by this observation, we first look for signatures of quantum critical $B/T$-scaling by analyzing magneto--resistivity data in Ref.~\cite{Beobinger-2018-Science-LSCO} at finite field and temperatures.
We find  magneto--resistivity can be decomposed into the following form (see \Autoref{sec:QCscaling}, Appendix \ref{app:srfxi}, and Ref.\cite{YYC-ROPP-2025}):

\begin{eqnarray}
\rho(B,T) &=& \rho_0(T) + \frac{k_BT}{\mu_B} \Phi(\frac{\mu_B B}{k_BT})\; ,\nonumber\\
\Phi(x) &=& \frac{x}{\gamma} \coth\frac{\zeta x}{2},
\label{eq:rho-tt}
\end{eqnarray}
with $x = \mu_BB/k_BT$, and $\rho_0(T)$ depends only on temperature but not fields and it shows $T$-linear behavior: $\rho_0(T) = c_0 + c_1 T$, and $\Phi(B/T)$ is a universal $B/T$-scaling function where $c_0$ (constant residual resistivity extrapolated to $T=B=0$); $c_1$, and $\zeta$ are considered as fitting parameters, and  $\gamma$ is fixed by high magnetic field ($\mu_BB\gg k_B T$) slope $\ddp \rho/\ddp B$.

To identify a possible common microscopic mechanism for the $T$-linear and $B$-linear Planckian behaviors in cuprates, we propose here a recently developed theoretical framework
based on Kondo-like charge fluctuations near local quantum criticality of heavy-fermion formulated slave-boson $t$-$J$ model \cite{YYC-ROPP-2025}.
This approach is motivated
by the striking similarity in strange metal phenomenology between cuprates and heavy-fermion Kondo lattice systems, in
which the Fermi surface volume reconstructs over the entire
strange metal region in both systems.
Therein, the electron is effectively fractionalized into charge-neutral fermionic spinon and a spinless charged holon (slave-boson).
The hopping $t$-term is mapped onto a Kondo-like coupling where disordered slave boson interacts with fermionic spinon band and the effective conduction electron band made of spinon-holon bound fermion living on the bonds connecting nearest-neighbor sites (see Appendix \ref{methods}). The charge Kondo hybridization is dictated by the condensation of slave-boson. The antiferromagnetic Heisenberg exchange coupling $J$ term is decomposed into Resonating-Valence-Bond spin liquid singlets in both
spinon hopping and preformed $d$-wave Cooper pairing channels. Via a controlled perturbative renormalization group (RG) approach, the Planckian metal phase appears as a result of local disordered bosonic charge Kondo fluctuations coupled to a fermionic spinon band and a heavy hole-like spinon-holon bound fermion band near
the Kondo-breakdown local
quantum critical point (QCP) due to the competition between the pseudogap phase (dominated by the Heisenberg $J$ term) and Fermi liquid (FL) phase (dominated by hopping $t$-term). It was further shown in Ref.~\cite{YYC-ROPP-2025} via RG renormalized 2nd-order perturbation in the hopping (effective Kondo) term that close to the predicted QCP an exact cancellation of the hopping coupling constant ($t$-term) in the self-energy of the fermionic spinons, spinon-holon bound fermions and slave-bosons occurs (see Appendix \ref{methods} and Appendix \ref{app:srfxi}). This leads to a universal local electron self-energy insensitive to couplings (or hole doping), reminiscent of but distinct from the Marginal-Fermi-Liquid (MFL) self-energy \cite{YYC-ROPP-2025}. This further leads to an extended coupling-constant independent (doping insensitive) quantum critical strange metal "phase"  showing a universal Planckian scattering rate with quantum-critical $\omega/T$-scaling (see below).

This approach offers an excellent description of the quantum critical Planckian metal phase observed in cuprates \cite{YYC-ROPP-2025}, including: universal Planckian scattering rate in DC-transport over an extended doping range \cite{Ramshaw-MottPlanckian-arxiv-2024}, bad metallic behavior with $T$-linear resistivity beyond MIR limit \cite{Takagi-PRL-1992}, universal quantum critical $\omega/T$-scaling in optical (AC) conductivity and scattering rate \cite{George-NatComm-SM}, the marginal Fermi-liquid (MFL) spectral function observed in ARPES \cite{zxshen-sicience-incoherent-cuprate} as well as Fermi surface reconstruction observed in Hall coefficients in various overdoped cuprates \cite{hussey-incoherent-nature-2021,Taillefer-annuphys-2019}.

At a zero field in the Planckian metal phase,
the conduction electron shows 
universal (coupling-constant independent) 
MFL-like self-energy:
$\Sigma_{c}^{\prime\prime}(\omega,T \to 0) = - \frac{1}{\pi} |\omega|$.
Via the general relation between self-energy and scattering rate: $1/\tau = -2\Sigma_c^{\prime\prime}$ and conformal transformation, the conduction electron self-energy at $T=0$ leads to the finite-temperature AC-scattering rate with universal quantum critical $\omega/T$-scaling:
$\frac{\hbar/\tau(\omega,T)}{k_BT}
= \frac{2}{\pi} x \coth\frac{x}{4}$
with $x \equiv \hbar\omega/k_BT$, which in the static limit gives the universal $T$--linear Planckian scattering rate: 
$\hbar/\tau(\omega \to 0,T) = \alpha_T k_B T$ with $\alpha_T = 8/\pi$
\cite{YYC-ROPP-2025,YYC-arXiv-2025}. 
At a finite field, the longitudinal magneto--conductivity via Drude formula is defined as:
$\sigma_{xx} = \sigma_{yy}
= (ne^2\tau_B/m^*) [1/(1+(\omega_c \tau_B)^2)]$,
where $\omega_c = eB/m^*$ is the cyclotron frequency. It was found experimentally that $\omega_c \tau_B \ll 1$ in the strange metal region of cuprates
\cite{Chien-PRL-1991,Chien-PRB-1991,Ando-PRL-2004}, the magnetic field effect is dominated by the Zeeman spin splitting of the fermionic bands; As a result, the Landau level and orbital motion of electrons are negligible, justifying the assumption of our approach.
The longitudinal magneto--resistivity therefore takes the approximate Drude form:
$\rho_{xx} \equiv \rho(B,T) \approx (m^*/ne^2) (1/\tau_B)$. We now generalize this approach by including the Zeeman term in magnetic field and focus on critical charge fluctuations near local criticality (Appendix \ref{methods}). 
Here, we restrict ourselves to the isotropic scattering where only the isotropic conduction electron states near the Fermi surface are considered (\Autoref{fig:alpha_B_T} Inset). At zero field, this  approximation is further justified and supported by the experimental observation that the Planckian scattering rate is contributed from the isotropic component of scattering rate \cite{2022-Gael-Nature-isotropic-Planckian}. At a finite field, the Zeeman splitting effect will contribute to the isotropic component of the magneto-scattering rate. 
We obtain the AC magneto--scattering rate $1/\tau(\omega,B,T)$ by computing the field-dependent electron self-energy $\Sigma_c^{\prime\prime}(\omega,B,T)$. We find that the effect of Zeeman field on the fermionic 
self-energy $\Sigma_{fermion}^{\sigma}(\omega,B)$ is equivalent to a shift in fermionic frequency $\omega$ in the zero-field self-energy $\Sigma_{fermion}^{\sigma}(\omega)$: 
$\Sigma_{fermion}^{\sigma}(\omega,B) = \Sigma_{fermion}^{\sigma}(\omega \to \omega-s_{\sigma}\mu_BB)$, which applies to both fermionic spinons and spinon-holon bound fermions (see Appendix \ref{app:srfxi}). 
At a finite field, our model leads to the field- and
spin-dependent AC magneto--scattering rate (after subtracting the purely $T$-dependent scattering rate $\hbar/\tau_0(T)$) as (see Appendix \ref{app:srfxi}):
\begin{eqnarray}
    && \frac{\hbar}{\tau_B^{\sigma}}-\frac{\hbar}{\tau_0(T)} = \nn
    && \frac{2}{\pi} k_BT
    \left[ \frac{\hbar\omega/2+s_{\sigma}\mu_BB}{k_BT} \right]
    \coth\left[\zeta \frac{\hbar\omega/2+s_{\sigma}\mu_B B}{2k_BT} \right]\label{eq:rates}
\end{eqnarray}
with $s_{\sigma} = \sgn(\sigma) = +1\,(-1)$ for $\sigma = \uparrow\,(\downarrow)$, respectively.
Here, a tuning parameter $\zeta$, allowed by conformal mapping of scattering rate in the scaling regime from zero temperature to finite temperatures \cite{YYC-arXiv-2025} (see Appendix \ref{app:srfxi}, \ref{AppE}), is introduced to
match the observed universal $B/T$--scaling in
magneto--scattering rate (see \Autoref{eq:rho-tt} and \Autoref{fig:rho}), whose value is constrained by the $T$--linear Planckian coefficient $\alpha_T= 8/\pi$ (see below and Appendix \ref{app:srfxi}, \ref{usffx}).

The total field dependent magneto--scattering rate (subtracted by a constant residual scattering rate at $\omega=B=T=0$) is given by
$\hbar/\tau_B - \hbar/\tau_0(T) =
\hbar/\tau_B^{\uparrow}+\hbar/\tau_B^{\downarrow}$,
where $\hbar/\tau_0(T) = b k_BT$ with $b = 8(1-1/\zeta)/\pi$ being the $T$-linear residual scattering rate at $\omega=B=0$.
In the DC ($\omega \to 0$) limit, this gives universal $B/T$-scaling in scattering rate
\begin{eqnarray}
\frac{\hbar/\tau_B - \hbar/\tau_0(T)}{k_BT} &=&
\frac{2}{\pi} \frac{\mu_B B}{ k_B T}
\coth\left( \zeta\frac{\mu_B B}{2 k_B T} \right).\label{eq:rate}
\end{eqnarray}

\begin{figure}[H]
	\begin{center}
	\begin{minipage}[b]{1\textwidth}
     {
 				\includegraphics[width=0.397\linewidth]{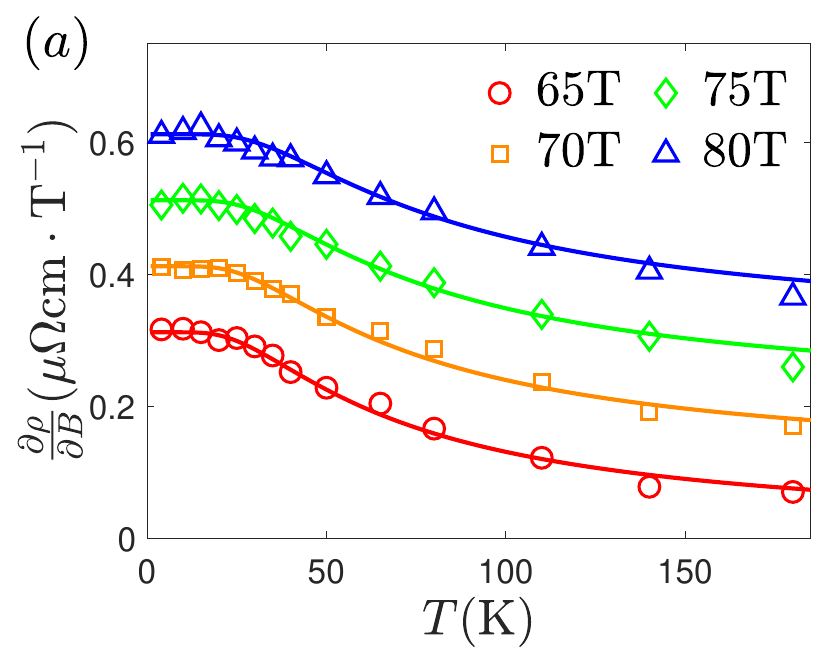}
			}\\
 	{
 				\includegraphics[width=0.47\linewidth]{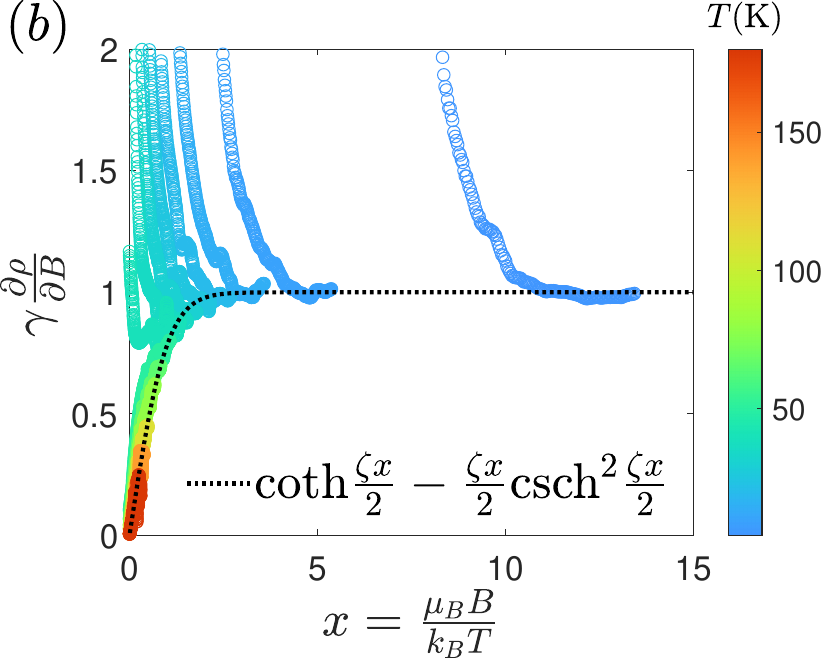}
			}
		\end{minipage}
	\end{center}
 	\caption{ (a)
    Slope of the magneto--resistivity with respect to magnetic field versus temperature at different magnetic fields for doping $p=0.19$. Colored symbols represent experimental data and colored solid lines correspond to the associated theoretical prediction [\protect \autoref{eq:slope}] obtained by taking the derivative of \protect \autoref{eq:rho-tt} , with same parameter values as in  \protect\autoref{fig:rho} : $\gamma=3.2 (\mu\rm{\Omega cm})^{-1}\rm{T}$ , $\zeta=3.06$.
    For visibility, the data points and curves have been shifted by $0.1 \mu\rm{\Omega cm \cdot T^{-1}}$ in the y-axis successively for each value of magnetic field.
     (b)~$B/T$--scaling  of the rescaled slope of the magneto--resistivity as a function of $x=\mu_B B/(k_B T)$  for doping $p=0.19$. The slope values are then rescaled by $\gamma_{p=0.19}=3.2 (\mu\rm{\Omega cm})^{-1}\rm{T}$ for  $p=0.19$, such that at high magnetic field values $\mu_B B>>k_B T$, the product $\gamma (\ddp \rho/\ddp B)$ saturates to unity. Note that $\gamma$ depends on the carrier density $n$  and is expected to slightly vary for different doping values. The black dotted line corresponds to the theoretical prediction of the derivative of the magneto--resistivity given in \protect \autoref{eq:slope} with $\zeta=3.06$.  Magneto--resistivity data are reproduced from the same data as in ~\cite{Beobinger-2018-Science-LSCO} for  $p=0.19$. 
	}
	\label{fig:drhoBT}
 \end{figure}

The universal scaling function in scattering rate (\Autoref{eq:rate}) from which \Autoref{eq:rho-tt} is deduced, is our key theoretical result. Remarkably, as we will show in the following section, it unifies the $T$-linear and $B$-linear Planckian strange metal state in magneto--resistivity observed in cuprates:
(i) It gives excellent analytic fits (within error bars) to the observed $T$-linear and $B$-linear Planckian scattering rates in the zero field and zero temperature limit, respectively:
$\hbar/\tau_B(B=0,T) = \alpha_T k_BT$
with $\alpha_T = 8/\pi$
and $\hbar/\tau_B(B,T=0) = \alpha_B\mu_BB$
as well as the correlation between them
$\alpha_B = \alpha_T/2 = 4/\pi$ (\Autoref{fig:alpha_B_T});
(ii) More generally at finite temperatures and fields,
it leads to the universal scaling function $f(x)$
well fitted to
$\rho(B,T)$ in \Autoref{fig:rho}
and $d\rho/dB$ in \Autoref{fig:drhoBT}.
The scaling function based on \Autoref{eq:rho-tt} and \Autoref{eq:rate} further undermine the Marginal Fermi Liquid (MFL) assumption \cite{varma-mfl-prl} 
(see \Autoref{fig:drhoBT_19_B_fit} in Appendix~\ref{usffx}), 
with empirical quadrature scaling form $\rho-\rho_0 \sim \sqrt{(a k_BT)^2 + (b \mu_BB)^2}$. This formula for MFL does not allow for a regime in which $T$--linear and $B$--linear resistivity co-exist, which is the most striking feature of the data for $T<$ 50K and $B>$ 60T. (indicated by the shaded regions in \Autoref{fig:rho}).


\begin{table}[h!]
\begin{center}
\begin{tabular}{|c c c c c|}
 \hline
 $p$ & $c_0$ ($\mu\Omega$.cm) & $c_1$ ($\mu\Omega$.cm/K) & $\zeta$ & $\gamma$ (T/$\mu\Omega$.cm) \\ [0.5ex]
 \hline\hline
 0.19 & 6.48 & 0.685 & 3.06 & 3.2 \\
 \hline
\end{tabular}
\end{center}
\caption{The fitting parameters used for the LSCO $p=0.19$ data throughout the manuscript. The values of $c_1$ and $\gamma$ can be calculated from the theory and agree well with these values (see \protect\Autoref{table:parameters} of the Appendix \protect\ref{sec:ana_slope}).}
\label{table:parameters-ms}
\end{table}

\section{FITTING THE DATA WITH THE THEORETICAL FRAMEWORK}\label{sec:QCscaling}

The form of \Autoref{eq:rho-tt} is derived from a recently developed theoretical framework based on Kondo-like charge fluctuations near a local quantum critical point within the slave-boson formulation of the $t$--$J$ model \cite{YYC-ROPP-2025}, as explained in detail in the previous section. This framework naturally accounts for a unified microscopic origin of the $T$--linear and $B$--linear Planckian behaviors in cuprates (see \Autoref{sec:theory}, Appendix~\ref{app:srfxi}).

As shown in \Autoref{fig:rho}, magneto--resistivity both in the $T$-linear (\Autoref{fig:rho}~(a)) and
$B$-linear (\Autoref{fig:rho}~(b)) limits are well fitted by \Autoref{eq:rho-tt}; and $B/T$-scaling behavior is clearly seen in $[\rho(B,T)-\rho_0(T)]/T \propto \Phi(x)$ (\Autoref{fig:rho}~(c)).
It is important to note  that we use the same values of these four fitting parameters for all fits for $p = 0.19$ throughout the manuscript (see \Autoref{table:parameters-ms}).

At fixed higher fields, the seemingly two $T$-linear behaviors of magneto--resistivity observed at lower and higher temperature regions (\Autoref{fig:rho}~(a)) can be well accounted for by \Autoref{eq:rho-tt} as a result of interplay between $T$-linear term $\rho_0(T)$ and the scaling function $\Phi(x)$ (see \Autoref{fig:fit_rho_B_tot} in Appendix \ref{sec:planck_coeff}). Meanwhile, in \Autoref{fig:rho}~(b), the deviation between theoretical predictions (dashed lines) and the data at low temperatures is likely due to superconducting fluctuations. Such a deviation vanishes at a higher temperature well above $T_c$ (see the fit to $T=180K$ data), where superconducting fluctuations are absent.
Moreover, the $B/T$-scaling function $\Phi(x)$ can be extracted directly by taking the slope of the magneto--resistivity $d\rho/dB$.
As shown in \Autoref{fig:rho}~(c), clear universal $B/T$--scaling exists in the slope of magneto–resistivity, which is a signature of quantum criticality. 
 We further find this quantum critical behavior exists over an extended doping ($p$) range
$0.16 < p < 0.19$ : $\gamma d\rho/dB = f(x)$, where
$f(x) = \coth(\zeta x /2) - \zeta x \csch^2(\zeta x /2)$
is a universal well-fitted scaling function with $\zeta \approx 3.06$ (see \Autoref{fig:drhoBT} and \Autoref{fig:drhoBT_appx}) \cite{Beobinger-2018-Science-LSCO,YYC-ROPP-2025}.
The temperature dependent slope of magneto--resistivity $\ddp\rho(B,T)/\ddp B$ observed in Ref.~\cite{Beobinger-2018-Science-LSCO} is also well--fitted by $f(x)$, coming from the same scaling function $\Phi(x)$ (\Autoref{fig:drhoBT}~(a)) for $\ddp\rho/\ddp B$ vs. $T$ at different fields, and \Autoref{fig:drhoBT}~(b) for $\gamma d\rho/dB$ vs. $x$.
Similar results have been reported in other cuprate compounds \cite{hussey-incoherent-nature-2021}. Note that the normalized scaling function $f(x)$ is identical to that for AC--scattering rate found in Ref.~\cite{YYC-ROPP-2025} provided $x \equiv \hbar\omega/k_BT$ and $\hbar\omega \equiv 2\zeta\mu_BB$.

\section{Correlation between co-existing $T$--linear and $B$-linear Planckian scattering rates near optimal doping}\label{sec:BT_scaling}

The $T$-linear Planckian scattering rate for LSCO in Ref.~\cite{Beobinger-2018-Science-LSCO} in the quantum critical scaling regime was recently studied in Ref.~\cite{Ramshaw-MottPlanckian-arxiv-2024}. Therein, the doping insensitive Planckian coefficient $\alpha_T \sim 2.6$ (for $0.16 < p < 0.19$) was found from the zero-field $T$-linear resistivity in the SM region via the Drude formula 
expressed in terms of plasma frequency ($\omega_p$)
(\Autoref{fig:rho}~(a)):
$\rho(T,B=0) = A_T^1 T = (4\pi /\omega^2_p)(1/\tau) = (m^*/ne^2) (1/\tau)$ with
$\omega_p^2 = 4 \pi n e^2/m^* $\cite{Ashcroft} , $m^*$ being electron effective mass, and $1/\tau = \alpha_T k_B T/\hbar$ being the  Planckian scattering rate. Here, $n/m*$ ratio is directly estimated via the plasma frequency through optical conductivity data as: $\omega_p^2 \approx 4 \pi e^2 a_w p / V m_e$ with $a_w \sim 0.42$ \cite{PhysRevB.72.060511,PhysRevB.106.195110,George-NatComm-SM,Ramshaw-MottPlanckian-arxiv-2024}, $V=A l$ is the volume of unit cell with $A=a^2$ being the planar area, $a\simeq 0.378~\rm{nm}$ being the in--plane lattice spacing, and $l \simeq$ 0.66 nm being the CuO$_2$ interlayer distance.

\begin{figure}[h]
    \begin{center}
    \includegraphics[width=1\linewidth]{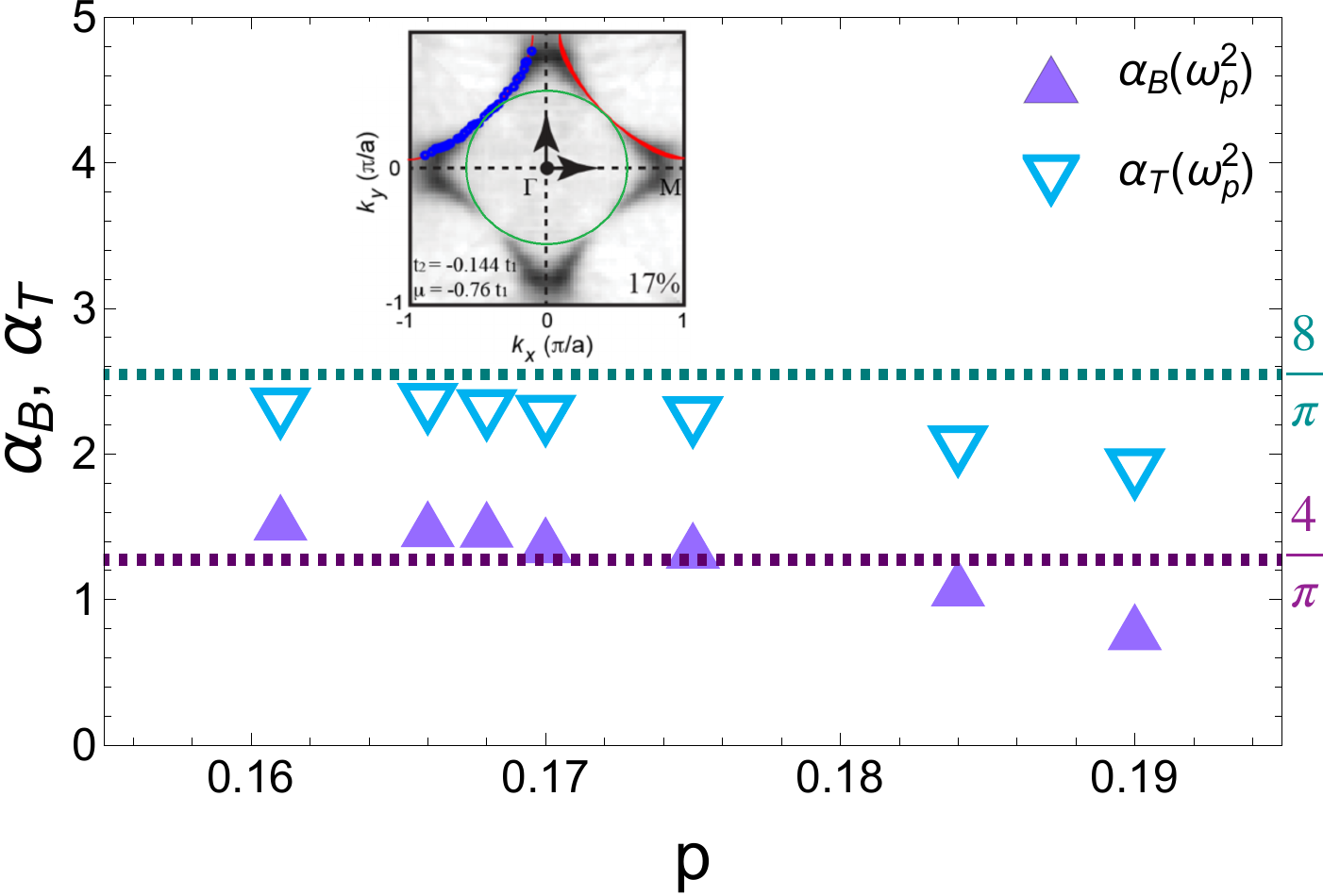}
    \end{center}
    \caption{The linear-in-temperature (down triangles) and linear-in-field (up triangles) Planckian coefficients versus doping for LSCO obtained from magneto--resistivity data reproduced from Ref.~\cite{Beobinger-2018-Science-LSCO} as detailed in Appendix \ref{sec:planck_coeff}. Horizontal purple and turquoise dashed lines correspond to theoretical prediction $\alpha_T=8/\pi\simeq 2.54$ and $\alpha_B=4/\pi\simeq 1.27$, respectively. Note that our theoretical framework finds quantitative agreement with the experimentally measured values of $\alpha_B$ and agrees within $\sim 25\%$ of the experimentally measured values of $\alpha_T$. The calculated ratio of $\alpha_T/\alpha_B =2$ is in close agreement with experimental data for $0.16<p<0.19$. Inset: A plot of two-dimensional Fermi surface of LSCO at $p=0.17$ (dark region guided by red and blue curves), as measured by ARPES and adapted from Ref. \cite{LSCO_FS}. The green circle is a schematic plot of isotropic electronic states near the Fermi surface that contribute to the isotropic scattering rate within our theoretical framework in the experimental doping range $0.16<p<0.19$.
    }
    \label{fig:alpha_B_T}
\end{figure}

We examine whether the similar Planckian behavior exists in the $B$-linear magneto--resistivity
in the same doping of the same material (LSCO) studied in Ref.~\cite{Beobinger-2018-Science-LSCO}. We first extract the slope $A_B^1$ of the low-temperature (averaged between $T$ = 20K and $T$ = 30K) linear-in-field magneto--resistivity:
$\rho(B,T=T_0)-\rho_0 = A_B^1 B$.
Via the Drude form of $\rho(B,T)$, we indeed find the similar doping-insensitive 
universal Planckian behavior in magneto--transport of the same compound in the same doping range
$\hbar/\tau_B = \alpha_B\mu_BB$
with the Planckian coefficient
$\alpha_B = \hbar (\omega_p^2/4\pi) A_B^1/\mu_B  =  \hbar (n e^2/m^*) A_B^1/\mu_B  \sim 1.4$ (with the same $m^*$ and $n$ as those at zero field) (\Autoref{fig:alpha_B_T}). Remarkably, we find the two Planckian coefficients $\alpha_T,\alpha_B$ are
correlated by the approximated relation: $\alpha_T \sim 2\alpha_B$, emerging from the correlation between the slopes of $T$--linear and
$B$--linear resistivities: $A^1_T \sim 2A^1_B$ \cite{Beobinger-2018-Science-LSCO}. Similar correlation be tween $A^1_T$ and $A^1_B$ slopes was observed in the strange metal state of different cuprate compounds \cite{Ayres-NatComm-HT}.

Note that plasma frequency approach from optical conductivity measurement is an  accurate and reliable approach to estimate the Planckian coefficients $\alpha_T$ and $\alpha_B$ since it gives directly the needed $n/m^*$ ratio.  Though the error bars in determining $\alpha_T$ and $\alpha_B$ is increased, $m^*$ and $n$ can also be estimated separately where $m^* \sim 1/p$ is obtained from specific heat and $n \sim n_{eff} \sim 1/V$ \cite{Ramshaw-MottPlanckian-arxiv-2024}. As shown in \Autoref{fig:alpha_B_T_wp}, we find that the values of $\alpha_T$ and
$\alpha_B$ estimated from both approaches agree very well. Here, the carrier density $n$ is taken as an effective density $n_{eff}$ by fitting to $\omega_p^2 \propto n/m^* \propto p$ with $m^* \sim 1/p$ \cite{Ramshaw-MottPlanckian-arxiv-2024} (see Appendix \ref{sec:planck_coeff}). To reconcile the discrepancy between the doping insensitive $n_{eff} \sim 1/V$ estimated here and the doping dependent carrier density $n$ via the Hall coefficient measurement goes beyond the scope of the present work and deserves further study elsewhere.

\section{Discussions}\label{sec:discussions}
Within our theoretical framework, we find in 2D limit the fitting parameters $\gamma$ and $c_1$, as defined in \Autoref{eq:rho-tt}, take the analytical forms: $\gamma_{\mbox{\tiny{2D}}}= (\hbar ne^2)/(m^*\alpha_B \mu_B)$ and
$c_1=(m^*/ne^2)(k_B/\hbar)(\alpha_T-\alpha_B/\zeta)$. Their values obtained via the above analytic forms (see Appendix \ref{sec:ana_slope}) are in good agreement with that by numerical fits in \Autoref{fig:rho} and \Autoref{fig:drhoBT}.

Since in the
quantum critical state the thermal energy is the only relevant
energy scale, $\omega/T$-scaling in observables can be in general
understood as the universal scaling in the dimensionless ratio between the external energy ($\hbar\omega$) and the thermal energy
($k_BT$). In this context, the $B/T$-scaling in cuprates is expected from the $\omega/T$-scaling since the Zeeman--energy from external magnetic fields ($\mu_BB$) constitutes one possible realization of the external energy in the quantum critical state. Note that our mechanism based on Zeeman splitting predicts that the Planckian scattering rate $1/\tau_B$ is insensitive to field orientation, as observed in Tl2201 and Bi2201 \cite{hussey-incoherent-nature-2021}.  However, clear field-angle dependence of magneto--resistivity has been observed in the overdoped region ($p=0.23$, $0.24$) of LSCO, possibly due to proximity of Fermi surface to the Van Hove singularity at those dopings \cite{Taillefer-Boltzmann-2022,LSCO23}. This could explain the deviation of our theoretical predictions for $\alpha_T$, $\alpha_B$ from the experimentally observed values (\Autoref{fig:alpha_B_T})
as $p$ increases. In this case of proximity to the Van Hove singularity, a correction to our analysis from the anisotropic orbital motion is to be  expected, which goes beyond the scope of this paper.

\begin{figure}[h]
    \centering
    \includegraphics[width=1\linewidth]{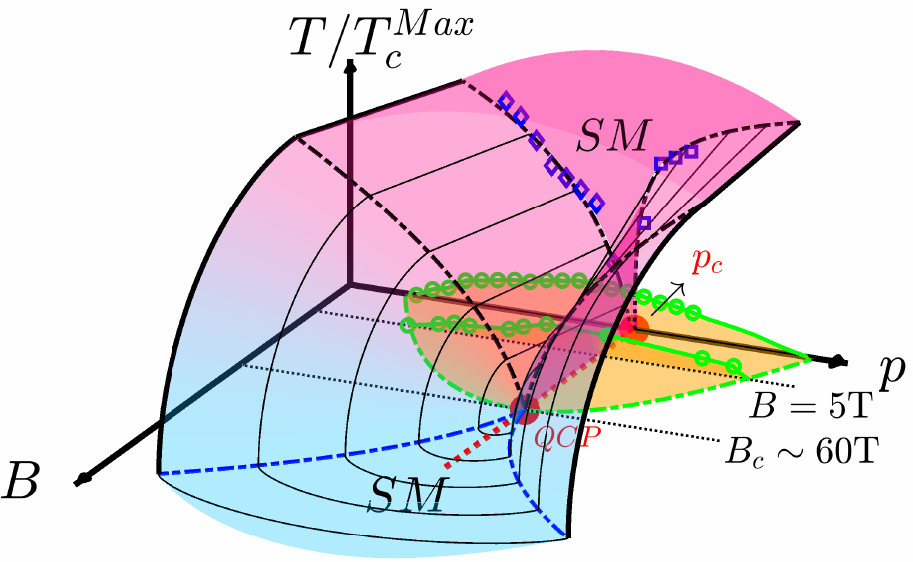}
    \caption{The doping-field-temperature $(p,B,T)$ phase diagram of LSCO near critical doping $p_{cr} \sim 0.19$ based on magneto--transport data in \cite{Beobinger-2018-Science-LSCO}. Green circles defining the superconducting  $T_c$ dome are experimental data obtained from \cite{Ando-PRL-2004} at zero magnetic field, and from \cite{Caprara_SciPost_2020} for finite magnetic field $B=5 \rm{T}$.  Blue diamonds and blue square are data from \cite{Ramshaw-MottPlanckian-arxiv-2024} and \cite{Ando-PRL-2004} respectively and correspond to the crossover temperature between the SM phase into the PSG phase and the FL phase. Note that the temperature axis is rescaled by $T^{Max}_c$, which is experimental group dependent, and is defined by the highest $T_c$ value of superconducting $T_c$ dome at zero field.}
    \label{fig:3Dplot}
\end{figure}

Our experimental and theoretical analysis strongly supports the scenario of a quantum critical point (QCP) hidden inside the superconducting dome in cuprates. It separates pseudogap from Fermi liquid phases and serves as a common origin of the observed $T$-linear and $B$-linear resistivity. We anticipate this QCP to emerge at $B = B_c \sim$ 60 T \cite{Ando-PRL-1995,Ando-PRB-1999}, $p = p_c \sim 0.19$ and $T \to 0$ where superconductivity is suppressed exactly by field near critical (optimal) doping, in LSCO. As shown in \Autoref{fig:3Dplot}, based on existing data and our theory, we propose a quantum critical fan-like volume, expanded from the single QCP in the general doping-field-temperature ($p, B, T$) phase diagram of LSCO, serving as a common origin to unify the universal $T$--linear and $B$--linear Planckian scattering rate as well as $B/T$--scaling in magneto--resistivity observed therein. Further experimental studies are needed to clarify this picture.

\textbf{Acknowledgments.}
This work is supported by the National Science and Technology Council (NSTC) Grants
110-2112-M-A49-018-MY3, the National Center for Theoretical Sciences of Taiwan, Republic of
China (C.H.C.). C.H.C. acknowledges the hospitality of Aspen Center for
Physics, USA and Kavli Institute for Theoretical Physics, UCSB, USA where part of the work was
done. This research was supported in part by grant NSF PHY-2309135 to the Kavli Institute for
Theoretical Physics (KITP).
The experimental data in this paper were collected at the NHMFL, supported by the NSF through DMR-1644779.

\appendix
\section{Methods}\label{methods}

Our starting point is the heavy-fermion formulated slave-boson $t$-$J$ (HFSBtJ) Hamiltonian on 2D lattice which was developed recently \cite{YYC-ROPP-2025,Punk-SBtJ-PRB}. We start from the original one-band $t$-$J$ model, reads:
\begin{align}\label{eq:SBtJH}
    H_{t} &=-t\sum_{\langle i,j \rangle , \sigma} c_{i\sigma}^{\dagger} c_{j\sigma}-\mu \sum_{i\sigma} c_{i\sigma}^{\dagger} c_{i\sigma}, \nn
     H_J  & = J_{H} \sum_{\langle i,j\rangle} \bm{S_{i}} \cdot\bm{S_{j}}\;,
\end{align}
where $(t,\, \mu, \, J_H)$ denotes the (hopping strength, chemical potential, Heisenberg coupling).
Under the slave-boson representation $c_{i\sigma}^\dagger \to  f^\dagger_{i\sigma} b_i$ with $f_{i\sigma} \, (b_i)$ being fermionic charged-neutral spinon (bosonic spinless charged holon) operators,
the hopping $t$-term is mapped via Hubbard-Stratonovich transformation onto a Kondo-like coupling where disordered slave boson (holon $b$-field) interacts with fermionic spinon ($f$-field) band and the effective conduction electron band made of spinon-holon bound fermion ($\xi$-field with $\xi_{ij}^{\sigma} = \sum_{\left<i,j\right>} b_i f_j^{\sigma}$) living on the bonds connecting nearest-neighbor sites $\left<i,j\right>$ \cite{Punk-PNAS-dimer,Punk-SBtJ-PRB}
\begin{align}
    H_{t} &\to \frac{t}{\sqrt{N}}\sum_{\langle i,j\rangle,\sigma} \left[\left(f_{i\sigma}^{\dagger}b_{j}^{\dagger}+f_{j\sigma}^{\dagger}b_{i}^{\dagger}\right)\xi_{ij,\sigma}+H.c.\right] \;.
\end{align}

The charge Kondo hybridization is dictated by the condensation of slave-boson.
The anti-ferromagnetic Heisenberg exchange coupling $J$ term is decomposed into Resonating-Valence-Bond (RVB) spin liquid singlets in both particle-hole $\left( \chi_{ij} = \chi =\left<f_{i\sigma}^{\dagger} f_{j\sigma}\right> \right)$ and particle-particle $\left( d-\text{wave Cooper pairing } \Delta_{ij} = \Delta = \left< f_{i\uparrow}^{\dagger} f_{j\downarrow}^{\dagger} - f_{j\uparrow}^{\dagger} f_{i\downarrow}^{\dagger} \right> \right)$ channels
\begin{align}
    H_{J} \to 	& -J_{H} \sum_{\langle i,j\rangle}\left(\sum_{\sigma}\chi_{ij}f_{i\sigma}^{\dagger}f_{j\sigma}+H.c.-\left|\chi_{ij}\right|^{2}\right) \nn
    		& +J_{H} \sum_{\langle i,j\rangle}\sum_{\sigma,\sigma^\prime}\left(\Delta_{ij}\varepsilon_{\sigma\sigma^\prime}f_{i\sigma}^{\dagger}f_{j\sigma^\prime}^{\dagger}+H.c.\right)  \nn
    		& +J_{H} \sum_{\langle i,j\rangle}\left|\Delta_{ij}\right|^{2}.
\end{align}

Four mean-field phases are realized depending on whether Bose-condensation of
the slave-boson $b$-field $\left<b_i\right>=\left<b\right>$ and $\Delta$-field $\left<\Delta\right>$
occurs:
the pseudogap phase, known as $Z_2$ fractionalized FL or $Z_2$ FL$^{\star}$ \cite{Senthil-prl-fractionalized-FL} (Landau Fermi liquid, FL) is reached
when $\left<\Delta\right> \neq 0$, $\left<b\right> = 0$ ($\left<\Delta\right> = 0$, $\left<b\right> \neq 0$); while the $U(1)$ FL$^{\star}$ ($d$-wave superconducting) phase arises when $\left<b\right> = \left<\Delta\right> = 0$ ($\left<b\right> \neq 0$, $\left<\Delta\right> \neq 0$).
The Planckian metal phase appears
in $U(1)$ FL$^{\star}$ phase beyond mean-field
as a result of
critical local charge Kondo fluctuations where disordered local bosons couple
to a fermionic
spinon band and a heavy-electron band near a localized-delocalized quantum critical point due to the competition
between the pseudogap phase (dominated by $J$) and FL phase (dominated by $t$).

The action of the model at zero field beyond mean-field in $U(1)$ FL$^*$ is given by Ref.~\cite{YYC-ROPP-2025}. 
We shall explore the phase diagram in U(1) FL$^*$ phase beyond mean-field by including fluctuations of both $t$- and $J$-terms. We will not include U(1) gauge fluctuations in our RG analysis since the spinons are deconfined and stable against them due to the presence of a spinon Fermi surface \cite{SSLee-PRB-2008-stablility}.  The leading  effective action in units of the half-bandwidth of the $f$--spinon band $D\approx 4\chi \approx 4J_H$ beyond the mean-field level reads ($\hbar = k_B = 1$) \cite{YYC-ROPP-2025}
     \begin{align}
         S=&-\sum_{k\sigma}f_{k\sigma}^{\dagger}\mathcal{G}^{-1}_f (k)f_{k\sigma}-\sum_{k}b_{k}^{\dagger}\mathcal{G}^{-1}_b (k)b_{k}  \nn 
         &+ \sum_{k\sigma a}\xi_{k\sigma}^{a\,\dagger} \mathcal{G}^{-1}_\xi(k) \xi_{k\sigma}^{a} + \sum_{k a}\varphi_{k}^{a\,\dagger}\mathcal{G}^{-1}_\varphi(k) \varphi_{k}^{a} \nn
         & +\frac{g}{\sqrt{\beta N_{s}}}\sum_{kp\sigma a}\left(f_{k\sigma}^{\dagger}b_{p}^{\dagger}\xi_{k+p,\sigma}^{a}+H.c.\right)  \nn
         & +\frac{J}{\sqrt{\beta N_{s}}} \sum_{kpa}\left[f_{k\uparrow}^{\dagger}f_{p \downarrow}^{\dagger}\varphi_{k+p}^{a}+H.c.\right] 
    \label{eq:action-S}
     \end{align}
with $k = (\bm{k},\omega)$ and $p = (\bm{p},\nu)$, $\varphi$ being the fluctuating field for the \textit{d}-wave RVB pairing order parameter $\Delta$. In the phases with $\langle b\rangle=0$, the bare hopping parameter $t$ is strongly suppressed by the disordered bosons, leading to an effective hopping $g \equiv 2t \sqrt{\delta}$. Here, $ J = 2J_H$ is the effective exchange, $\mathcal{G}_f (k)= (i\omega-\varepsilon_{\bm{k}})^{-1}$, $\mathcal{G}_\xi (k) = \zeta^{-1} \left( i\omega-\xi_{\bm{k}} \right)^{-1}$, $\mathcal{G}_b = (i\omega - \lambda)^{-1}$, and $\mathcal{G}_\varphi = (J/2)^{-1}$ denote the bare Green's functions (see Appendix B in Ref.~\cite{YYC-ROPP-2025}). The momenta relevant for the bosonic $\varphi$ field are near the antinode. Our perturbative expansion in bare couplings $g$, $J$ is controlled since $g/D, \, J/D <1$ (for an  estimated $J_H/t \sim 0.3$).
To study the charge dynamics and transport properties, we go beyond static mean-field level of $\xi$ field by generating its dynamics and dispersion via second-order hopping process at fixed $g=g^*_0$ such that $\zeta^{-1} \equiv ( g^2 \rho_0/D)^{-1}$ appears as a prefactor in $\mathcal{G}_\xi$, with $\rho_0 = 1/D$ being the constant density of states at Fermi level for $f$ spinon band. The $f$-spinon band is approximated by a linear-in-momentum dispersion, namely $\varepsilon_{\bm{k}}=h_{\bm{k}}-\mu_{f} \approx v |\bm{k}|$ with  $h_{\bm{k}}\equiv-2\chi\left(\cos k_{x}+\cos k_{y}\right)$ and $\mu_{f}=\mu-\lambda$ being the effective chemical potential for $f$ spinon ($\mu$ is the chemical potential for the original conduction electron, see Appendix A in Ref.~\cite{YYC-ROPP-2025}). Here, the spinon band is fixed at half-filling, $\mu_f=0$.  The band structure for the $\xi$ fermion shows a hole-like dispersion, $\xi_{\bm{k}}=-\zeta\varepsilon_{\bm{k}}-\mu_\xi$. The Lagrange multiplier $\lambda >0$ is introduced to enforce the local constraint for slave boson. The slave boson ($b$ field) is effectively treated as a local boson with a flat band of energy $\lambda$ and with a negligible dispersive band (see Appendix B in Ref.~\cite{YYC-ROPP-2025}). In the pseudogap and U(1) FL$^\star$ phases, the chemical potential of the $\xi$ band $\mu_\xi$ fixes hole doping $\delta$ for the system such that  $N_s^{-1} \sum_{ij,\sigma}  \langle \xi^\dagger_{ij, \sigma}\xi_{ij, \sigma} \rangle = \delta$. 
Our model in Eq.~(\ref{eq:action-S}) shows the U(1) gauge symmetry: $f_{i\sigma} \to f_{i\sigma} e^{i\theta_i},\,b_i \to b_i e^{i\theta_i},\,\chi_{ij} \to \chi_{ij} e^{i(\theta_i-\theta_j)}$ and $\xi_{ij,\sigma} \to \xi_{ij,\sigma} e^{i(\theta_i+\theta_j)}$.

In Ref.~\cite{YYC-ROPP-2025}, the one-loop perturbative renormalization group (RG) analysis was applied to study the quantum phase transition of the action (Eq.~(\ref{eq:action-S}) and Fig.~2 in Ref.~\cite{YYC-ROPP-2025}) due to competition between charge fluctuations of the hopping $t$ (effective Kondo) term and the RVB fluctuations (the $J$ term). 
It is convenient to define the dimensionless bare coupling constants $g \to g/D$, $J \to J /D$, and absorb $1/\zeta$ in $\mathcal{G}_\xi$ into $g$ by a rescaling, $g \to \bar{g} = g/\sqrt{\zeta}, \, \xi \to \sqrt{\zeta} \xi $. Our RG analysis is perturbatively controlled as $\bar{g}<1$ 
(the bare value $g<g_0^*$ is set), and $J<1$. Our RG approach can be further controlled by the $\epsilon$-expansion technique with a small parameter $\epsilon=d-z$ within the  convergence radius $|\epsilon|\leq 1$. We set the dynamical exponent $z=1$ due to the linearized dispersion of $f$, and spatial dimension $d=2$ here. 

Following the above RG approach, a non-trivial quantum critical point (QCP) was predicted, separating all the above-mentioned four mean-field phases (U(1) FL*, pseudogap, superconducting and FL phases). 
In the strange metal state with $\langle b\rangle=\Delta=0$, the gauge-invariant (physical) electron operator can be constructed from $\xi_{ij}$ and $\varphi_{ij}$ as $c= \xi \sqrt{2} \varphi_{ij}^* \sqrt{J}$. The dynamical scattering (relaxation) rate at $T = 0$ is calculated via the imaginary part of the electron self-energy: $\tau(\omega) =-\hbar /2\Sigma_c^{\prime\prime}(\omega)$ and $\Sigma_c^{\prime\prime}(\omega)=(1/2) \Sigma^{\prime\prime}_{\xi}(\omega)$,  where $\Sigma^{\prime\prime}_{\xi}(\omega)$ at $T=0$ is obtained via RG renormalized second-order perturbation close to the QCP, where the bare couplings are replaced by the renormalized ones i.e. $\bar{g}, \, \bar{g}^*_0\to \bar{g}(l)$ with $l \equiv -\ln(\bar D) > 0$ being the RG running cutoff length scale and $\bar{D} < D$ being the RG running cutoff energy scale \cite{YYC-ROPP-2025}. 
As a result, close to the predicted QCP an exact
cancellation of the coupling constant $g$ in the 
self-energy $\Sigma_{f/\xi}$ occurs between $1/g^2$ in $G_{\xi}$ and $g^2$ in the corresponding Feynman diagram of the self-energy, i.e. the coupling constant dependence of the prefactor $g^2/\zeta$ in $\Sigma_{f/\xi}$ gets exactly cancelled under RG:  $g^2/\zeta \to D^2 \bar{g}^2(l)/\bar{g}^2(l) \to D^2$ (see \Autoref{eq:Sigma_xi} in Appendix \ref{app:srfxi} and Fig.~5). This leads to a universal local Marginal-Fermi-Liquid-like self-energy insensitive to couplings, including a constant and a linear-in-$|\omega|$ term, i.e.  $\Sigma_{\xi}^{\prime\prime}(\omega)= \alpha-\varsigma\left|\omega\right|$ with $\alpha \approx \frac{5}{2}\bar{g}^{2}\rho_{0} = \frac{5}{2} D$ and $ \varsigma=\frac{2}{\pi}$; and further leads to a coupling-constant independent Planckian scattering rate with universal quantum-critical $\omega/T$-scaling \cite{YYC-ROPP-2025} (see the main text). This Planckian metal state becomes a quantum critical "phase", extending to the entire U(1) FL* phase, in excellent agreement with universal Planckian scattering rate observed in cuprates in DC-transport over an extended doping range \cite{Ramshaw-MottPlanckian-arxiv-2024}.

We now generalize Eq.~(\ref{eq:action-S}) in magnetic fields by including a Zeeman term as
\begin{eqnarray}
    S(B) &=& S(B=0) + H_B
\end{eqnarray}
with $S(B=0) = S$ given by Eq.~(\ref{eq:action-S}) and $H_B$ being a Zeeman Hamiltonian
\begin{eqnarray}
    H_B &=& - \mu_BB_z \sum_k \left[
    \left( f_{k\uparrow}^{\dagger} f_{k\uparrow} -
    f_{k\downarrow}^{\dagger} f_{k\downarrow} \right) \right. \nonumber \\
    && + \left.\left( \xi_{k\uparrow}^{\dagger} \xi_{k\uparrow} - \xi_{k\downarrow}^{\dagger} \xi_{k\downarrow} \right) \right].
    \label{eq:HB}
\end{eqnarray}

With \Autoref{eq:HB} included,
the energies of the $f$-spinon band ($\epsilon_f$) and
$\xi$-fermion band ($\epsilon_{\xi}$) are shifted by the Zeeman energy: $\epsilon_{f,\xi}^{\sigma} \to \epsilon_{f,\xi}^{\sigma} -s_{\sigma}\mu_BB$ with $s_{\sigma} = \pm 1$ for spin-up and spin-down fermion, respectively.

\section{Magneto--scattering rate for conduction $\xi$-band and physical $c$-electron band}\label{app:srfxi}

\begin{figure}[H]
	\begin{center}
	\begin{minipage}[b]{1\textwidth}
     {
 				\includegraphics[width=0.45\linewidth]{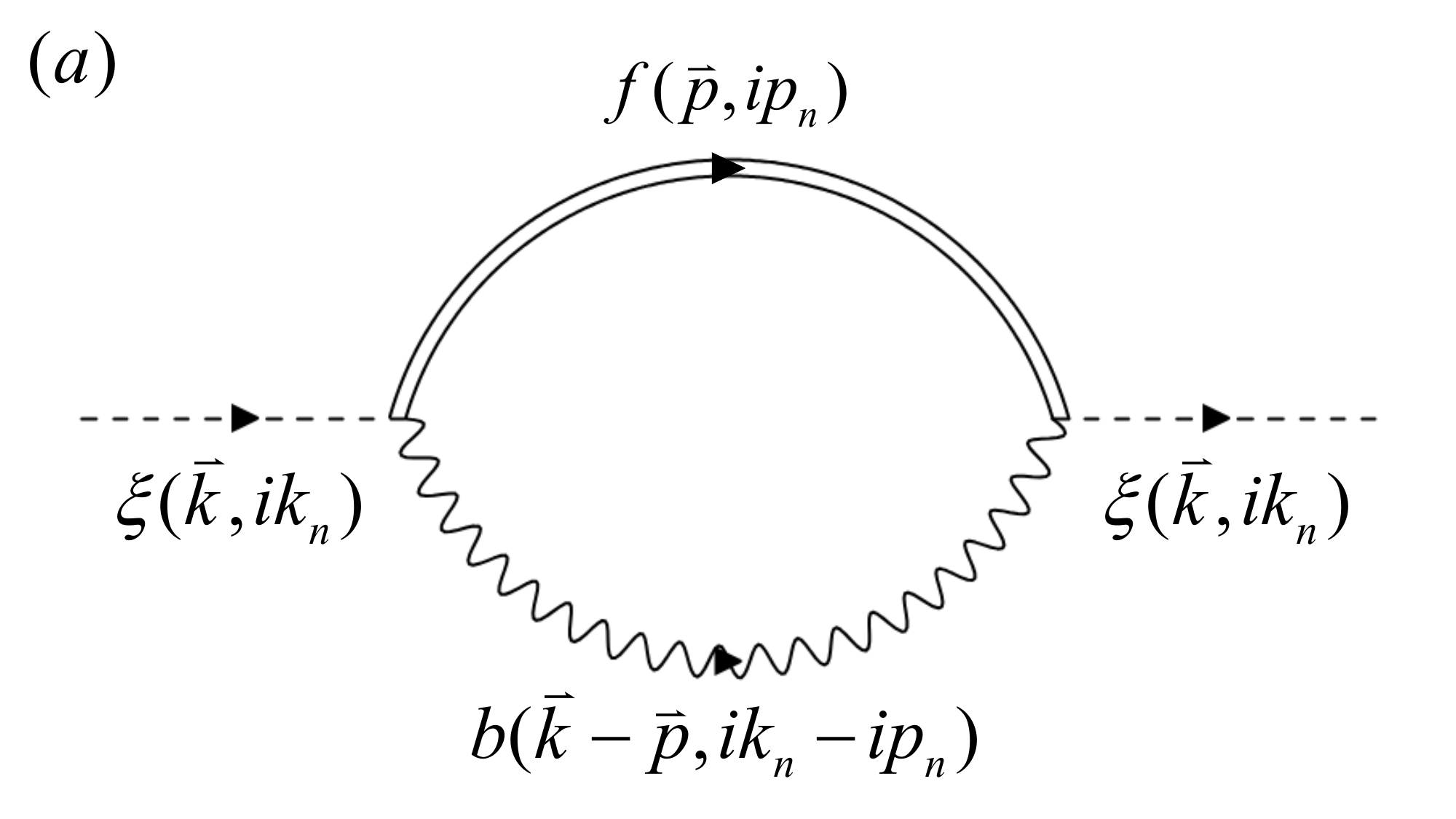}
			}\\
 	{
 				\includegraphics[width=0.48\linewidth]{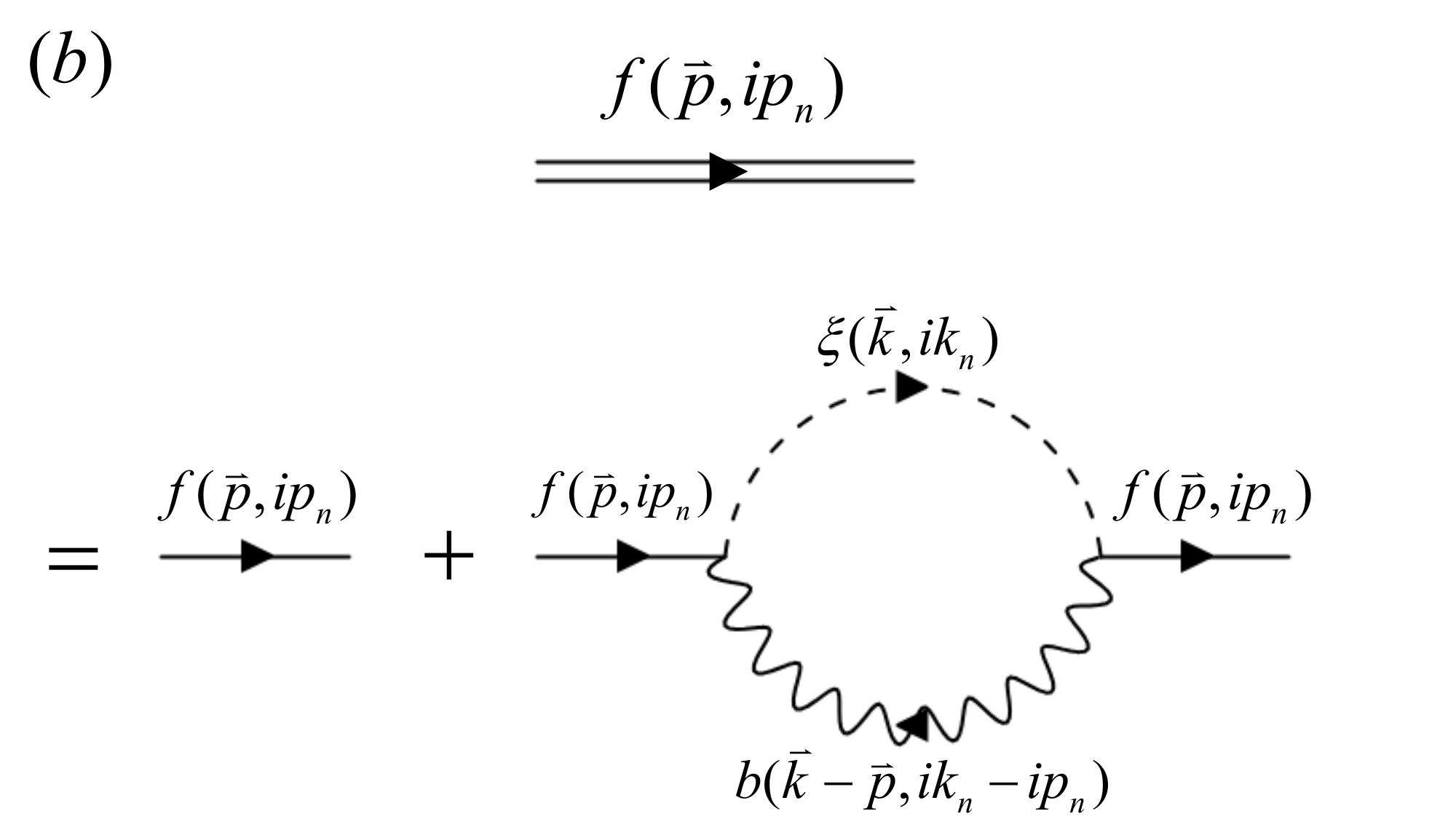}
			}
		\end{minipage}
	\end{center}
 	\caption{The self-energies (a) $\Sigma_{\xi}$ and (b) $\Sigma_f$, where the solid lines represent the spinon fields $G_{f}^{\sigma}$, the dashed lines the slave--boson fields ${G}_{b}$, and the wavy lines the $\xi$--fermions fields $G_{\xi}^{\sigma}$.
	}
	\label{fig:fse}
 \end{figure}

In this section, following RG renormalized second-order perturbation in Ref.~\cite{YYC-ROPP-2025}, we calculate the magneto--scattering rate for conduction $\xi$-band and the corresponding physical $c$-electron band by computing the local self-energy of the $\xi$-fermions near local quantum criticality. 
Here, we restrict ourselves to the isotropic scattering where only the isotropic conduction electron states near the Fermi surface are considered (\Autoref{fig:alpha_B_T} Inset). At zero field, this  approximation is further justified and supported by the experimental observation that the Planckian scattering rate is contributed from the isotropic component of scattering rate \cite{2022-Gael-Nature-isotropic-Planckian}. At a finite field, the Zeeman splitting effect will contribute to the isotropic component of the magneto-scattering rate.
The local self-energy of the $\xi$ field is given by (see \Autoref{fig:fse}) 
\begin{align}
    \Sigma_{\xi}(ik_{n})	& =
    \left(\frac{g^{2}}{N_{s}\zeta\beta}\right) \frac{1}{2}
    \sum_{p,\sigma}G_{f}^{\sigma}(p) {G}_{b}(k-p) \nn
	& \approx \left(\frac{g^{2}}{\zeta\beta}\right)\sum_{ip_{m}}\left(\frac{1}{N_{s}}\sum_{\bm{p}}G_{f}(p)\right) \mathcal{G}_{b}(k-p),
 \label{eq:Sigma_xi}
\end{align}
where $G_f$ denotes the self-energy corrected Green's function of $f$-spinon and $\mathcal{G}_b$ stands for the bare Green's function of the slave-boson $b$-field; 
and the momentum and Matsubara frequency are defined via the compact notation $p \equiv (ip_m,\bold{p})$. 
Here, the slave-boson $b$-fields are initially (at the bare level) treated as local impurity-like boson fields with no Bose condensation. In
the wide-band limit the narrow band of the $b$-fields $(\epsilon_b(k))$ perturbatively generated through self-energy correction is negligible $(\epsilon_b(k) \ll D,\,\lambda)$, and under one-loop RG, the local self energy of the $b$-field $\Sigma_b(\omega)$ does not lead to logarithmic divergence in running cutoff scale $(\Sigma_b \propto [l]^0)$ \cite{YYC-ROPP-2025}. Therefore, 
within the one-loop RG scheme, the slave-boson Green's function is approximated as the bare Green's function $\mathcal{G}_b$ in \Autoref{eq:Sigma_xi}.
Note that as mentioned above, under RG, an exact cancellation of coupling constant dependence in $\Sigma_{\xi/f}$ occurs: 
$g^2/\zeta \to D^2 \bar{g}^2(l)/\bar{g}^2(l) \to D^2$.

Following perturbative analysis shown in Ref.~\cite{YYC-ROPP-2025}, in the wide-band limit $D/\omega \gg 1$,
the Zeeman splitting of the self-energies
effectively leads to shifts of fermionic frequencies $\omega \to \omega + s_{\sigma}\mu_BB$ for both $f$-spinons and $\xi$-fermions.

To proceed with the calculation, we first compute the self-energy $\Sigma_f^{\sigma}$ in $G_f$ in \Autoref{eq:Sigma_xi} as follows.
\begin{align}
    &\Sigma_f^{\sigma}(k_n,k,B,\omega_n) \nonumber\\
    &= -\frac{g^2}{2N_s\beta}\sum_{p_n,p} 
    \mathcal{G}_{\xi}(p_n,p,B)
    \mathcal{G}_b(p_n-\omega_n,p-k),
    \label{eq:Sigma_f0}
\end{align}
where $\mathcal{G}_{\xi}=\frac{1}{\zeta} 
\frac{1}{ip_n-\left(\epsilon_{\xi}(p)-\mu_{\xi}\right)
-s_{\sigma}\mu_BB-i\Sigma_{\xi}^{\prime\prime}(p_n)}$ and 
$\mathcal{G}_b(p)=\frac{1}{ip_n-\lambda}$ are bare (initial) Green's functions of $\xi$-field and $b$-field, respectively, with $\zeta=g^2/D^2$, $\epsilon_{\xi}(p)-\mu_{\xi}=-\left(\epsilon_p-\bar{\mu}_{\xi}\right)$ with 
$\bar{\mu}_{\xi}=-\mu_{\xi}$, 
and $\Sigma_{\xi}^{\prime\prime}(p_n)=\pi D\sgn(p_n)$. 
Thus we obtain 
\begin{align}
    &\Sigma_f^{\sigma}(k_n,k,B,\omega_n) 
    = -\frac{g^2}{2\zeta N_s\beta}\sum_{p_n,p} 
    \frac{1}{ip_n-i\omega_n-\lambda} \nonumber\\
    &\times
    \frac{1}{ip_n+\left(\epsilon_p-\bar{\mu}_{\xi}\right)
    -s_{\sigma}\mu_BB-i\pi D\sgn(p_n)}.
    \label{eq:Sigma_f1}
\end{align}
Due to the $\sgn$ function in $\Sigma_{\xi}^{\prime\prime}$, 
\Autoref{eq:Sigma_f1} can be expressed as 
$\Sigma_f^{\sigma} = \Sigma_f^{\sigma >}+\Sigma_f^{\sigma <}$, where $\Sigma_f^{\sigma\gtrless}$ corresponds to 
$p_n \gtrless 0$ branch in the Matsubara sum.
\begin{align}
    &\Sigma_f^{\sigma\gtrless}(k_n,k,B,\omega_n) 
    = -\frac{g^2}{2\zeta N_s\beta}\sum_{p_n \gtrless 0} \sum_{p} \frac{1}{ip_n-i\omega_n-\lambda} \nonumber\\
    &\times
    \frac{1}{ip_n+\left(\epsilon_p-\bar{\mu}_{\xi}\right)
    -s_{\sigma}\mu_BB \mp i\pi D}.
    \label{eq:Sigma_f2}
\end{align}
Following the analysis in Appendix E of Ref.~\cite{YYC-ROPP-2025}, we have
\begin{align}
    &\Sigma_f^{\sigma >}=\frac{D^2\rho_0}{2}\nonumber\\
    &\times
    \int_{-D+\bar{\mu}_{\xi}}^{D+\bar{\mu}_{\xi}} d\epsilon'
    \frac{n_F(\epsilon'+s_{\sigma}\mu_BB+i\pi D)}
    {i\omega_n-\epsilon'+\lambda-s_{\sigma}\mu_BB-i\pi D},
    \label{eq:Sigma_fgt}
\end{align}
which in the $T \rightarrow 0$ limit becomes
\begin{align}
    \Sigma_f^{\sigma >}=\frac{D^2\rho_0}{2}
    \int_{-D+\bar{\mu}_{\xi}+s_{\sigma}\mu_BB}^{s_{\sigma}\mu_BB} 
    \frac{d\bar{\epsilon}}
    {i\omega_n-\bar{\epsilon}+\lambda-i\pi D},
    \label{eq:Sigma_fgt0K0}
\end{align}
where $\epsilon' \equiv -\epsilon+\bar{\mu}_{\xi}$ 
and $\bar{\epsilon} \equiv \epsilon'+s_{\sigma}\mu_BB$. 
Note that under RG, 
$g^2/\zeta \to D^2 \bar{g}^2(l)/\bar{g}^2(l) \to D^2$. 
Performing the analytical continuation by absorbing $\lambda$ into $\omega$, 
$i\omega_n \rightarrow \omega-\lambda+i\eta$ for 
$\eta \rightarrow 0^+$, we get (in the $T \rightarrow 0$ limit) 
\begin{align}
    &\left[\Sigma_f^{\sigma >}(\omega>0)\right]^{\prime\prime} \nonumber\\
    &=\frac{\pi D^2}{2}
    \int_{-D+\bar{\mu}_{\xi}+s_{\sigma}\mu_BB}^{s_{\sigma}\mu_BB} 
    \frac{d\epsilon}
    {(\omega-\epsilon)^2+(\pi D)^2} \nonumber\\
    &\cong -\frac{1}{2\rho_0} \left[
    \frac{\omega-s_{\sigma}\mu_BB}{\pi D}-
    \tan^{-1}\frac{1}{\pi}\right] \nonumber\\
    &= \frac{\alpha}{2}-\frac{1}{2\pi}
    (\omega-s_{\sigma}\mu_BB),
    \label{eq:Sigma_fgt0K}
\end{align}
where $\alpha = D\tan^{-1}\frac{1}{\pi} \approx D/3$. 
Similarly, we find 
\begin{align}
    &\left[\Sigma_f^{\sigma <}(\omega<0)\right]^{\prime\prime} 
    =-\left[\frac{\alpha}{2}-\frac{1}{2\pi}
    (\omega-s_{\sigma}\mu_BB)\right].
    \label{eq:Sigma_flt0K}
\end{align}
Combining $\Sigma_f^{\sigma\gtrless}(\omega\gtrless 0)$, we obtain (in the $T \rightarrow 0$ limit)  
\begin{align}
    &\left[\Sigma_f^{\sigma}(\omega)\right]^{\prime\prime} 
    =\frac{\alpha}{2}\sgn(\omega)-\frac{1}{2\pi}
    \left|\omega-s_{\sigma}\mu_BB\right|.
    \label{eq:Sigma_f3}
\end{align}
Finally, the total self-energy in the $T \rightarrow 0$ limit is the sum of contributions from both spins 
\begin{align}
    &\Sigma_f^{\prime\prime}(\omega,B) =\sum_{s_{\sigma}=\pm}
    \left[\Sigma_f^{\sigma}(\omega,B)\right]^{\prime\prime} 
    \nonumber\\
    &=\alpha\sgn(\omega)-\frac{1}{2\pi}
    \left(|\omega-\mu_BB|+|\omega+\mu_BB|\right).
    \label{eq:Sigma_f4}
\end{align}
Above the unit of $\hbar=1$ has been used; we will restore $\hbar$ when necessary. 
Based on \Autoref{eq:Sigma_f3}, we conclude that the effect of Zeeman field on the self-energy $\Sigma_f^{\sigma}(\omega,B) = \Sigma_f^{\sigma}(\omega \to \omega-s_{\sigma}\mu_BB)$ is equivalent to a shift in fermionic frequency $\omega$: $\omega \to \omega-s_{\sigma}\mu_BB$. 
This shift also applies to the self-energy $\Sigma_{\xi}^{\sigma}(\omega)$.

We now compute $\Sigma_{\xi}^{\prime\prime}(\omega)$ via \Autoref{eq:Sigma_xi} at T=0. Following the analysis in Appendix E of Ref.~\cite{YYC-ROPP-2025}, we have the following relation 
\begin{align}
    \Sigma_{\xi}^{\prime\prime}(\omega,B=0) 
    =-A+2\Sigma_f^{\prime\prime}(\omega,B=0) 
    \label{eq:Sigma_xi1}
\end{align}
with $A$ being an un-important constant. Based on the above result for $\Sigma_f^{\prime\prime}(\omega,B)$, we can generalize \Autoref{eq:Sigma_xi1} at a finite field $B \neq 0$ as 
\begin{align}
    \left[\Sigma_{\xi}^{\sigma}(\omega,B)\right]^{\prime\prime} 
    =\frac{1}{2}\Sigma_{\xi}^{\prime\prime}
    (\omega \to \omega-s_{\sigma}\mu_BB). 
    \label{eq:Sigma_xi2}
\end{align}
The total self-energy $\Sigma_{\xi}^{\prime\prime}(\omega,B)$ reads 
\begin{align}
    &\Sigma_{\xi}^{\prime\prime}(\omega,B) =\sum_{s_{\sigma}=\pm}
    \left[\Sigma_{\xi}^{\sigma}(\omega,B)\right]^{\prime\prime} 
    \nonumber\\
    &=-A+2\sum_{s_{\sigma}=\pm}
    \left[\Sigma_{f}^{\sigma}(\bar{\omega} = \omega-s_{\sigma}\mu_BB,B)\right]^{\prime\prime}.
    \label{eq:Sigma_xi3}
\end{align}
Plugging \Autoref{eq:Sigma_f4} into \Autoref{eq:Sigma_xi3}, 
since $\left[\Sigma_{f}^{\sigma}(\bar{\omega},B)\right]^{\prime\prime}
=\alpha\sgn(\omega)/2-
|\bar{\omega}-s_{\sigma}\mu_BB|/2\pi$ (\Autoref{eq:Sigma_f3}), we finally arrive (by restoring $\hbar$) 
\begin{align}
    &\Sigma_{\xi}^{\prime\prime}(\omega,B) \nonumber\\
    &=-\frac{A}{\hbar}+2\sum_{s_{\sigma}=\pm}
    \left(\frac{\alpha}{2\hbar}\sgn(\omega)-
    \frac{1}{2\pi}\left|\omega
    -\frac{2s_{\sigma}\mu_BB}{\hbar}\right|\right) \nonumber\\
    &=-\frac{A}{\hbar}+\sum_{s_{\sigma}=\pm}
    \left(\frac{\alpha}{\hbar}\sgn(\omega)-
    \frac{2}{\pi}\left|\frac{\omega}{2}
    -\frac{s_{\sigma}\mu_BB}{\hbar}\right|\right).
    \label{eq:Sigma_xi4}
\end{align}

Note that in deriving \Autoref{eq:Sigma_xi4}, the frequency $\omega$ in $\Sigma_{\xi}^{\prime\prime}$ has been shifted twice by the Zeeman splitting: one in \Autoref{eq:Sigma_f3} by $f$-fermions, and another in \Autoref{eq:Sigma_xi2} by $\xi$-fermions.

The physical gauge-invariant electron operator $c_i$ is defined
via $\xi_{ij}$-fermion as $c_i = \sum_j \xi_{ij} \phi_{ij}\sqrt{J}$
where $\phi_{ij}$-fields represent fluctuating spinon pairing $\Delta_{ij}$-fields and $G_{\phi} = 1/J$ \cite{YYC-ROPP-2025}.
It has been shown in Ref.~\cite{YYC-ROPP-2025} that
$\Sigma_{c}^{\prime\prime} = \Sigma_{\xi}^{\prime\prime}/2$.
As a result, at $T=0$, we have
\begin{align}
\frac{1}{\tau^c_B} = -2\Sigma_{c}^{\prime\prime}
=-\Sigma_{\xi}^{\prime\prime}(\omega,B).\label{eq:scat_T0}
\end{align}

At finite temperatures and fields, we perform conformal transformation on $\Sigma_c$, similar to the case at zero-field\cite{YYC-ROPP-2025, YYC-arXiv-2025}, leading to the scaling part of magneto--scattering rate:
\begin{align}
   \frac{\hbar/\tau_B^{s_{\sigma}}(\omega/T,B/T)}
   {\frac{2}{\pi}k_BT} = x^{s_{\sigma}}
   \coth\left(\zeta \frac{x^{s_{\sigma}}}{2}\right),
   \label{eq:mag-res-sca}
\end{align}
where $x^{s_{\sigma}} = (\hbar\omega/2 + s_{\sigma}\mu_BB)/k_BT$.

Note that in deriving \Autoref{eq:mag-res-sca} from \Autoref{eq:scat_T0} via conformal transformation, a tuning parameter $\zeta$, allowed by conformal mapping of scattering rate in the scaling regime from zero temperature to finite temperatures \cite{YYC-ROPP-2025,YYC-arXiv-2025} (see Appendix \ref{AppE}) is introduced to match the observed universal $B/T$--scaling in $\omega/T$ and $B/T$ in magneto--scattering rate (see \Autoref{eq:rho-tt} and \Autoref{fig:rho} of maintext) whose value is constrained by the $T$-linear Planckian coefficient $\alpha_T$  via 
via \Autoref{eq:tauB0},  \Autoref{eq:tauB}, and \autoref{eq:c1-vs-gamma2} \cite{YYC-ROPP-2025}.

After summing over the spin index $s_{\sigma} = \pm 1$ of \Autoref{eq:mag-res-sca}, the total magneto--scattering rate $\hbar/\tau_B$ (after subtracting the field independent $T$--linear residual scattering
rate $\hbar/\tau_0(T) = (8/\pi) (1-1/\zeta) k_B T$) is given by
\begin{align}
\frac{\hbar}{\tau_B} -\frac{\hbar}{\tau_0(T)}&=\nonumber\\
+\frac{2}{\pi}k_BT&\left((\frac{\hbar\omega/2 + \mu_BB}{k_BT})\coth(\zeta\frac{\hbar\omega/2 + \mu_BB}{2k_BT})\right.\nonumber\\
&+\left.(\frac{\hbar\omega/2 - \mu_BB}{k_BT})\coth(\zeta\frac{\hbar\omega/2 - \mu_BB}{2k_BT})\right)
\label{eq:Im-Sigma-xi-final5}
\end{align}
We now discuss the scaling behaviors of \Autoref{eq:Im-Sigma-xi-final5} in various limits.

(i) In the $B \to 0$ limit, \Autoref{eq:Im-Sigma-xi-final5} reduces to
\begin{align}
\frac{\hbar/\tau_B(B \to 0,\omega,T)}{k_BT} =
\frac{\hbar/\tau_0(T)}{k_BT} +
\frac{2}{\pi} \frac{\hbar\omega}{k_BT}
\coth\frac{\kim{\zeta}\hbar\omega}{4k_BT}.
\end{align}
This result may further reduce to the expressions in the following two limits:

(i)(a) The DC-limit ($B \to 0,\omega \to 0$) of $T$-linear Planckian scattering rate
\begin{align}
\frac{\hbar}{\tau_B}(B \to 0,\omega \to 0,T) =
\frac{\hbar}{\tau_0(T)} + \frac{8}{\pi\zeta} k_BT =
\alpha_T k_BT
\end{align}
with $\alpha_T = 8/\pi$.

(i)(b) The AC-limit ($B \to 0,T \to 0$) of $\omega$-linear scattering rate
\begin{align}
\frac{\hbar}{\tau_B}(B \to 0,T \to 0,\omega) =
\frac{2}{\pi} \hbar|\omega|.
\end{align}

(ii) In DC ($\omega \to 0$) limit, \Autoref{eq:Im-Sigma-xi-final5} reduces to \Autoref{eq:rate} of the main text
\begin{align}
\frac{\hbar/\tau_B(\omega \to 0,B,T)}{k_BT} =
\frac{\hbar/\tau_0(T)}{k_BT} +
\frac{4}{\pi} \frac{\mu_BB}{k_BT}
\coth\frac{\zeta\mu_BB}{2k_BT}.
\end{align}
This result may further reduce to the expressions in the following two limits:

(ii)(a) The zero-field, DC-limit ($\omega \to 0,B \to 0$) of $T$-linear Planckian scattering rate
\begin{align}\label{eq:tauB0}
\frac{\hbar}{\tau_B}(\omega \to 0,B \to 0,T) =
\frac{\hbar}{\tau_0(T)} + \frac{8}{\pi\zeta} k_BT =
\alpha_T k_BT.
\end{align}

(ii)(b) The $T \to 0$, DC-limit ($\omega \to 0,T \to 0$) of $B$-linear Planckian scattering rate
\begin{align}\label{eq:AppB_31}
\frac{\hbar}{\tau_B}(\omega \to 0,T \to 0,B) =
\alpha_B \mu_BB
\end{align}
with $\alpha_B = 4/\pi$.

In the exact $B=0$ limit, the $\omega/T$-scaling of the AC-scattering rate of LSCO at $p=0.24$ has been observed in optical conductivity \cite{George-NatComm-SM} and theoretically studied in Ref.~\cite{YYC-ROPP-2025}, the result is
\begin{align}\label{eq:AppB_32}
\frac{\hbar/\tau(B=0,\omega,T)}{k_BT} =
\frac{2}{\pi} \frac{\hbar\omega}{k_BT}
\coth\frac{\hbar\omega}{4k_BT}\; .
\end{align}

This result reduces to the following limiting cases:

(a) The DC-limit ($\omega \to 0$) of $T$-linear Planckian scattering rate
\begin{align}\label{eq:AppB_33}
\frac{\hbar}{\tau}(B=0,\omega \to 0,T) = \alpha_T k_BT.
\end{align}

(b) The AC-limit ($B=0,T \to 0,\omega$) of $\omega$-linear scattering rate
\begin{align}\label{eq:AppB_34}
\frac{\hbar}{\tau}(B=0,T \to 0,\omega) = \frac{2}{\pi} \hbar|\omega|.
\end{align}

It is clear that all the above results in various limits are consistent with each other. 

Note that the value of the tuning parameter $\zeta$ only modify the transport properties at finite temperatures and finite fields, in particular the $B/T$--scaling behaviors in scattering rate and magneto--resistivity; however, it would not affect the transport coefficients both in the $T=0$ limit (see \Autoref{eq:AppB_31}, and \Autoref{eq:AppB_34}) and the exact zero-field ($B=0$) limit (see \Autoref{eq:AppB_32}, and \Autoref{eq:AppB_33}).

\section{Universal scaling function $f(x)$}\label{usffx}

In this Appendix, we explain how we obtained the slope of the magneto--resistivity data as well as how we calculated the theoretical B/T--scaling of the slope.

\subsection{Slope derivation for the experimental data of the magneto--resistivity in LSCO}
Here we provide details concerning the data analysis performed to obtain the $B/T$-scaling of the slope of the magneto--resistance as shown in \Autoref{fig:drhoBT}.
Raw data for doping $p\in [0.161,0.19]$ is taken from the available resources of \cite{Beobinger-2018-Science-LSCO}.

Simply taking the discrete derivative of the raw data leads to very important fluctuations in the slope values. In order to avoid these large fluctuations, we decided to perform a slope window average analysis. The slope window average analysis consists in taking a regression line for points within an interval set by the window width $w$.
\begin{equation}
\frac{\dd \rho(B_{i})}{\dd B_{i}}=\textrm{a of Fit}[a* B+c=\rho(B): B\in [B_{i}-w,B_{i}+w]]\; . \label{eq:w_av}
\end{equation}

\begin{figure}[H]
	\begin{center}
	\begin{minipage}[b]{1\textwidth}
 			{
 				\includegraphics[width=0.48\linewidth]{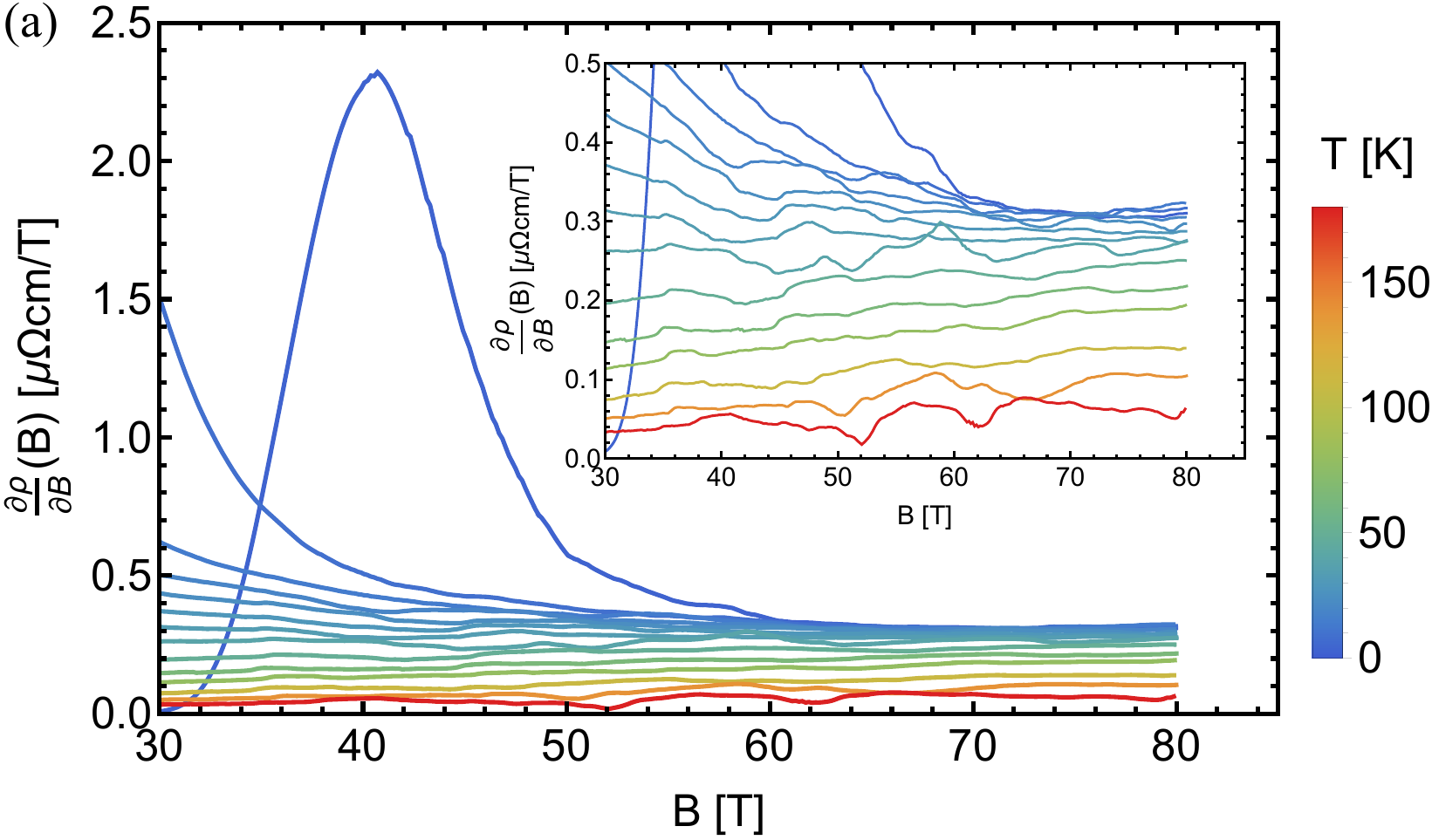}
			}\\
			{
				\includegraphics[width=0.48\linewidth]{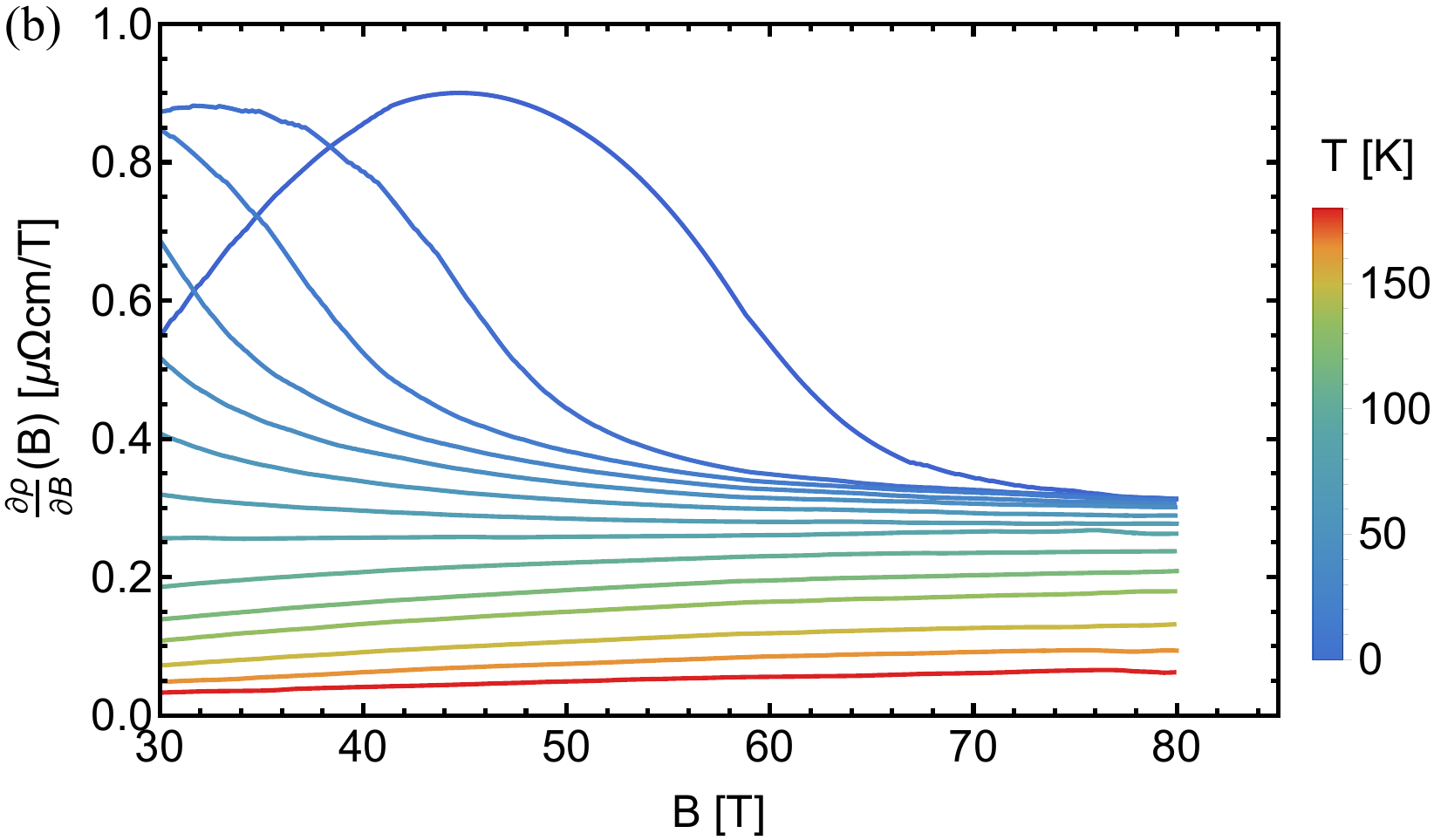}				
			}
		\end{minipage}
	\end{center}
 	\caption{
    Slope window analysis of the magneto--resistivity data for $p=0.19$  using \protect\Autoref{eq:w_av} with (a) $w=5T$ and (b) $w=22T$ While smaller window values still contain fluctuations, larger values exaggerate the superconductivity fluctuations at higher magnetic field values. The inset in (a) shows a zoom withing the invervall $[0,0.5] \mu\Omega \rm{cm}/T$ of the ordinate axis, where the slope fluctuations are visible
    }
	\label{fig:drhoBT_19_w}
 \end{figure}

To illustrate this method, we plot in \Autoref{fig:drhoBT_19_w} the slope window of the magneto--resistivity using \Autoref{eq:w_av} for $p=0.19$ with $w=5T$ (\Autoref{fig:drhoBT_19_w}~(a)) and $w=22T$ (\Autoref{fig:drhoBT_19_w}~(b)). We easily see that for smaller window widths some fluctuations persist, whereas for $w=22T$ (\Autoref{fig:drhoBT_19_w}~(b)) the slope values are very smooth. Additionally, taking a very large window width shifts the peak in the derivatives, sign of the transition towards superconductivity, towards higher values of magnetic fields and reduces the amplitude of the peak. Indeed, for windows widths which are very wide and include the transition to superconductivity,  the values obtained from the linear regression, which is only valid in the strange metal phase (linear regime), will be affected by the inclusion of data which are in the superconducting state.

\begin{figure}[H]
	\begin{center}
	\begin{minipage}[b]{1\textwidth}
                {
 				\includegraphics[width=0.48\linewidth]{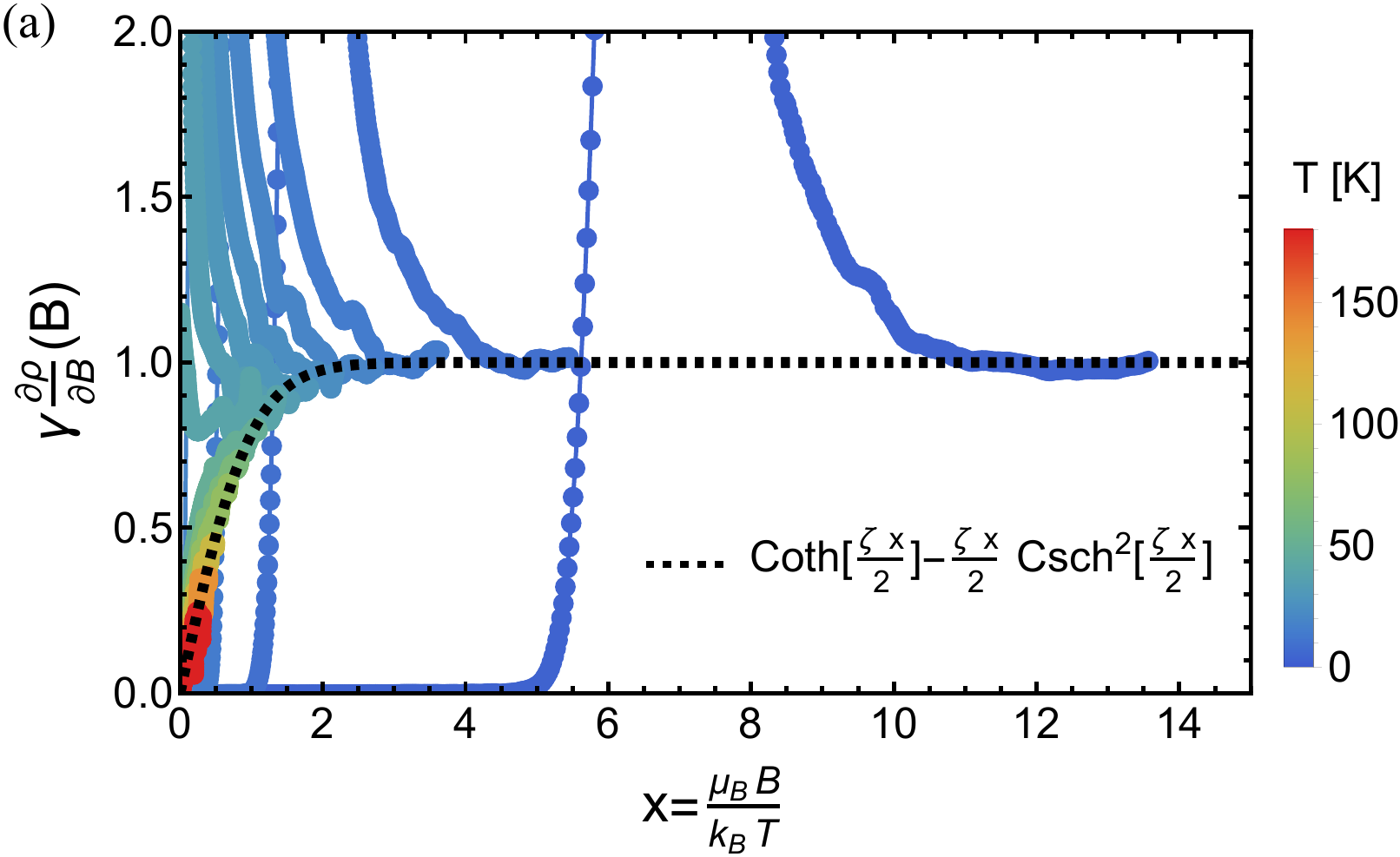}}
                \\
                {
				\includegraphics[width=0.48\linewidth]{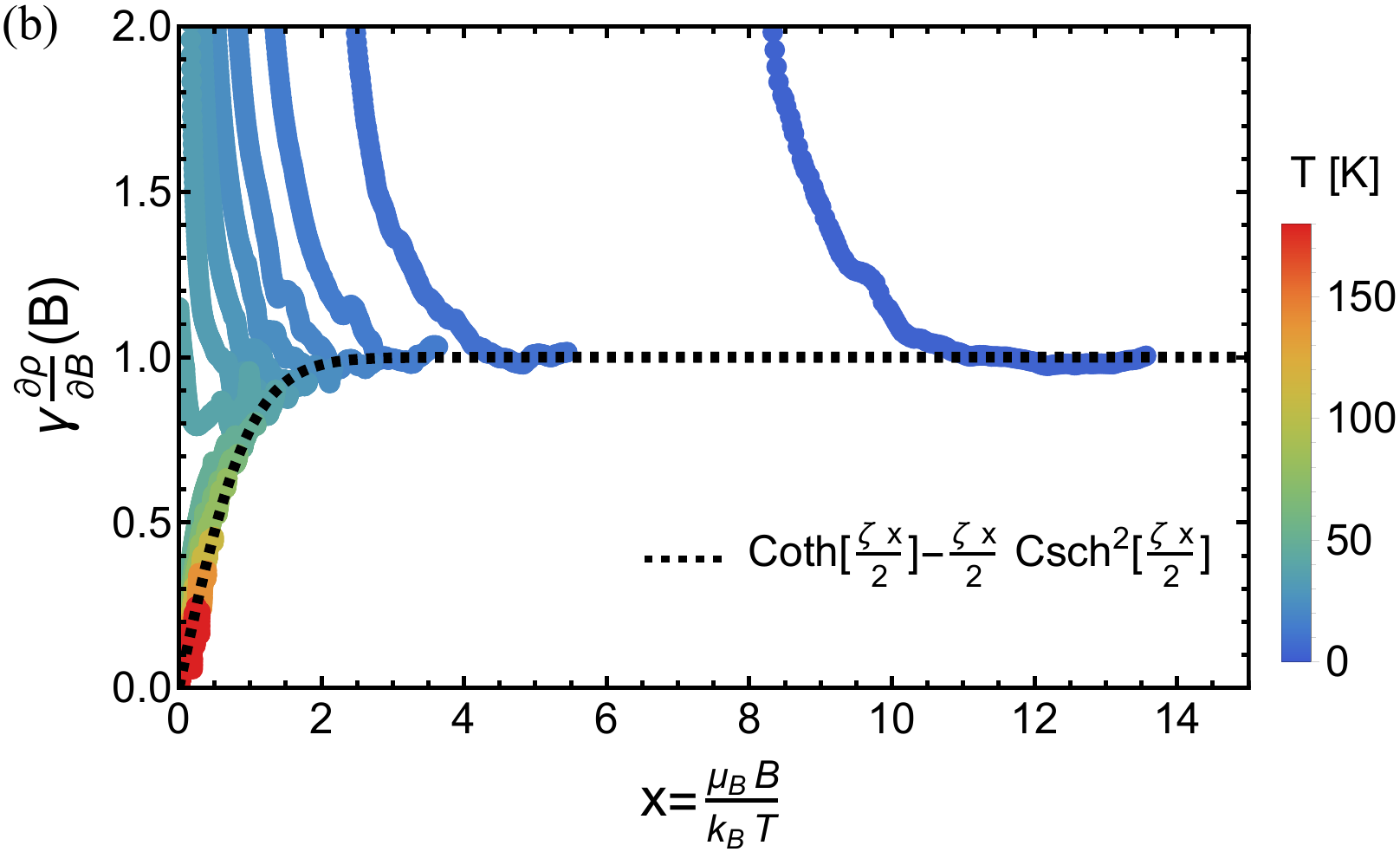}
                }
		\end{minipage}
	\end{center}
 	\caption{$B/T$-scaling  of the rescaled slope of the magneto--resistivity for $p=0.19$ where in the panel (a) is shown the slope of the magneto--resistivity with a window averaging of $w=5~T$ [\protect\autoref{eq:w_av}] with identical parameter as in \protect\autoref{fig:rho} and \protect\autoref{fig:drhoBT}: $\gamma=3.2 (\mu\rm{\Omega cm})^{-1}\rm{T}$, and $\zeta=3.06$.  The red dotted line corresponds to the theoretical prediction of the derivative of the magneto--resistivity given in \protect \autoref{eq:slope} with $\zeta=3.06$. In (b) is represented the same data as in (a) but points at lower magnetic field values in the superconducting phase have been truncated.
    }
	\label{fig:drhoBT_19}
 \end{figure}

For \Autoref{fig:drhoBT} of the main text and \Autoref{fig:drhoBT_19}, we choose to take the slope between values separated by a window width of  $w=5 T$, which corresponds to an interval of about 57 data points.
The slope of the magneto--resistivity is rescaled by a factor $\gamma$ and plotted against the rescaled energy $\frac{\mu_{B}B}{k_B T}$, where \mbox{$\gamma_{p=0.19}=3.2 (\mu\Omega cm/T)^{-1}$} and $\beta_{p=0.19}=3.06$  are fitting parameters. More precisely in  \Autoref{fig:drhoBT_19}~(a), we show the same data as in \Autoref{fig:drhoBT_19_w}~(a), but rescaled as explained above. For clarity, we only plot the higher magnetic field data points where the curves have a linear slope or start to become superconducting, but we discard the parts in the superconducting phase as shown in \Autoref{fig:drhoBT_19}~(b) for $p=0.19$. \Autoref{fig:drhoBT}~(b) is reproduced from the same data as \Autoref{fig:drhoBT_19}~(b).

\begin{figure}[H]
	\begin{center}
	\begin{minipage}[b]{1\textwidth}
 			{
 				\includegraphics[width=0.48\linewidth]{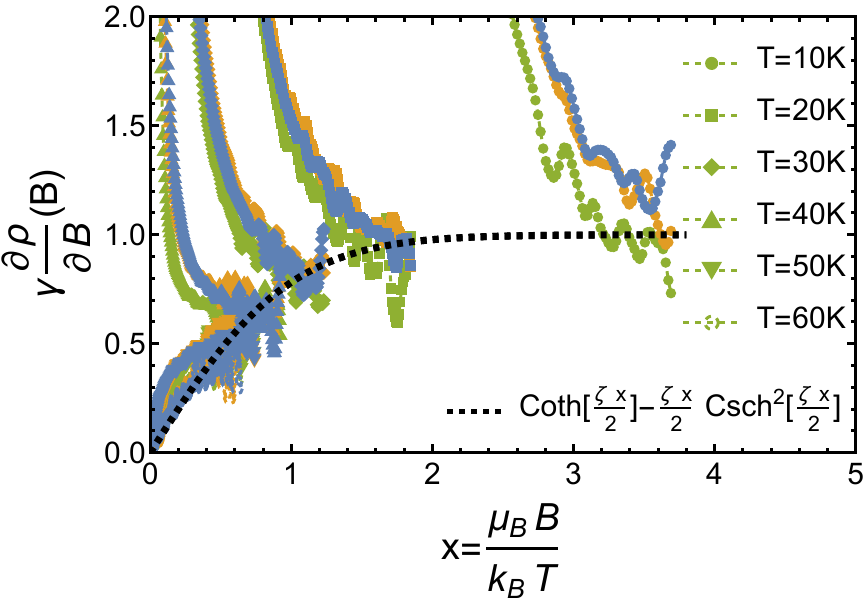}
			}
		\end{minipage}
	\end{center}
 	\caption{$B/T$--scaling  of the rescaled slope of the magneto--resistivity in function of $x=\mu_B B/(k_B T)$  for doping $p=0.161, 0.170, \textrm{and } 0.184$, using the window averaging procedure  with $w=1.5~T$ [\protect\autoref{eq:w_av}] for parameters given by  $\gamma_{p=0.161}=1.3 (\mu\rm{\Omega cm})^{-1}\rm{T}$,  $\gamma_{p=0.170}=1.6 (\mu\rm{\Omega cm})^{-1}\rm{T}$, $\gamma_{p=0.184}=1.9 (\mu\rm{\Omega cm})^{-1}\rm{T}$, and $\zeta=3.06$ for all three doping values.  Note that $\gamma$ depends on the carrier density $n=1/V$ as well as the effective mass [\protect\Autoref{eq:gamma}]  and is expected to vary for different doping values. The red dotted line corresponds to the theoretical prediction of the derivative of the magneto--resistivity given in \protect \autoref{eq:slope} with $\zeta=3.06$.
	}
	\label{fig:drhoBT_appx}
 \end{figure}

For $p=0.184$, $p=0.170$, and $p=0.161$, we follow a similar procedure, and we obtain the slope by window averaging with $w=1.5~T$ (which corresponds to about 55 data points) for the three doping values as shown in \Autoref{fig:drhoBT_appx}.

Unfortunately, the magneto--resistivity data for doping $p\in[0.161,\dots,0.184]$ were only taken up to 55~T, and for a temperature interval of $T\in [10,\dots,60]$. This makes the scaling of the experimental data harder to see. Indeed as shown in \Autoref{fig:drhoBT_appx}, the data just saturate on the scaling function (red curve). Lower temperature and higher magnetic field data would allow to recover the plateau as it is the case for $p=0.19$ (see \Autoref{fig:drhoBT_19}~(b) or \Autoref{fig:drhoBT}~(b)) .

\begin{figure}[H]
	\begin{center}
	\begin{minipage}[b]{1\textwidth}
 			\subfloat[$KHtJ$ prediction	\label{fig:drhoBT_19_B_fit_KHtJ}]{
 				\includegraphics[width=0.48\linewidth]{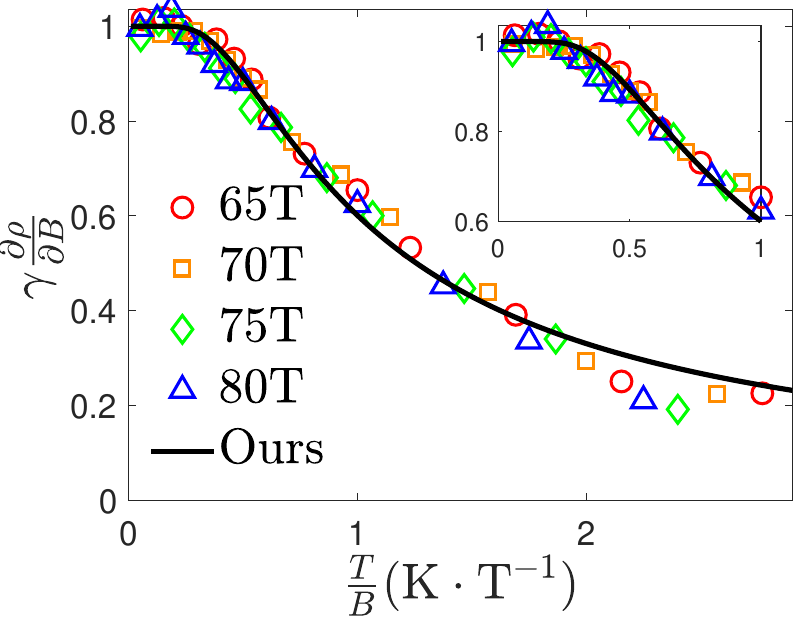}
			}\\
			\subfloat[$MFL$ prediction	\label{fig:drhoBT_19_B_fit_MFL}]{
				\includegraphics[width=0.48\linewidth]{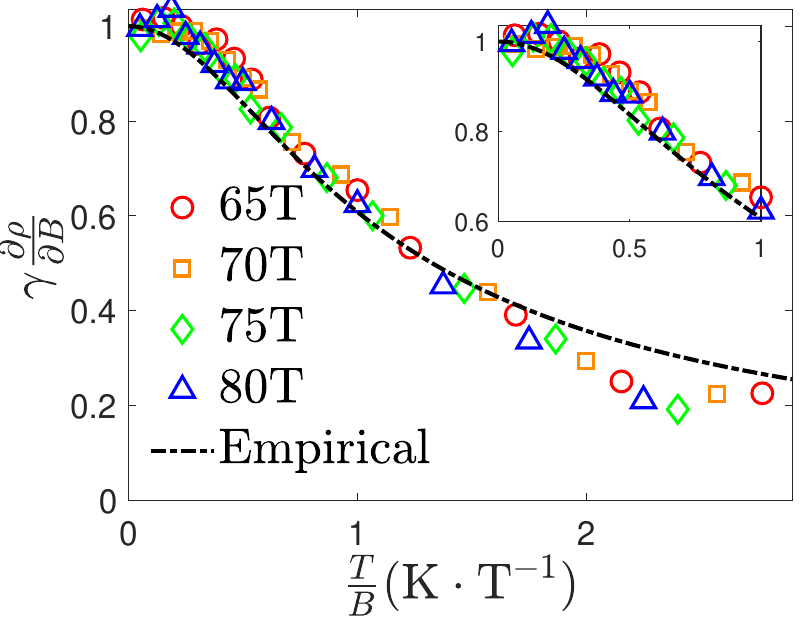}				
			}
		\end{minipage}
	\end{center}
 	\caption{(a) Rescaled magneto--resistivity slope values compared to our theoretical prediction [\protect\Autoref{eq:slope}], where the black line corresponds to $\left(\coth(\zeta x/2) - \zeta x/2 \csch^2(\zeta x/2)\right)$ with $x=\mu_B B/(k_B T)$,  $\gamma_{p=0.19}=3.2$  $(\mu\Omega \rm{cm/T})^{-1}$, and $\zeta=3.06$. (b) Same data set of the rescaled magneto--resistivity slope values as in (a), but compared to the marginal Fermi--liquid (MFL) approximation \cite{hussey-incoherent-nature-2021,Ayres-NatComm-HT} [\protect\Autoref{eq:rho_MFL}], where the dashed black line corresponds to
     $\gamma\dd \rho^{MFL}/\dd B=\sqrt{(a_1 \gamma T/B)^2+1}^{-1}$ with $\gamma=3.2$ $(\mu\Omega \rm{cm/T})^{-1}$ and $a_1=0.407$  $\mu\Omega \rm{cm/K}$. We note that the scaling function arising from our prediction allows to account for the small plateau observed at small $T/B$ values, while the MFL approximation shows a constant increase when approaching zero, as shown in the insets of (a) and (b). 
     }
	\label{fig:drhoBT_19_B_fit}
 \end{figure}

In \Autoref{fig:drhoBT_19_B_fit}, we show the values of the slope obtained by window average for $p=0.19$ at different magnetic field rescaled by $\gamma$ in function of $T/B$. In \Autoref{fig:drhoBT_19_B_fit}~(a) we compare the rescaled magneto--resistivity slope values with our theoretical prediction [\Autoref{eq:slope}], where the black line corresponds to $\left(\coth(\zeta x/2) - \zeta x/2 \csch^2(\zeta x/2)\right)$ with $x=\mu_B B/(k_B T)$,  \mbox{$\gamma_{p=0.19}=3.2 (\mu\Omega cm/T)^{-1}$}, and $\zeta=3.06$. In \Autoref{fig:drhoBT_19_B_fit}~(b), we show the same data set for the rescaled magneto--resistivity slope values as in (a), but we compare the data to the marginal Fermi--liquid (MFL)  empirical quadratic scaling approximation \cite{hussey-incoherent-nature-2021,Ayres-NatComm-HT}:
\be
\rho^{MFL}(B,T)=\rho_0^{MFL}(T)+\sqrt{(a_1T)^2+(B/\gamma)^2}\;.\label{eq:rho_MFL}
\ee
Taking the derivative with respect to magnetic field leads to
\be
\dd \rho^{MFL}/\dd B=\frac{1}{\gamma}\frac{1}{\sqrt{(a_1 \gamma T/B)^2+1}}\;,
\ee
where we set \mbox{$\gamma=3.2 (\mu\Omega \rm{cm/T})^{-1}$} (same value as for the scaling function) and where we find \mbox{$a_1=0.407 \mu\Omega \rm{cm/K}$} as the optimum parameter value. The dashed black line in \Autoref{fig:drhoBT_19_B_fit}~(b) corresponds to $\gamma\dd \rho^{MFL}/\dd B=\sqrt{(a_1 \gamma T/B)^2+1}^{-1}$

The insets in \Autoref{fig:drhoBT_19_B_fit}~(a) and (b) show a zoom where the data form a plateau at small $T/B$ values.  We note that the scaling function arising from our prediction allows to account well for this feature (inset of \Autoref{fig:drhoBT_19_B_fit}~(a)), whereas the MFL approximation shows a constant increase when approaching zero (inset of \Autoref{fig:drhoBT_19_B_fit}~(b)).

\begin{figure}[H]
    \centering
    \includegraphics[width=0.8\linewidth]{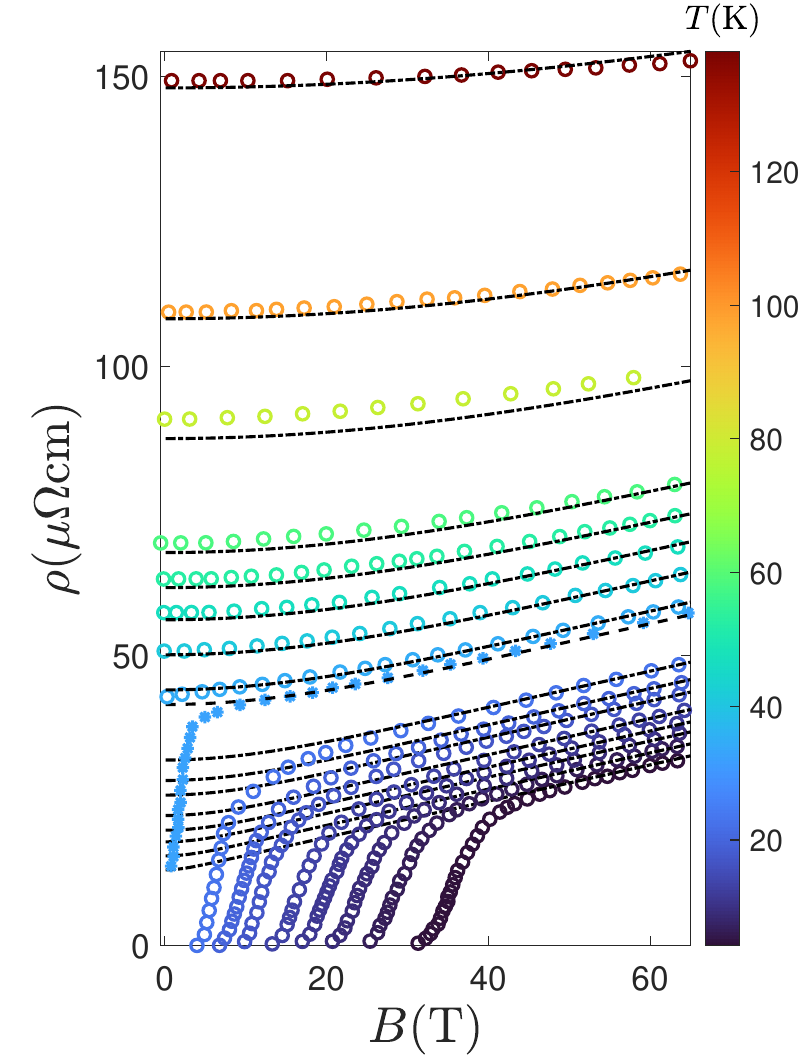}
    \caption{Magneto--resistivity data for $p=0.20$ versus magnetic field at different temperatures reproduced from \cite{shekhter2022_arXiv}.
    Colored symbols represent experimental data and black dotted lines correspond to the theoretical prediction given in \protect \autoref{eq:rho-tt} with parameter given by $c_0=8.82 \mu\rm{\Omega cm}$, $c_1=0.86\mu\rm{\Omega cm K^{-1}}$, $\gamma=3.2
    (\mu\rm{\Omega cm})^{-1}\rm{T}$ , $\zeta=6.27$.
    }
    \label{fig:rho_B_arkady}
\end{figure}

For completeness, in \Autoref{fig:rho_B_arkady}, we show magneto--resistivity for $p=0.20$, where the raw data is taken from \cite{shekhter2022_arXiv}. We follow a similar approach to what we performed for $p=0.19$ in \Autoref{fig:rho}~(b). We note that for $p=0.20$ magneto--resistivity data in the $B$-linear limits are well-fitted by \Autoref{eq:rho-tt}, with parameter given by $c_0=8.82 \mu\rm{\Omega cm}$, $c_1=0.86\mu\rm{\Omega cm K^{-1}}$, $\gamma=3.2 (\mu\rm{\Omega cm})^{-1}\rm{T}$ , $\zeta=6.27$. We note that some of these parameters are different from the $p=0.19$ case, which is not surprising since these two experiments were done independently and with different samples.
Deviations at low magnetic field values between theoretical predictions (dashed lines) and the data at low temperatures are interpreted as coming from the crossover into the superconducting state. Indeed, the deviation vanishes at higher magnetic field values and at a higher temperature well above $T_c$ where  superconductivity is absent.

\subsection{Analytical B/T--scaling of the slope of the magneto--resistivity} \label{sec:ana_slope}
Here we provide the theoretical calculation of the slope of the magneto--resistivity. We start by considering the results that we obtained for the scattering rate within the strange metal phase in the DC limit ($\omega\to0$)[\Autoref{eq:rate}]

\be
\frac{\hbar}{\tau(B,T)}-\frac{\hbar}{\tau_0(T)}=\frac{4}{\pi}k_BT\frac{\mu_B B}{k_BT}\coth(\zeta\frac{\mu_B B}{2k_BT})
\ee
In the zero temperature limit this leads to
\begin{align}
\lim_{T\to 0}\left[\frac{\hbar}{\tau(B,T)}-\frac{\hbar}{\tau_0(T)}\right]&=\frac{4}{\pi}k_BT\frac{\mu_B B}{k_BT}\\
&=\frac{4}{\pi}\mu_B B\\
&\equiv \alpha_B \mu_B B
\end{align}
where the linear--in--field Planckian coefficient is given by
\be
\alpha_B=\frac{4}{\pi} \label{eq:alphaB}
\ee
Similarly, In the zero field limit this leads to
\begin{align}
\lim_{B\to 0}\left[\frac{\hbar}{\tau(B,T)}-\frac{\hbar}{\tau_0(T)}\right]=\frac{8}{\pi}\frac{1}{\zeta}k_B T\\
\end{align}

Here, the purely temperature-dependent T-linear scattering rate $\hbar /\tau_0(T)$ corresponds to the purely temperature-dependent T-linear resistivity $\rho_0(T)$ observed in experiment (see \Autoref{eq:rho-tt} of main text) with $\rho_0(T) = m^*/(n e^2) 1/\tau_0$.

We therefore consider the temperature only contribution scattering rate $\hbar /\tau_0(T)$ to also be purely temperature-dependent with a T-linear dependence:
\be
\frac{\hbar}{\tau_0(T)} = b k_BT
\ee
 with $b = 8(1-1/\zeta)/\pi$ in order to retrieve the  the linear--in--temperature Planckian coefficient as follows:
\begin{align}\label{eq:tauB}
\lim_{B\to 0}\left[\frac{\hbar}{\tau(B,T)}\right]&=(1-\frac{1}{\zeta})\frac{8}{\pi} k_BT +\frac{8}{\pi}\frac{1}{\zeta}k_B T \\
&=\frac{8}{\pi} k_BT\\
&\equiv\alpha_T k_B T
\end{align}
with
\be
\alpha_T=\frac{8}{\pi} =2\alpha_B\label{eq:alphaT}
\ee
Using the DC Drude formula for resistivity, we have
\begin{align}
	(	\rho(B,T)-\rho_0(T))=&\frac{m^{\ast}}{\hbar n e^2}(\frac{\hbar}{\tau(B,T)}-\frac{\hbar}{\tau_0(T)})\nonumber\\
	=&\frac{m^{\ast}}{\hbar n e^2}\alpha_B k_BT\frac{\mu_B B}{k_BT}\coth(\frac{\zeta\mu_B B}{2k_BT})
\end{align}
where we used \Autoref{eq:alphaB}. Defining,
\be
\rho_1(B,T)\equiv\rho(B,T)-\rho_0(T)
\ee
\begin{align}
	\frac{\ddp	\rho_1(B,T)}{\ddp B})=&\frac{\mu_B}{k_BT}\frac{\ddp(	\rho_1(B,T))}{\ddp \frac{\mu_B B}{k_BT}}) \nonumber \\
	=&\frac{\mu_B m^{\ast}}{\hbar n e^2}\alpha_B \frac{\ddp}{\ddp \frac{\mu_B B}{k_BT}}\left[\frac{\mu_B B}{k_BT}\coth(\frac{\zeta\mu_B B}{2k_BT})\right] \nonumber\\
	=&\frac{\mu_B m^{\ast}}{\hbar n e^2}\alpha_B \frac{\ddp}{\ddp x}\left[x\coth(\frac{\zeta x}{2})\right] \nonumber\\
	=&\frac{\mu_B m^{\ast}}{\hbar n e^2}\alpha_B [\coth(\frac{\zeta x}{2})-\frac{\zeta x}{2}\textrm{csch}^2(\frac{\zeta x}{2})]
\end{align}
with $x=\frac{\mu_B B}{k_BT}$,
\begin{align}
&\gamma \frac{\ddp	\rho_1(B,T)}{\ddp B}=f(x)\; ,\nonumber\\
&f(x)=\coth(\zeta\frac{x}{2})-\zeta\frac{x}{2}\textrm{csch}^2(\zeta\frac{x}{2})\label{eq:slope}
\end{align}
where we defined
\be
\gamma\equiv\frac{\hbar n e^2}{\mu_B m^{\ast}\alpha_B}\;. \label{eq:gamma}
\ee

We rewrite the empirical formula for resistivity (Eq.~(\ref{eq:rho-tt})) as
\begin{eqnarray}
\rho(B,T) &=& c_0 + c_1 T
    +\frac{B}{\gamma}
    \coth\frac{\zeta\mu_B B}{2k_BT}\;, \label{eq:rho-tt-appxb}
\end{eqnarray}
which in the zero field ($B \to 0$) limit becomes
\begin{eqnarray}
\rho(B \to 0,T) &=& c_0
    +\left[ c_1 + \frac{2k_B}{\gamma\zeta\mu_B} \right] T\;. \label{eq:rho-tt-appxb0}
\end{eqnarray}

On the other hand, the AC resistivity is given by \cite{YYC-ROPP-2025,YYC-arXiv-2025}
\begin{eqnarray}
\rho(\omega,T) &=& c_0
    +\frac{m^*}{ne^2\hbar} \frac{2\hbar\omega}{\pi}
    \coth\frac{\hbar\omega}{2k_BT}\;, \label{eq:rho-tt-appxw}
\end{eqnarray}
which in the static ($\omega \to 0$) limit takes
\begin{eqnarray}
\rho(\omega \to 0,T) &=& c_0
    +\frac{m^*}{ne^2\hbar} \alpha_Tk_BT\;.
    \label{eq:rho-tt-appxw0}
\end{eqnarray}

Comparing Eq.~(\ref{eq:rho-tt-appxb0}) with Eq.~(\ref{eq:rho-tt-appxw0}) results in
\begin{eqnarray}
c_1 &=& \frac{m^*}{ne^2} \frac{k_B}{\hbar}
    \left[ \alpha_T - \frac{2\alpha_B}{\zeta} \right]\;,
    \label{eq:c1-vs-gamma1}
\end{eqnarray}
where we have used Eq.~(\ref{eq:gamma}).
For $\alpha_T = 2\alpha_B$ (Eqs.~(\ref{eq:alphaB}) and (\ref{eq:alphaT})), we obtain
\begin{eqnarray}
c_1 &=& \frac{m^*}{ne^2} \frac{k_B}{\hbar}
    \left[ 1 - \frac{1}{\zeta} \right] \alpha_T\;,
    \label{eq:c1-vs-gamma2}
\end{eqnarray}
which can be rewritten in terms of $\gamma$ and $\zeta$ as
\begin{eqnarray}
c_1 &=& \frac{2k_B}{\gamma\mu_B}
    \left[ 1 - \frac{1}{\zeta} \right]\;.
    \label{eq:c1-vs-gamma3}
\end{eqnarray}

\Autoref{table:parameters} below shows the estimated values of $\gamma$ (using Eq.~(\ref{eq:gamma})) and $c_1$ (taking Eq.~(\ref{eq:c1-vs-gamma3})) compared to the fitting ones (in parentheses).

\begin{table}[h!]
\centering
\begin{tabular}{|c c c c|}
 \hline
 $p$ & $m^*/m_e$ & $\gamma$ (T/$\mu\Omega$.cm) & $c_1$ ($\mu\Omega$.cm/K) \\ [0.5ex]
 \hline\hline
 0.19 & 14.86 (13.81) & 3.7 (3.2) & 0.513 (0.598) [0.654] \\
 0.20 & 13.75 (12.47) & 4.0 (3.5) & 0.475 (0.547) \\ [1ex]
 \hline
\end{tabular}
\caption{The estimated (fitted) [optimized] parameters used for the doping range $0.19 \le p \le 0.2$. Note that the estimated values of the effective mass are taken from Ref.~\cite{Ramshaw-MottPlanckian-arxiv-2024}.}
\label{table:parameters}
\end{table}

We actually use 3 different ways to estimate the various parameter values:

1) we use an estimate of $m^*$ from Ramshaw, use $n=1/V$, and we use $\alpha_T=8/\pi$ and $\alpha_B=4/\pi$, as well as the fitted value for $\zeta=3.06$. With this we can make the analytical estimated values for $\gamma$ and $c_1$. This corresponds to the values not in parentheses nor brackets.

2) we use the fitted value of $\gamma$ such that $\gamma \frac{\ddp \rho}{\ddp B}(\mu B\gg k_B T)\to 1$ as well as  $n=1/V$ and $\alpha_B=4/\pi$ to get a "fitted" estimate of $m^*$, we then use this fitted value of $m^*$ to further estimate the "fitted" value of $c_1$, with $\alpha_T=8/\pi$ and $\zeta=3.06$.  This corresponds to the values in parentheses.

3) for $p=0.19$, we have an optimized value of $c_1$, such that the fit $\rho(B,T) =c_0+ c_1T+  \frac{B}{\gamma}  \coth(\zeta  \frac{\mu_B B}{2 k_BT} )$ follows nicely $\rho$ vs $T$ for different $B$ as plotted in \Autoref{fig:rho}~(a), where we use the fitted value of $\gamma$ (see point 2 above) which gave us $c_1= 0.654$. This corresponds to the value in brackets.

\section{Planckian coefficients}\label{sec:planck_coeff}
Here we describe how we obtained the Planckian coefficients $\alpha_T$  and $\alpha_B$ from the experimental magneto--resistivity data from \cite{Beobinger-2018-Science-LSCO}. These coefficients are the proportionality factors between the scattering rate and the energy scales $k_B T$ in the DC and zero--field limit or $\mu_B B$ in the DC limit and zero temperature limit respectively
\begin{subequations}
\begin{align}
\lim_{B\to 0}\left[\frac{\hbar}{\tau(B,T)}\right]&=\alpha_T k_B T\;,\\
\lim_{T\to 0}\left[\frac{\hbar}{\tau(B,T)}\right]&=\alpha_B \mu_B B\;,
\end{align}\label{eq:def_alpha_T_B_tau}
\end{subequations}
where our theoretical predictions yield $\alpha_T=\frac{8}{\pi}$ and $\alpha_B=\frac{4}{\pi}$.

In order to estimate experimentally the Planckian coefficients $\alpha_T$  and $\alpha_B$, we start from the DC Drude formula for the resistivity
\begin{align}
	\rho(B,T)=&\frac{m^{\ast}}{\hbar n e^2}\frac{\hbar}{\tau(B,T)}\;.\label{eq:DC_Drude}
\end{align}
Combining with \Autoref{eq:def_alpha_T_B_tau}, we can relate the slope of resistivity with respect to either temperature $A_T^1$ or with respect to magnetic field $A_B^1$
\begin{subequations}
\begin{align}
\lim_{B\to 0}\left[\rho(B,T)]\right]&=\frac{m^{\ast}}{\hbar n e^2}\alpha_T k_B T\equiv A_T^1 T\;,\\
\lim_{T\to 0}\left[\rho(B,T)\right]&=\frac{m^{\ast}}{\hbar n e^2}\alpha_B \mu_B B\equiv A_B^1 B\;,
\end{align}\label{eq:def_A1_T_B_rho}
\end{subequations}
where we identify
\begin{subequations}
\begin{align}
A_T^1 &=\frac{m^{\ast}}{\hbar n e^2}\alpha_T k_B\;,\label{eq:def_A1T}\\
A_B^1&=\frac{m^{\ast}}{\hbar n e^2}\alpha_B \mu_B\;.\label{eq:def_A1B}
\end{align}\label{eq:def_A1_T_B}
\end{subequations}
Inverting these relations allows us to determinate the Planckian coefficients $\alpha_T$  and $\alpha_B$ from the values of the slope the resistivity
\begin{subequations}
\begin{align}
\alpha_T &=\frac{\hbar n e^2}{m^{\ast}} k^{-1}_B A_T^1\;,\label{eq:def_alphaT}\\
\alpha_B&=\frac{\hbar n e^2}{m^{\ast}} \mu^{-1}_B A_B^1 \;. \label{eq:def_alphaB}
\end{align}\label{eq:def_alpha_T_B}
\end{subequations}From \Autoref{eq:def_alpha_T_B}, we see that in order to determinate the values of the Planckian coefficients $\alpha_T$  and $\alpha_B$, we also need to estimate the ratio of effective electron mass $m^*$ to the carrier density $n$, $n/m^*$. There are two ways to estimate this ratio: one by plasma frequency $\omega_p$, the other by separating estimating $m^*$ via specific heat coefficient and $n$ via other transport measurements, as described below.

\subsubsection{Estimation of $n/m^*$ from plasma frequency measurements} 

 To estimate $n/m^*$ shown in \Autoref{eq:def_alpha_T_B}, we follow 
 the measurements of the plasma frequency $\omega_p^2$ \cite{Ramshaw-MottPlanckian-arxiv-2024}, where the plasma frequency is given by
\be
\omega_p^2=4\pi e^2\frac{n}{m^*}\; .\label{eq:w_p_2}
\ee
The $n/m^*$ derived from experimental  data of the plasma frequency  at low doping $p<0.2$ is well approximated by a linear regression given by\cite{Ramshaw-MottPlanckian-arxiv-2024} 
\be
\frac{n}{m^*}=\frac{0.42 p}{la^2 m_e}\; ,\label{eq:w_p_2_fit}
\ee
where $p$ is the doping percentage, $a\simeq 0.378~\rm{nm}$ is in the in--plane lattice spacing, and $l\simeq0.66~\rm{nm}$  corresponds to the half--interlayer spacing, and $m_e$ to the bare electron mass. See \cite{Ramshaw-MottPlanckian-arxiv-2024} and Fig. 4B therein. 

Replacing \Autoref{eq:w_p_2_fit} in \Autoref{eq:def_alpha_T_B}, we get
\begin{subequations}
\begin{align}
\alpha_T &=\frac{A_T^1\hbar e^2}{k_B }\frac{0.42 p}{la^2 m_e}\;,\label{eq:def_alphaT_wp}\\
\alpha_B&=\frac{A_B^1\hbar e^2}{\mu_B }\frac{0.42 p}{la^2 m_e} \;. \label{eq:def_alphaB_wp}
\end{align}\label{eq:def_alpha_T_B_wp_2}
\end{subequations}
Similarly to the results using effective mass estimations from specific heat measurements, the parameters $A_T^1$ and $A_B^1$ correspond to the slope of the resistivity curves obtained from \cite{Beobinger-2018-Science-LSCO} for LSCO, as details above in Appendix~\ref{sec:alpha_T} and ~\ref{sec:alpha_B}.
The values of the Planckian coefficients obtained from plasma frequency measurements through \Autoref{eq:def_alpha_T_B_wp_2} are shown in \Autoref{fig:alpha_B_T}.

\subsubsection{Estimation of $n/m^*$ by separately estimating $m^*$ and $n$} 

The $n/m^*$ ratio can also be estimated by separately estimating $m^*$ and $n$. Following Ref. \cite{Ramshaw-MottPlanckian-arxiv-2024}, the electron effective mass $m^*$ is taken to be inversely proportional to doping p by specific heat measurements, $ m^* \propto 1/p$. More precisely, we use the fact $m^{\ast}\propto p^{-1}$ in the overdoped regime, where effects from the pseudogap phase are absent, and extrapolate to lower doping values \cite{Ramshaw-MottPlanckian-arxiv-2024,Loram_2001}. To estimate the carrier density $n$, we follow the argument in  \cite{Ramshaw-MottPlanckian-arxiv-2024} by plasma frequency measurement, where the square of plasma frequency $\omega_p^2$ obtained from optical conductivity measurements grows linearly with doping for doping withing the interval of interest $p\in[0,0.3]$ and that for an parabolic band it is proportional to $n/m^{*}$, $\omega_p^2 \propto n/m^{*}$. Since $\omega_p^2 \propto n/m^{*}~\propto p$ and $m^{\ast}\propto p^{-1}$ according to specific heat measurement, we conclude that in our doping range of interest $p\in[0.161,0.19]$, the carrier density is close to an effective constant value \cite{Ramshaw-MottPlanckian-arxiv-2024}  
\be
n\approx n_{eff} \approx\frac{1}{a^2l},\label{eq:def_n}
\ee
where $a\simeq 0.378~\rm{nm}$ is in the in--plane lattice spacing, and $l\simeq0.66~\rm{nm}$ such that $c=2l$ corresponds to the interlayer spacing. We use $l$ instead of $c$, because there are 2 superconducting copper oxide planes separated by $l$ within a single unit cell \cite{Ramshaw-MottPlanckian-arxiv-2024}.

  On the other hand, the Fermi surface reconstruction of cuprates was observed from Hall measurement \cite{hussey-incoherent-nature-2021}, where the carrier density is proportional to doping at low doping values $n[p<0.15]\propto p$ and goes as $1+p$ for large doping values when the system reaches the Fermi liquid phase
phase on the overdoped side $n[p>0.3]\propto 1+p$. It seems that in between these two regions, the carrier density still grows linearly with doping  $n[p<0.15]\propto p$ with the overdoped $n[p>0.3]\propto 1+p$ regimes. 
To reconcile the discrepancy between the
doping insensitive $n_{eff} \sim 1/V$ estimated here and the doping
dependent carrier density $n$ via the Hall coefficient measurement goes beyond the scope of the present work and deserves
further study elsewhere.

\begin{figure}[H]
\begin{center}
	\begin{minipage}[b]{1\textwidth}
    {
 				\includegraphics[width=0.48\linewidth]{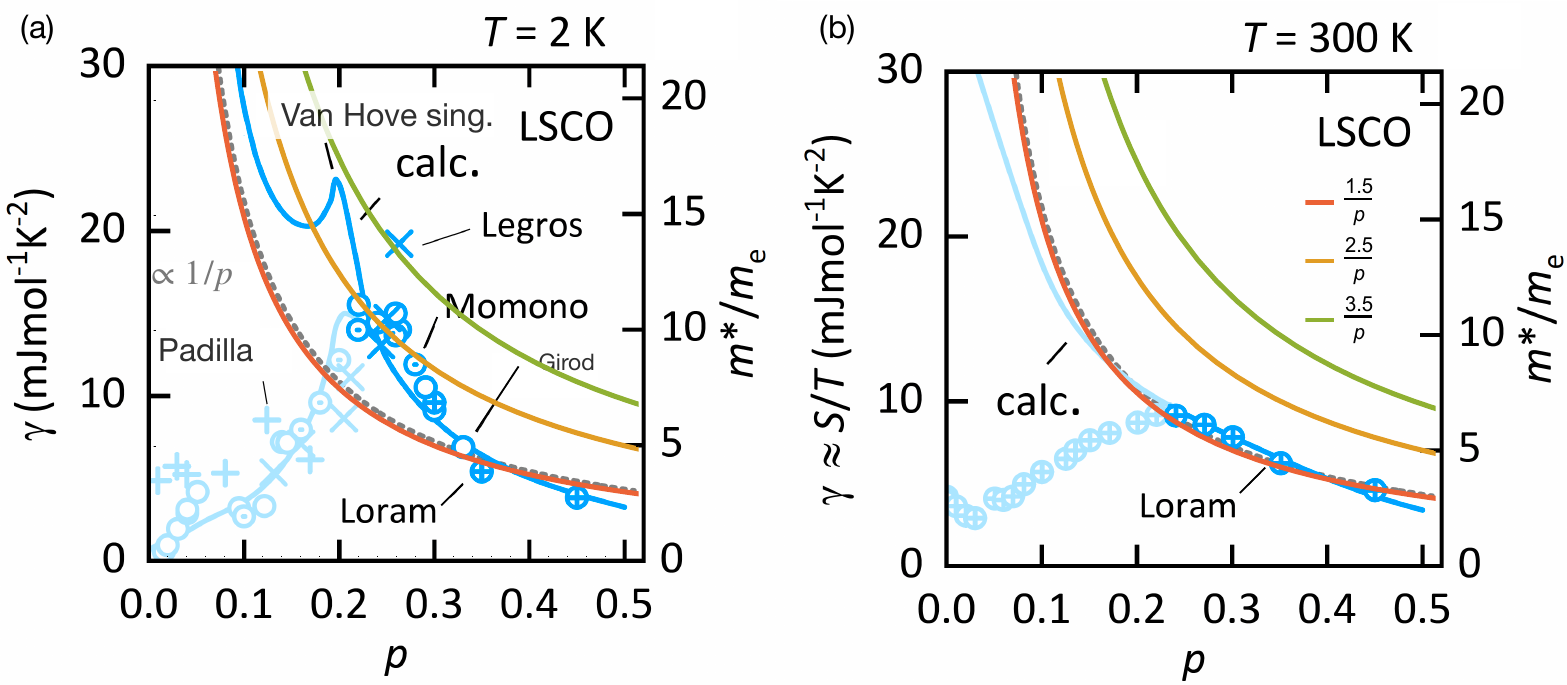}
			}
		\end{minipage}
	\end{center}
 	\caption{Estimation of $m^{*}/m_e$ from specific heat measurements \cite{Ramshaw-MottPlanckian-arxiv-2024}.
    A plausible estimate of the effective mass is illustrated by the middle orange curve $(m^{\ast}_{av}/m_e= 2.5/p)$. Additionally, upper bound and lower bound estimates are parametrize by $(m^{\ast}_{max}/m_e=3.5/p)$ and $(m^{\ast}_{min}/m_e=1.5/p )$ respectively. These values are used to determine the Planckian coefficients in \protect\Autoref{fig:alpha_B_T} of the main text.
    Figures are reproduced from \cite{Ramshaw-MottPlanckian-arxiv-2024}
    }
	\label{fig:fit_mstar}
 \end{figure}

As illustrated in \Autoref{fig:fit_mstar}, we choose a wide band of plausible estimate of the effective mass, given by the area between the green $(m^{\ast}_{max})$ and red curves  $(m^{\ast}_{min})$, such that the estimate $(m^{\ast}_{av})$ depicted as the orange curve is given by the average of $(m^{\ast}_{max})$ and $(m^{\ast}_{min})$. For simplicity, we parametrize these in function of doping $p$ as
\begin{subequations}
\begin{align}
m^{\ast}_{min}/m_e&=1.5/p \;,\label{eq:mstar_min}\\
m^{\ast}_{av}/m_e&= 2.5/p\;, \label{eq:mstar_av}\\
m^{\ast}_{max}/m_e&=3.5/p\;. \label{eq:mstar_max}
\end{align}\label{eq:mstar_function}
\end{subequations}
These values are used to determine the Planckian coefficients shown in \Autoref{fig:alpha_B_T_wp}. In \Autoref{fig:alpha_B_T_wp}, we compare the Planckian coefficients estimated by the two approaches mentioned above, they show excellent agreement. 

In the next two sections we explain in details how we calculate the slope of experimentally measured resistivity with respect to either temperature $A_T^1$ or with respect to magnetic field $A_B^1$. Once these values are known, we use \Autoref{eq:def_alpha_T_B}, with \Autoref{eq:mstar_av} and \Autoref{eq:def_n}, to evaluate the Planckian coefficients  $\alpha_T$  and $\alpha_B$, which corresponds to the blue circles and purple squares in \protect\Autoref{fig:alpha_B_T} respectively. The error bars on the Planckian coefficients  $\alpha_T$  and $\alpha_B$ are given by the upper $m^{\ast}_{max}$ and lower $m^{\ast}_{min}$ bound estimates. From \Autoref{eq:def_alpha_T_B}, it is clear the upper bound $m^{\ast}_{max}$ will give the lower bound estimates of the Planckian coefficients, and lower $m^{\ast}_{min}$ bound will give the upper estimation.

\subsection{Linear--in--temperature  Planckian coefficients }\label{sec:alpha_T}
\begin{figure}[H]
	\begin{center}
	\begin{minipage}[b]{0.9\textwidth}
 			{
 				\includegraphics[width=0.5\linewidth]{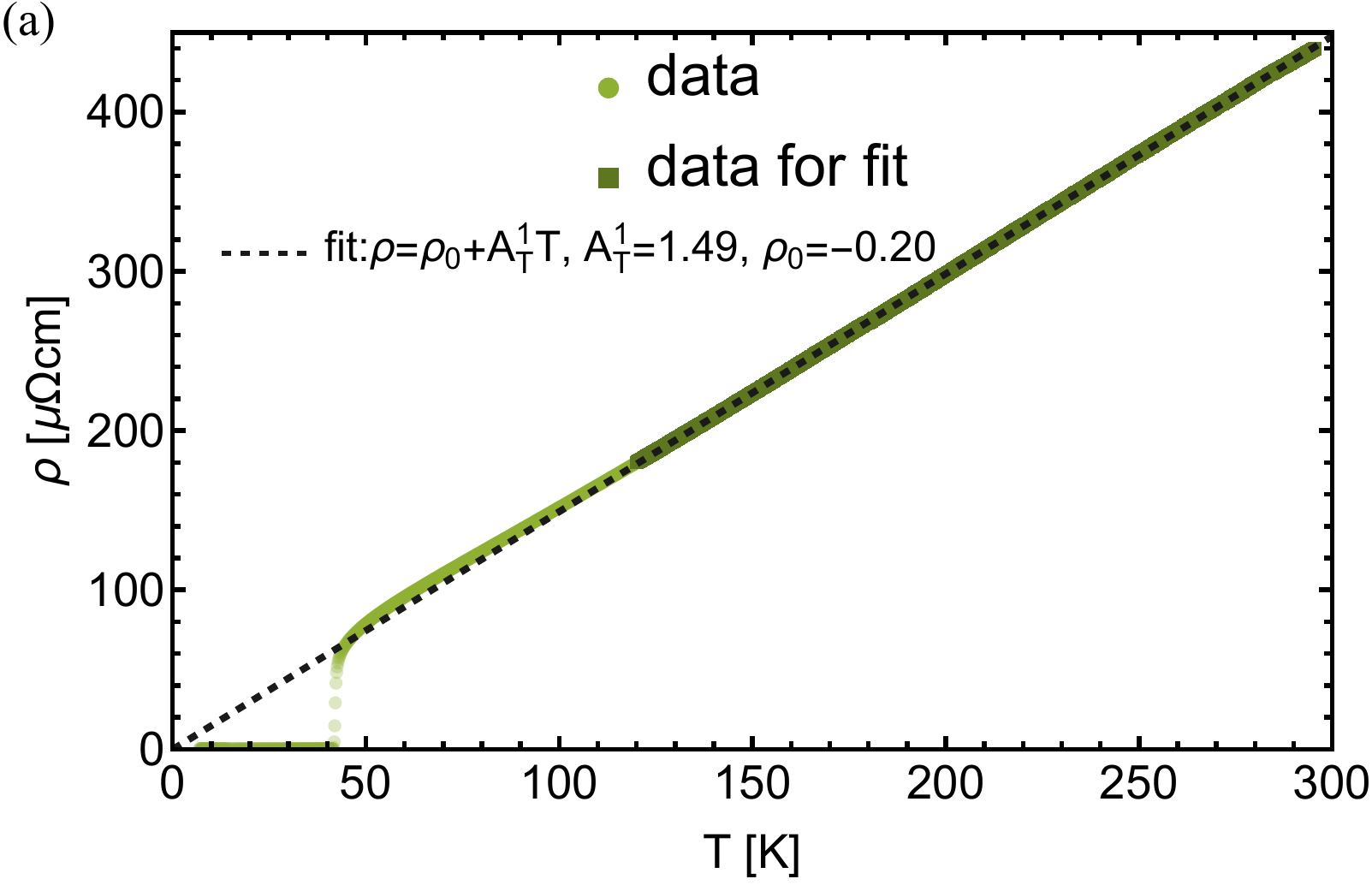}
			}\\
			{
				\includegraphics[width=0.5\linewidth]{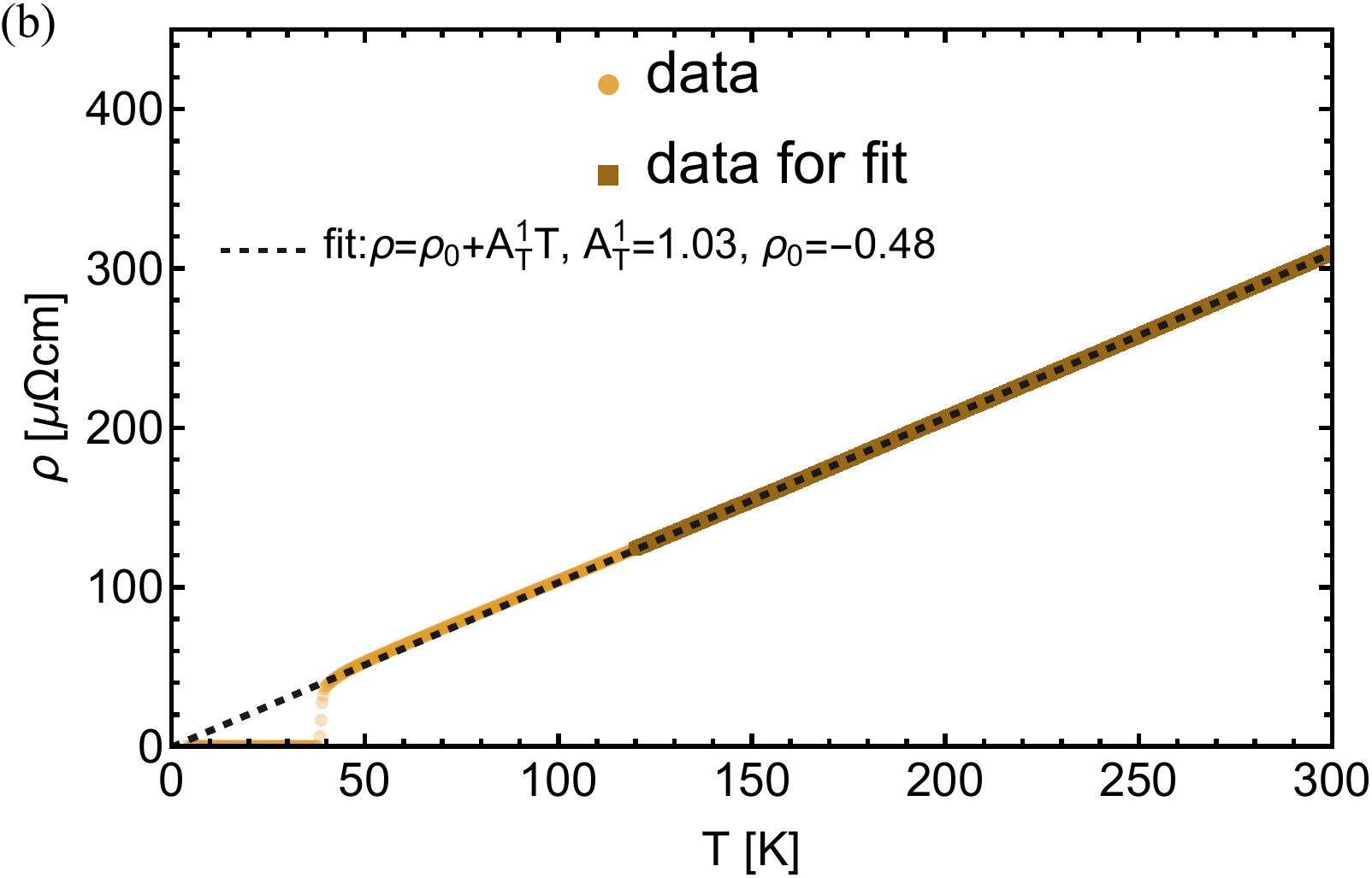}				
			}
		\end{minipage}
	\end{center}
 	\caption{ Resistivity data with respect to temperature at zero magnetic field reproduced from \cite{Beobinger-2018-Science-LSCO} for (a) $p=0.161$, and (b) $p=0.190$. Colored lines correspond to the experimental data, darker parts of the data are used to perform the linear regression, and the black dashed line correspond to the obtained linear fit from which we extract the slope value $A_T^1$ used to estimate the linear--in--temperature Planckian coefficients shown in \protect\Autoref{fig:alpha_B_T}. The value obtained of the parameters $A_T^1$ and $\rho_0$ from the linear fit [\protect\Autoref{eq:fit_rho_T}] for all doping values are shown in \protect\Autoref{fig:fit_rho_T}
    }
	\label{fig:rhoT_16_19}
 \end{figure}

\begin{figure}[H]
	\begin{center}
	\begin{minipage}[b]{0.9\textwidth}
 			{
 				\includegraphics[width=0.4\linewidth]{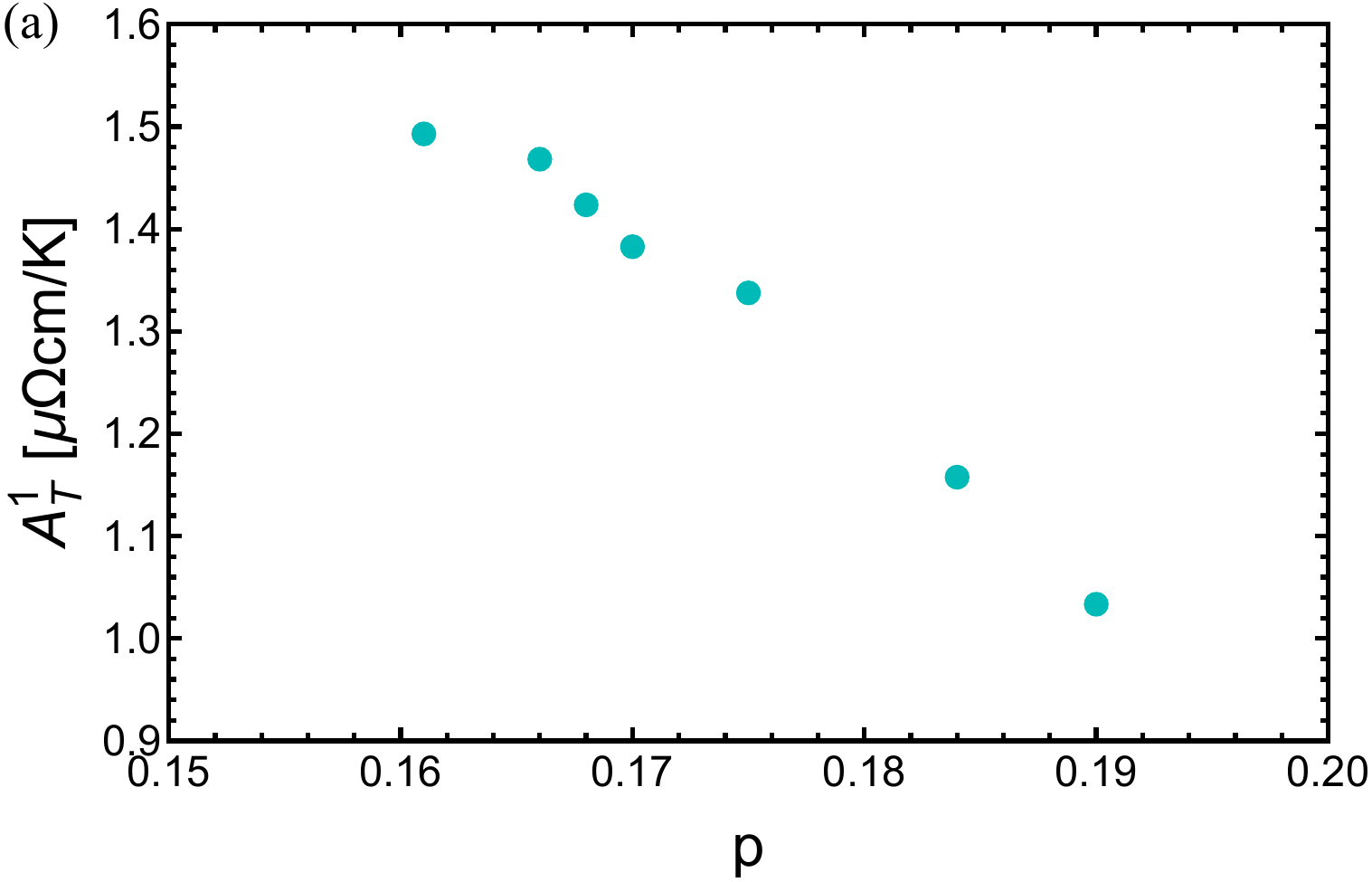}
			}\\
			{
				\includegraphics[width=0.4\linewidth]{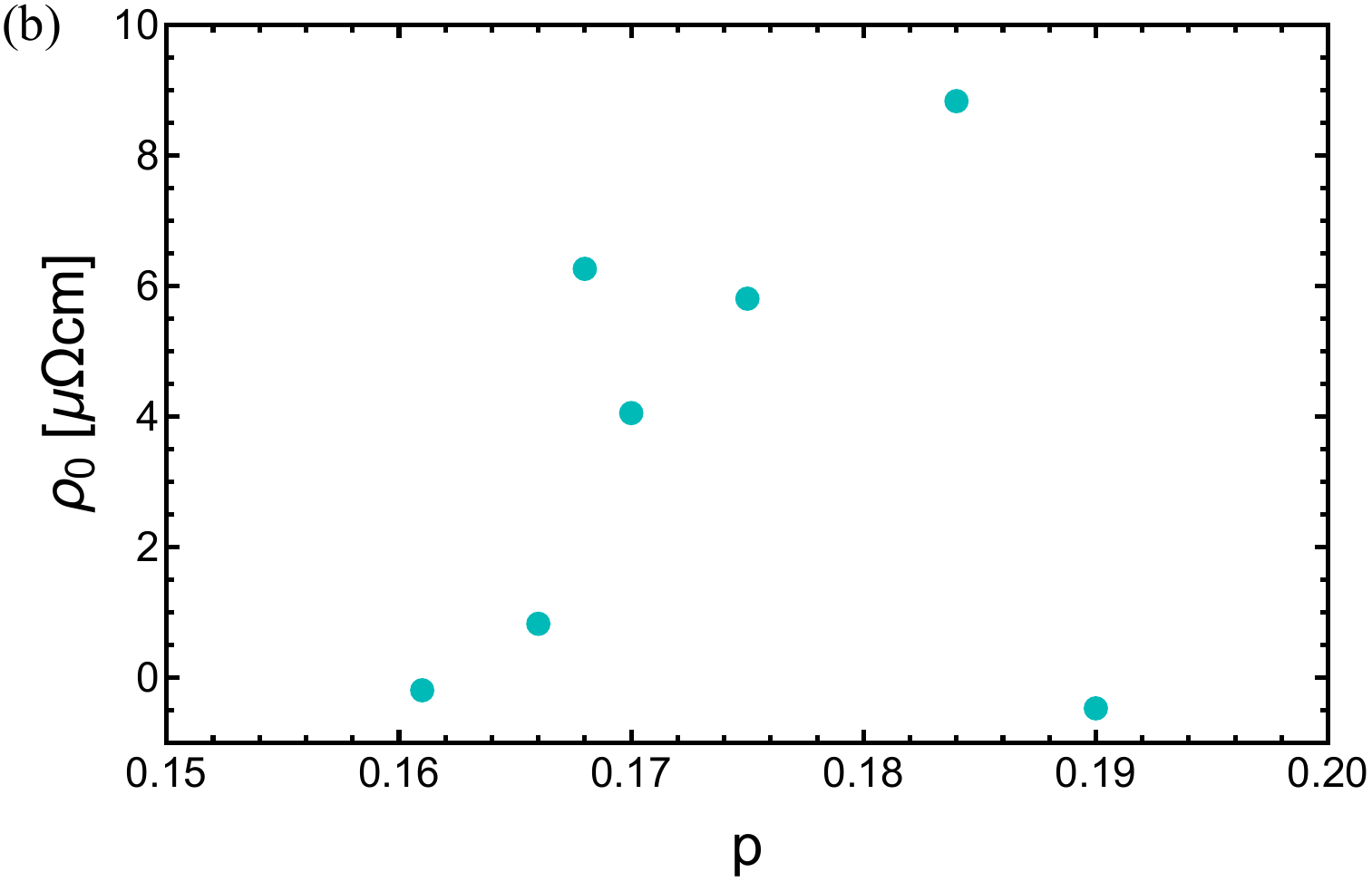}				
			}
		\end{minipage}
	\end{center}
 	\caption{  Parameters values for (a) the slope $A_T^1$ and (b) the residual zero--field resistivity $\rho_0$  obtained from the linear fit [\protect\Autoref{eq:fit_rho_T}] plotted versus doping values. $A_T^1$ shows a monotonic, almost linear decrease with doping, while $\rho_0$ does not show any specific doping dependence. The slope values $A_T^1$ are further used through \protect\Autoref{eq:def_alpha_T_B}--\protect\Autoref{eq:mstar_function} to estimate the linear--in--temperature Planckian coefficients shown in \protect\Autoref{fig:alpha_B_T}
    }
	\label{fig:fit_rho_T}
 \end{figure}

In this section, we explain how we obtained the linear--in--temperature  Planckian coefficients $\alpha_T$ from the experimental data in \cite{Beobinger-2018-Science-LSCO}. As explain above, one need to determine the slope of resistivity with respect to  temperature $A_T^1$. For this we use the zero--field data and parametrize the resistivity as
\be
\rho(T)=\rho_0+A_T^1 T\label{eq:fit_rho_T}
\ee
and proceed to a linear fit of the experimental data in order to extract $A_T^1$. As an illustration in \Autoref{fig:rhoT_16_19} for $p=0.161$ in (a) and $p=0.190$ in (b), we show the experimental data of the resistivity with respect to  temperature at zero magnetic field in blue, the data used to perform the linear fitting in orange and the obtained numerical linear regression as the black dashed line. The value obtained of the parameters $A_T^1$ and $\rho_0$ from the linear fit [\Autoref{eq:fit_rho_T}] for all doping values are shown in \Autoref{fig:fit_rho_T}.  We note that $A_T^1$ shows a monotonic, almost linear decrease with doping, consistent with the fact that $1/A_T^1$ exhibits a linear dependence with respect to doping (see \cite{Ramshaw-MottPlanckian-arxiv-2024} and Fig. 1B therein). Whereas $\rho_0$ does not show any regular pattern and is relatively small compare to the values of the resistivity in the temperature range of interest.

Finally, using \protect\Autoref{eq:def_alpha_T_B}--\protect\Autoref{eq:mstar_function}, the slope values $A_T^1$ are used to estimate the linear--in--temperature $\alpha_T$ Planckian coefficients shown in \protect\Autoref{fig:alpha_B_T}.


\subsection{Linear--in--field  Planckian coefficients }\label{sec:alpha_B}
\begin{figure}[H]
	\begin{center}
	\begin{minipage}[b]{1\textwidth}
 			{
 		 		\includegraphics[width=0.48\linewidth]{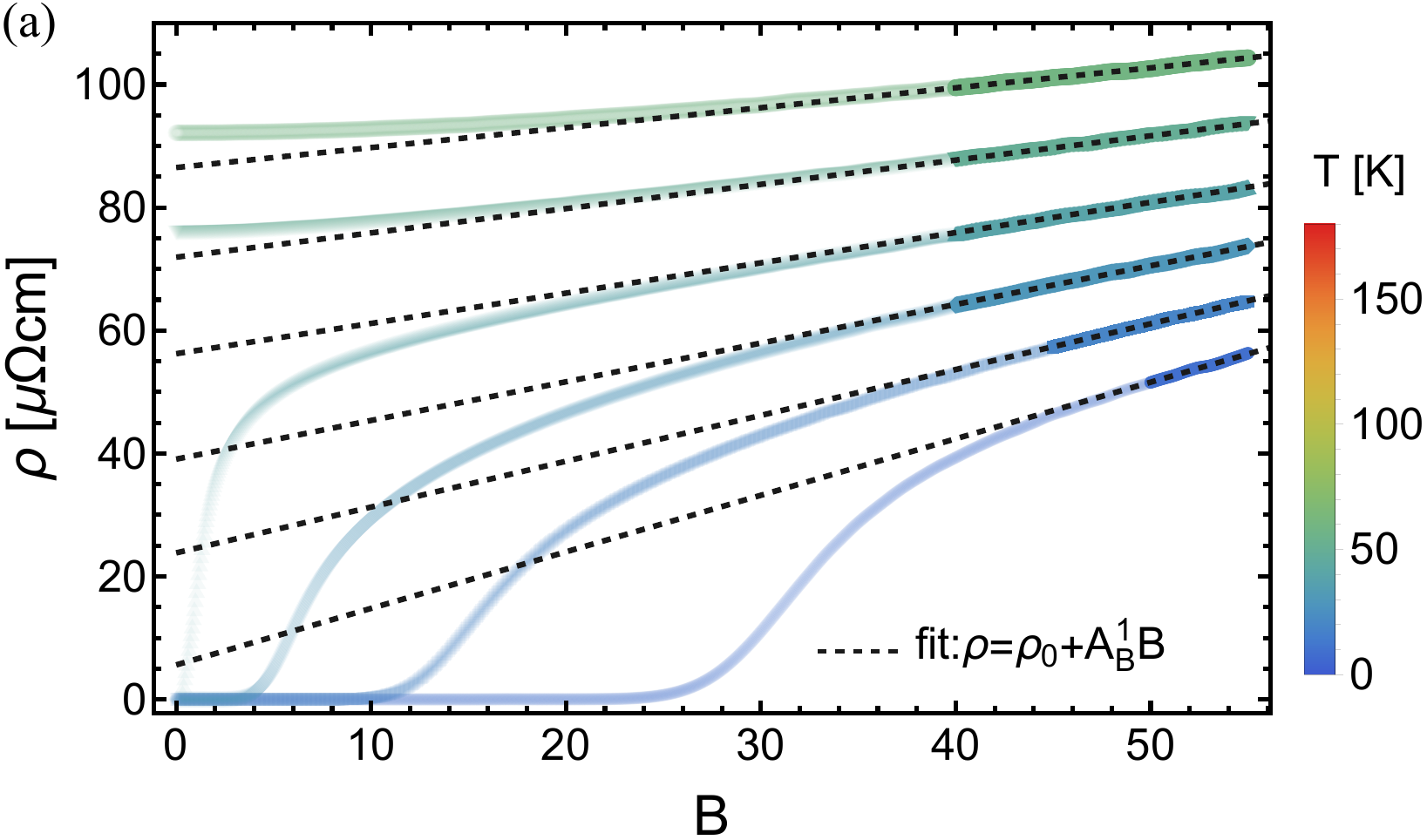}
			}\\
			{
				\includegraphics[width=0.48\linewidth]{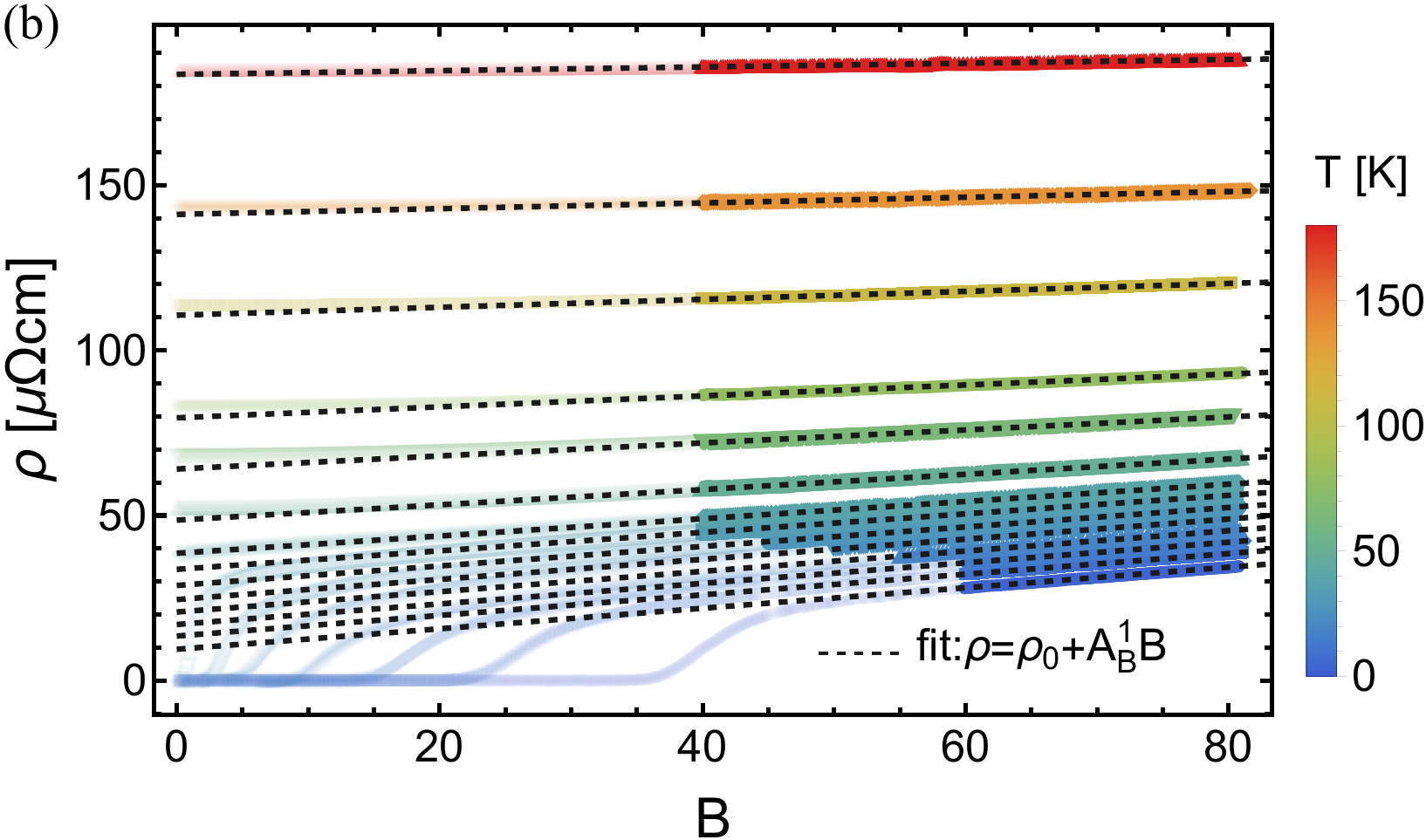}				
			}
		\end{minipage}
	\end{center}
 	\caption{ Resistivity data with respect to magnetic field at various temperature values reproduced from \cite{Beobinger-2018-Science-LSCO} for (a) $p=0.161$, and (b) $p=0.190$. Colored lines correspond to the experimental data and are associated to the temperature color scheme, the darker parts of the data are used to perform the linear regression, and the black dashed lines are the obtained linear fits from which we extract the slope value $A_B^1$ used to calculate the Planckian coefficients. The obtained fitting parameters  \protect\Autoref{fig:fig_A1B_rho0_p} (a) and (b) respectively, plotted versus doping for the different temperatures ; and they are shown plotted versus temperature for the different doping values in \protect\Autoref{fig:fig_A1B_rho0_T} (a) and (b) respectively.
    }
	\label{fig:rhoB_16_19}
 \end{figure}

Here we explain how we obtained the linear--in--field Planckian coefficients $\alpha_B$ from the experimental data in \cite{Beobinger-2018-Science-LSCO}.
As explain above, one need to determine the slope of resistivity with respect to  magnetic field $A_B^1$. We follow a similar procedure to the determination of the in--temperature linear coefficient $A_T^1$ described above, to the difference that for a given doping we will obtain various values $A_B^1$ depending on temperature. Indeed, as shown in \Autoref{fig:rhoB_16_19} (a) for $p=0.161$ and (b) for $p=0.190$, magneto--resistivity curves were taken at various temperature. In order to determine $A_B^1$ from the experimental data, we parametrize the resistivity as
\be
\rho(B)=\rho_0+A_B^1 B\label{eq:fit_rho_B}
\ee
and proceed to a linear fit of the experimental data at the available temperatures and for all doping. For simplicity, we here use a linear regression to determine the slope $A_B^1$ instead of performing the window slope analysis described in Appendix~\ref{usffx}. In \Autoref{fig:fit_drho_B_comp}, we later show that the two methods are almost equivalent.

\begin{figure}[H]
	\begin{center}
	\begin{minipage}[b]{1\textwidth}
 			{
 				\includegraphics[width=0.48\linewidth]{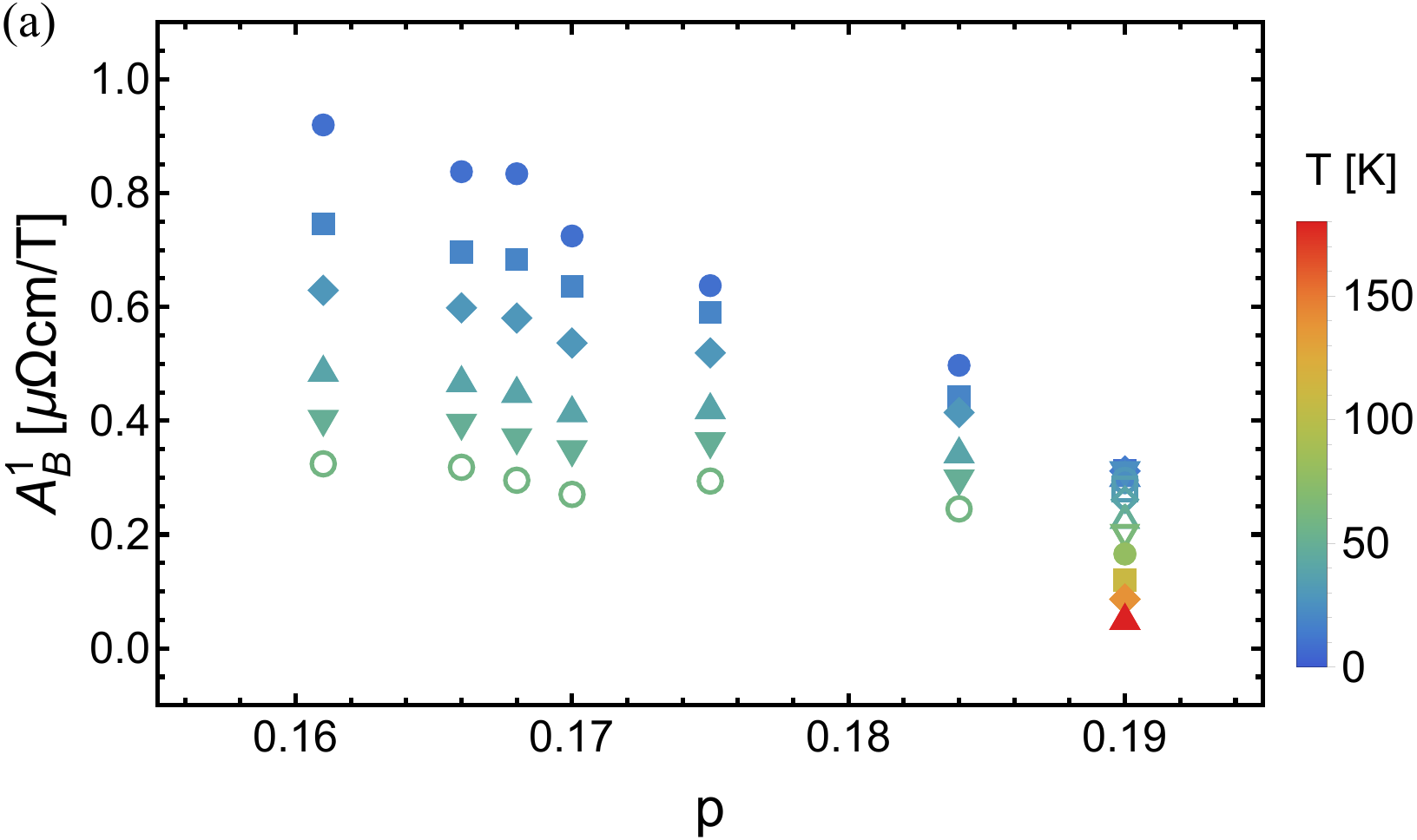}
			}\\
			{
				\includegraphics[width=0.48\linewidth]{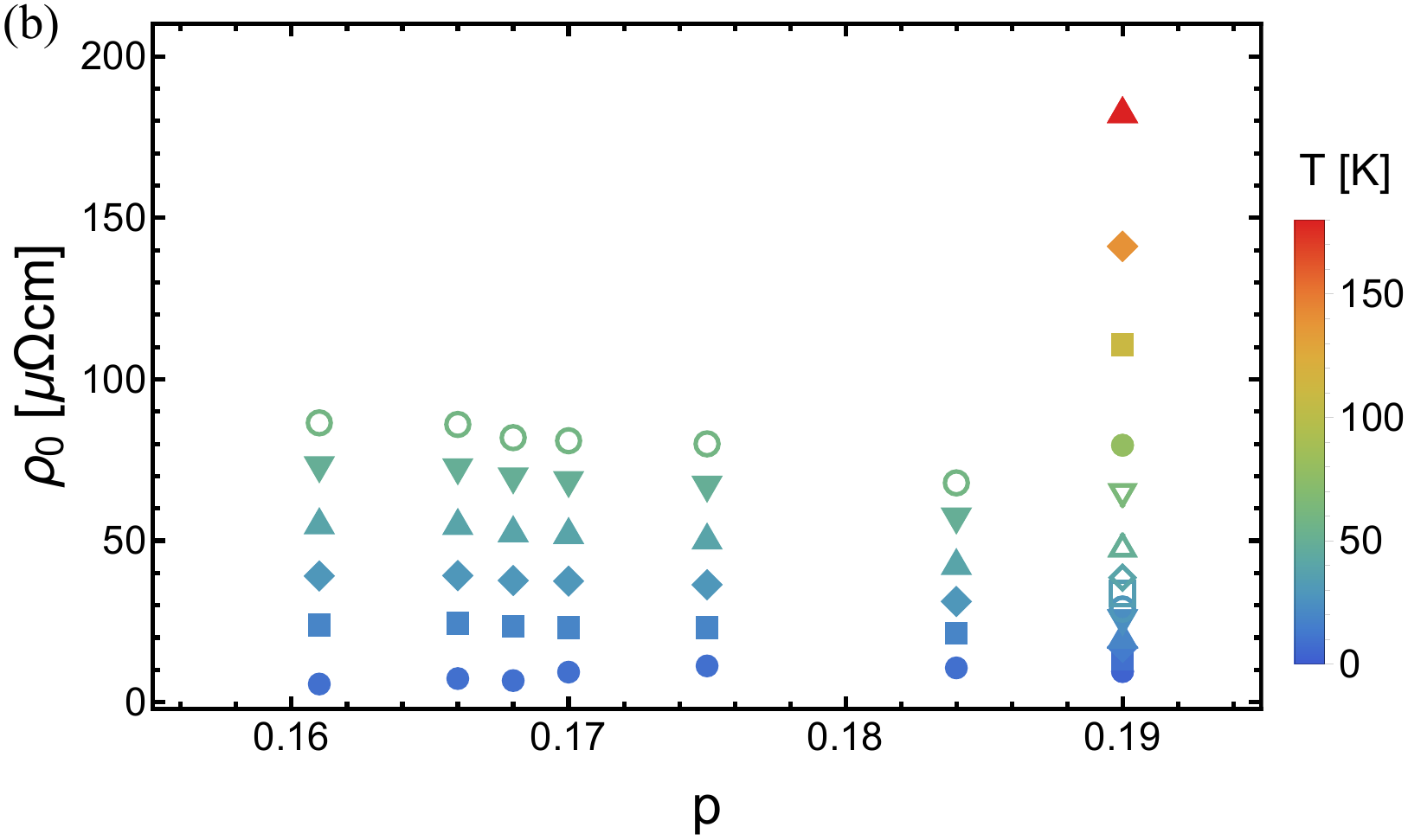}				
			}
		\end{minipage}
	\end{center}
 	\caption{ (a) Slope $A^1_B$ values and (b) zero--field residual resistivity $\rho_0$ obtained from the linear fit [\protect\Autoref{eq:fit_rho_B}] and plotted versus doping for the different temperature values. The color scheme corresponds to the temperature gradient shown in the bar legend. $A_B^1$ versus doping shows a monotonic, almost linear, decrease with doping for all temperature values as shown in \protect\Autoref{fig:fig_A1B_rho0_p} (a), which is consistent with the behavior of the linear--in--temperature coefficient $A_T^1$ [\protect\Autoref{fig:fit_rho_T} (a)]. In \protect\Autoref{fig:fig_A1B_rho0_p} (b), we see that $\rho_0$ versus doping does not show any specific doping dependence.
    }
	\label{fig:fig_A1B_rho0_p}
 \end{figure}

\begin{figure}[H]
	\begin{center}
	\begin{minipage}[b]{1\textwidth}
            {
				\includegraphics[width=0.48\linewidth]{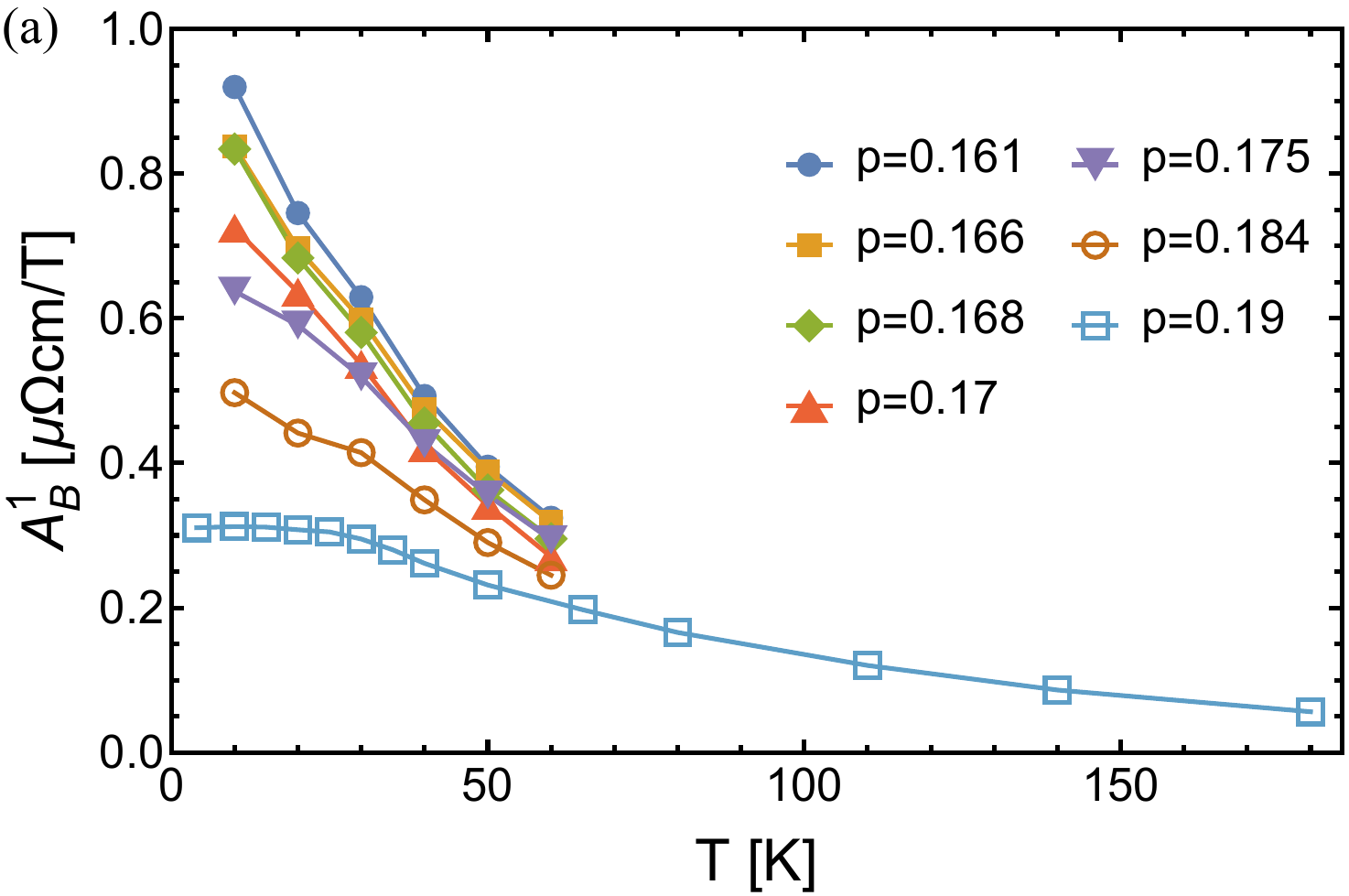}				
			}\\
 		{
 				\includegraphics[width=0.48\linewidth]{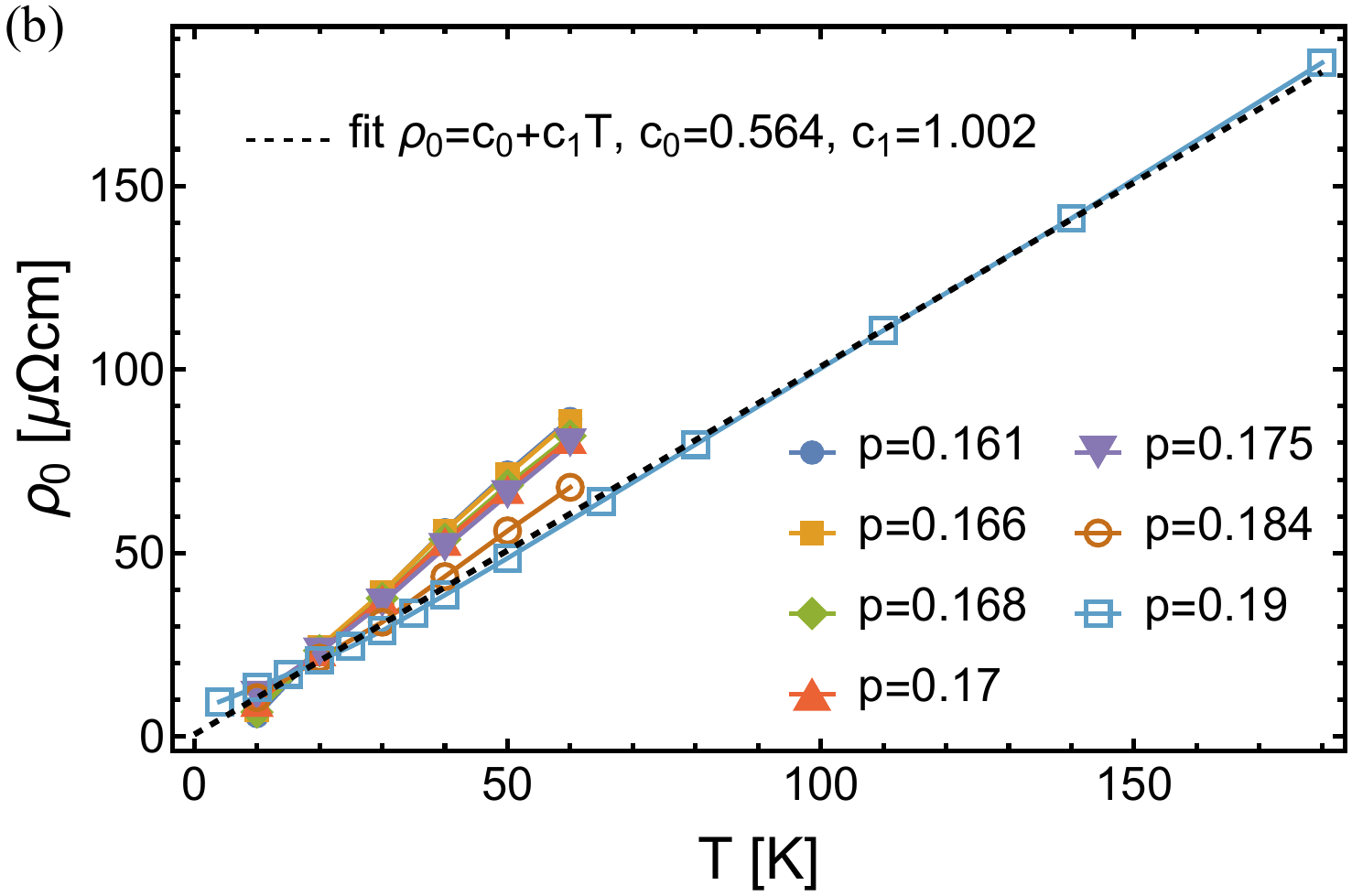}
			}
		\end{minipage}
	\end{center}
 	\caption{
    (a) Slope $A^1_B$ values and (b) zero--field residual resistivity $\rho_0$ obtained from the linear fit [\protect\Autoref{eq:fit_rho_B}] and plotted versus temperature for the different doping values.
    (a) $A_B^1$ versus temperature shows a monotonic decrease for all doping values. For $p=0.19$, we observe a saturation plateau at low temperature, which is well described by our theoretical scaling function as shown in \protect\Autoref{fig:drhoBT_19_B_fit} and \protect\Autoref{fig:drhoBT} (a) for different magnetic field values. For other doping values, magneto--resistivity data were only taken down to $10~\rm{K}$, which is unfortunately too high to see the appearance of the saturation plateau.
    (b) $\rho_0$ shows a linear temperature dependence for all doping. This implies that the zero--field resistivity contains a linear--in--temperature only dependence, introduced in the expression of the resistivity in \protect\Autoref{eq:rho-tt} as $\rho_0(T) = c_0 + c_1 T$ , which through the DC formula [\protect\Autoref{eq:DC_Drude}] leads to an extra explicit temperature only dependence of the scattering rate expressed as $\hbar/\tau_0(T)$. This extra explicit temperature only dependence of the resistivity plays an important role when explaining the experimental data. Indeed as shown in \protect\Autoref{fig:fit_rho_B_tot} and \protect\Autoref{fig:rho}~(a) in the main text, it allows to account for the slight change of the slope of the resistivity with respect to temperature.
    }
	\label{fig:fig_A1B_rho0_T}
 \end{figure}

In \Autoref{fig:rhoB_16_19} (a) for $p=0.161$ and (b) for $p=0.190$, we show the experimental data of the resistivity with respect to magnetic field at various temperatures in the temperature scaled color scheme. The data used to perform the linear fitting are shown in orange and the obtained numerical linear regressions are represented by the black dashed lines. The value obtained for the parameters $A_B^1$ and  $\rho_0$ from the linear fit [\Autoref{eq:fit_rho_B}]  are shown in \Autoref{fig:fig_A1B_rho0_p} (a) and (b) respectively, plotted versus doping for the different temperatures ; and they are shown plotted versus temperature for the different doping values in \Autoref{fig:fig_A1B_rho0_T} (a) and (b) respectively.

We note that $A_B^1$ versus doping shows a similar monotonic, almost linear decrease with doping for all temperature values as shown in \Autoref{fig:fig_A1B_rho0_p} (a), which is consistent with the behavior of the linear--in-- temperature coefficient $A_T^1$ [\Autoref{fig:fit_rho_T} (a)]. In \Autoref{fig:fig_A1B_rho0_p} (b), we see that $\rho_0$ versus doping does not show any specific doping dependence.

However, when plotted against temperature we note that $\rho_0$ shows a linear temperature dependence for all doping as illustrated in \Autoref{fig:fig_A1B_rho0_T} (b). This implies that the zero--field resistivity contains a linear--in--temperature dependence. For this reason, we introduce an extra explicit temperature only dependence of the resistivity given by $\rho_0(T) = c_0 + c_1 T$ in \Autoref{eq:rho-tt}, which through the DC formula [\Autoref{eq:DC_Drude}] leads to an extra explicit temperature only dependence of the scattering rate expressed as $\hbar/\tau_0(T)$.

\begin{figure}[H]
\begin{center}
	\begin{minipage}[b]{1\textwidth}
   {				\includegraphics[width=0.48\linewidth]{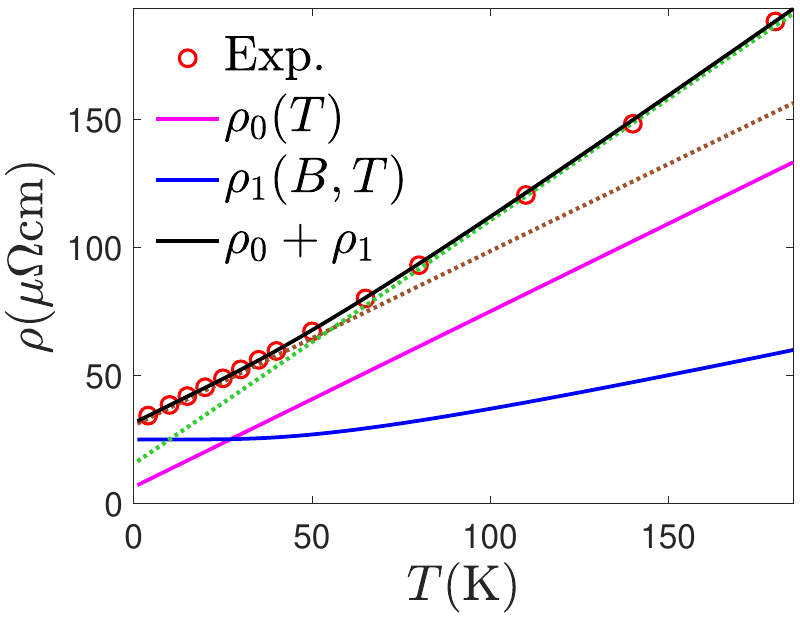}
			}
		\end{minipage}
	\end{center}
 	\caption{
    Different contributions of each of the terms in the theoretical expression for the resistivity in \protect\Autoref{eq:rho-tt}, compared to the experimental data at $p=0.19$ and $B=80T$ represented by the red circles. The pink line corresponds to the added temperature only dependence $\rho_0(T) = c_0 + c_1 T$, the blue line, which saturates at low temperatures, by the theoretical scaling function $\rho_1(B,T)=\frac{B}{\gamma}\coth\frac{\zeta\mu_B B}{2k_BT}$ and the total contribution by the black line. The parameters used in the theoretical prediction are the same as in \protect\Autoref{fig:rho} with $c_0=7.2 \mu\rm{\Omega cm}$, $c_1=0.654\mu\rm{\Omega cm K^{-1}}$, $\gamma=3.2 (\mu\rm{\Omega cm})^{-1}\rm{T}$, $\zeta=3.06$. The combination of the two contributions  $\rho_0(T)$ and $\rho_1(B,T)$ results in a slight change of slope of the calculated resistivity around $T_{kink}\sim 60~\rm{K}$, made more visually evident by the two dashed lines. This feature is also present in the experimental data,in excellent agreement with the theory.
    }
	\label{fig:fit_rho_B_tot}
 \end{figure}

This extra explicit temperature only dependence of the resistivity given by $\rho_0(T) = c_0 + c_1 T$ in \Autoref{eq:rho-tt}  turns out to be of crucial importance when explaining the experimental data. Indeed as shown in \Autoref{fig:fit_rho_B_tot} and \Autoref{fig:rho}~(a) in the main text, it allows to account for the slight change of the slope of the resistivity with respect to temperature. In \Autoref{fig:fit_rho_B_tot} we clearly show the contribution of each of the terms in the expression for the resistivity in \Autoref{eq:rho-tt}. The added temperature only dependence $\rho_0(T) = c_0 + c_1 T$  is represented by the pink line, the theoretical scaling function $\rho_1(B,T)=\frac{B}{\gamma}\coth\frac{\zeta\mu_B B}{2k_BT}$ by the blue line, which saturates at low temperatures, and the total contribution by the black lines. This results in a slight change of slope of the resistivity enhanced by the black dashed lines in \Autoref{fig:fit_rho_B_tot} around $T_{kink}\sim 60~\rm{K}$. This feature is likewise observed in the experimental data, and the theoretical predictions are in excellent agreement with the measurements.

In \Autoref{fig:fig_A1B_rho0_T} (a), we see that $A_B^1$ versus temperature shows a monotonic decrease for all doping values. For $p=0.19$, we observe a saturation plateau at low temperature. For other doping values, magneto--resitivity data were only taken down to $10~\rm{K}$, which is unfortunately too high to see the appearance of the saturation plateau. We also note that these features are well described by our theoretical scaling function as shown in \Autoref{fig:drhoBT_19_B_fit} and \Autoref{fig:drhoBT} (a) for different magnetic field values for $p=0.19$.

In \Autoref{fig:fit_drho_B_comp}, we show the slope of the resistivity data versus temperature for $p=0.19$ using the values obtained by linear regression $A_B^1$ (purple circles) and the values obtained from the window slope analysis $\dd_B\rho(B=80T)$ method described in Appendix~\ref{usffx} at high magnetic field (80 T) with a window width of $w=22\rm{T}$, that is the saturation values of the slope in \Autoref{fig:drhoBT_19_w}~(b) at $B=80~\rm{T}$. We see that the two methods are in good agreement.

Because for lower doping values, the zero temperature limit of $A_B^1$ is not available experimentally, and in order to avoid effects arising from the neighboring superconducting fluctuations at low temperature, we choose to estimate the slope of resistivity at finite temperature by taking the average between $T\in[20,30]\rm{K}$. Indeed, we observed that for instance in the $p=0.161$ case in  \Autoref{fig:rhoB_16_19}~(a), the resistivity data at $T=10\rm{K}$ are just barely linear with magnetic field at the highest measure magnetic field values (55T), and are cleary affected by superconducting fluctuations below $B=45\rm{T}$. We therefore believe that our estimation at low temperature is not optimal and measurements at higher magnetic field values would be necessary for a more accurate estimation, as was performed for $p=0.19$. Therefore, we choose to take average between $T\in[20,30]\rm{K}$,  this means
\footnotesize
\begin{align}
A_B^1=&\frac{A_B^1(T=20\rm{K})+A_B^1(T=25\rm{K})+A_B^1(T=30\rm{K})}{3}\;, p=0.19\\
A_B^1=&\frac{A_B^1(T=20\rm{K})+A_B^1(T=30\rm{K})}{2}\;, p\in [0.161,\dots, 0.184]
\end{align}
\normalsize

The slope values obtained from the linear regression $A_B^1$ divided by $\mu_B$ are shown in \Autoref{fig:fig_A1B_T_rho0B_T}~(a), alongside the previously obtained linear--in--temperature coefficient $A_T^1$ divided by $k_B$ (see Appendix~\ref{sec:alpha_T}). We note that both the in--temperature  $A_T^1/k_B$ and in--field $A_B^1/\mu_B$ rescaled coefficients follow a similar trend with respect to doping,  such that their ratio $A^1_T \mu_B/(A^1_B k_B)$ illustrated in \Autoref{fig:fig_A1B_T_rho0B_T}~(b) is almost doping independent, and close to 2 for all doping values. From \Autoref{eq:def_A1_T_B}, it is easy to see that this ratio corresponds to the ratio of the Planckian coefficients $\alpha_T/\alpha_B$, for which our theoretical prediction yields exactly 2, since our theoretical scaling predicts $\alpha_T=\frac{8}{\pi}$ and $\alpha_B=\frac{4}{\pi}$ (see Appendices \ref{app:srfxi} and  \ref{sec:ana_slope}).

Finally, using \protect\Autoref{eq:def_alpha_T_B}--\protect\Autoref{eq:mstar_function}, the slope values $A_T^1$ and  $A_B^1$ are used to estimate the linear--in--temperature $\alpha_T$ and linear--in--field $\alpha_B$ Planckian coefficients shown in \protect\Autoref{fig:alpha_B_T_wp}.

\begin{figure}[H]
\begin{center}
	\begin{minipage}[b]{1\textwidth}
    {
 				\includegraphics[width=0.48\linewidth]{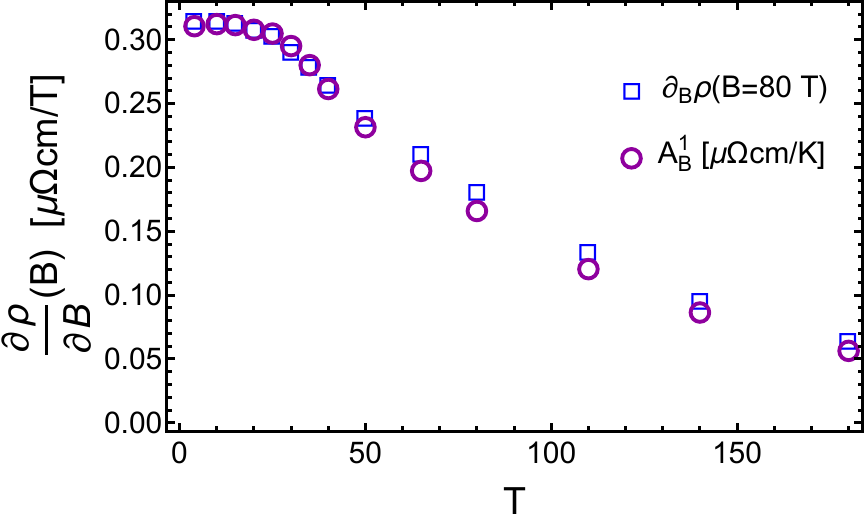}
			}
		\end{minipage}
	\end{center}
 	\caption{Slope of the resistivity data versus temperature for $p=0.19$ using the values obtained by linear regression (purple circles) $A_B^1$ (purple circles) and the values obtained from the window slope analysis $\dd_B\rho(B=80T)$  method described in Appendix~\protect\ref{usffx} at $B=80 T$ with a window width of $w=22\rm{T}$, corresponding to the saturation values of the slope in \protect\Autoref{fig:drhoBT_19_w}~(b) at $B=80~\rm{T}$. The two methods are in good agreement.
    }
	\label{fig:fit_drho_B_comp}
 \end{figure}

\begin{figure}[H]
	\begin{center}
	\begin{minipage}[b]{1\textwidth}
            {
				\includegraphics[width=0.48\linewidth]{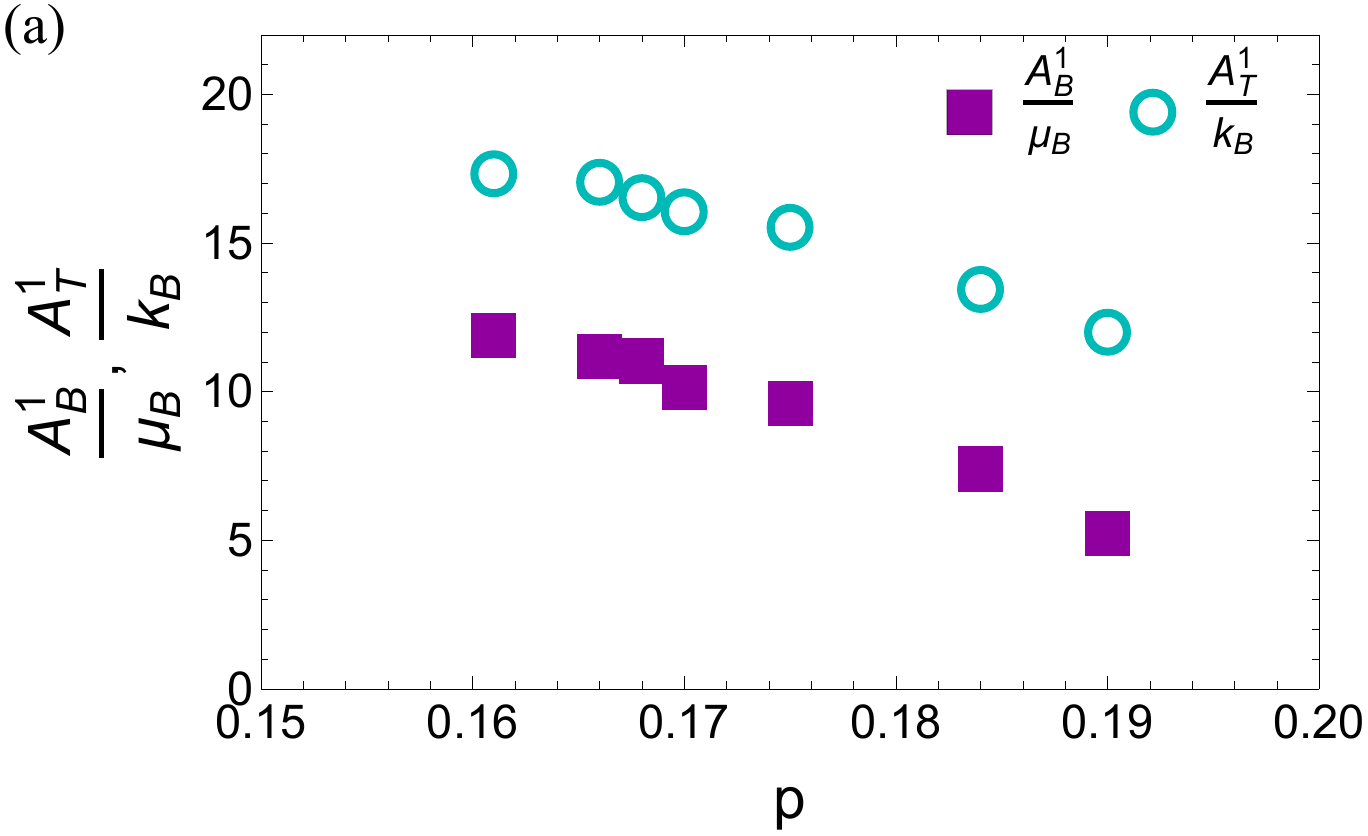}				
			}\\
 		{
 				\includegraphics[width=0.48\linewidth]{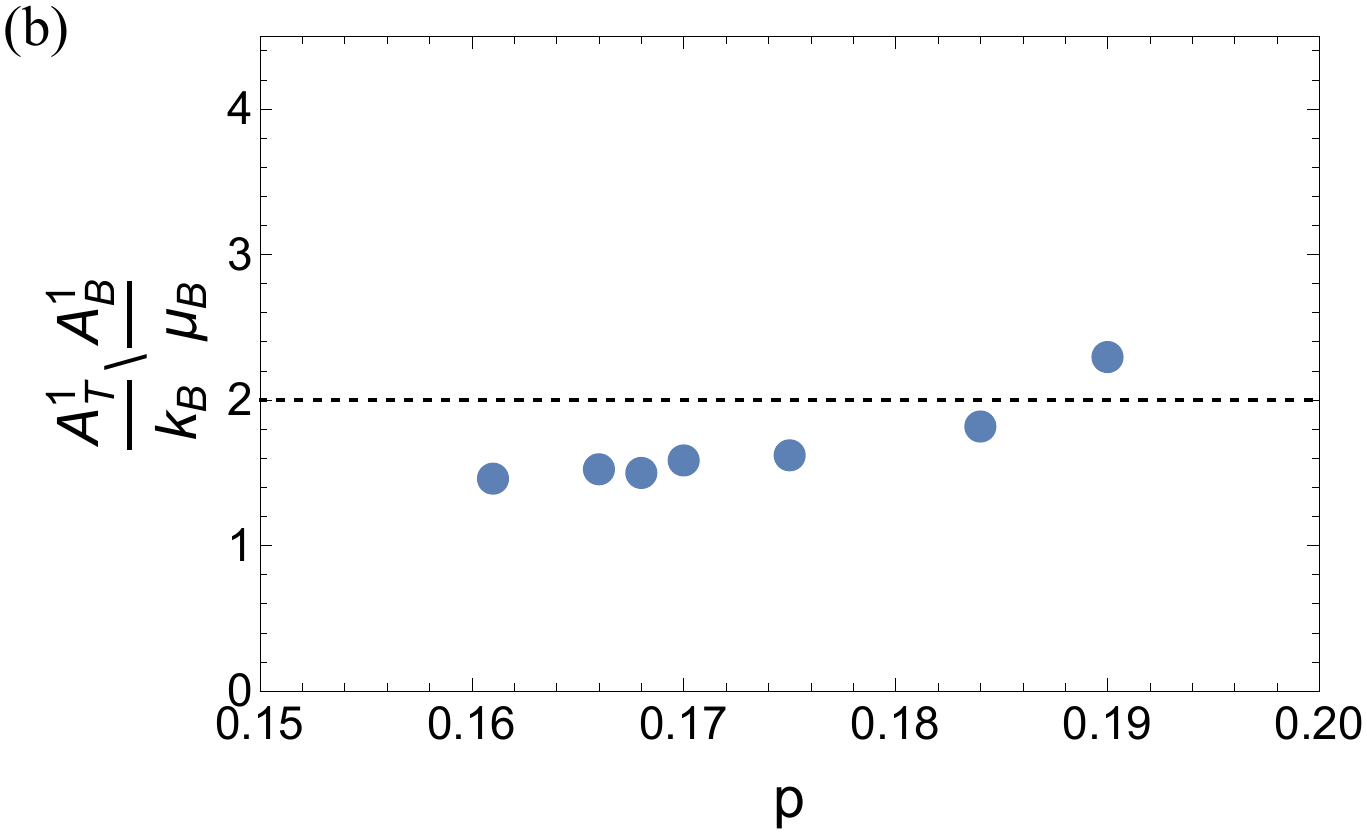}
			}
		\end{minipage}
	\end{center}
 	\caption{(a) Rescaled in--temperature  $A_T^1/k_B$ and in--field $A_B^1/k_B$ slope coefficients obtained from linear regression of the experimental resistivity data plotted versus doping. Both the in--temperature  $A_T^1/ k_B$ and in--field $A_B^1/\mu_B$ rescaled coefficients follow a similar trend with respect to doping, such that their ratio in (b) is almost doping independent. (b) ratio $A^1_T \mu_B/(A^1_B k_B)$ corresponding to the ratio of $\alpha_T/\alpha_B$ [\protect\Autoref{eq:def_A1_T_B}] versus doping in good agreement with our theoretical  prediction yielding $\alpha_T/\alpha_B=2$ (see Appendices \protect\ref{app:srfxi} and  \protect\ref{sec:ana_slope}), denoted by the dashed black line. The slope values $A_T^1$ are further used through \protect\Autoref{eq:def_alpha_T_B}--\protect\Autoref{eq:mstar_function} to estimate the linear--in--temperature Planckian coefficients shown in \protect\Autoref{fig:alpha_B_T_wp}
    }
	\label{fig:fig_A1B_T_rho0B_T}
\end{figure}

\begin{figure}[h]
    \begin{center}
    \includegraphics[width=1\linewidth]{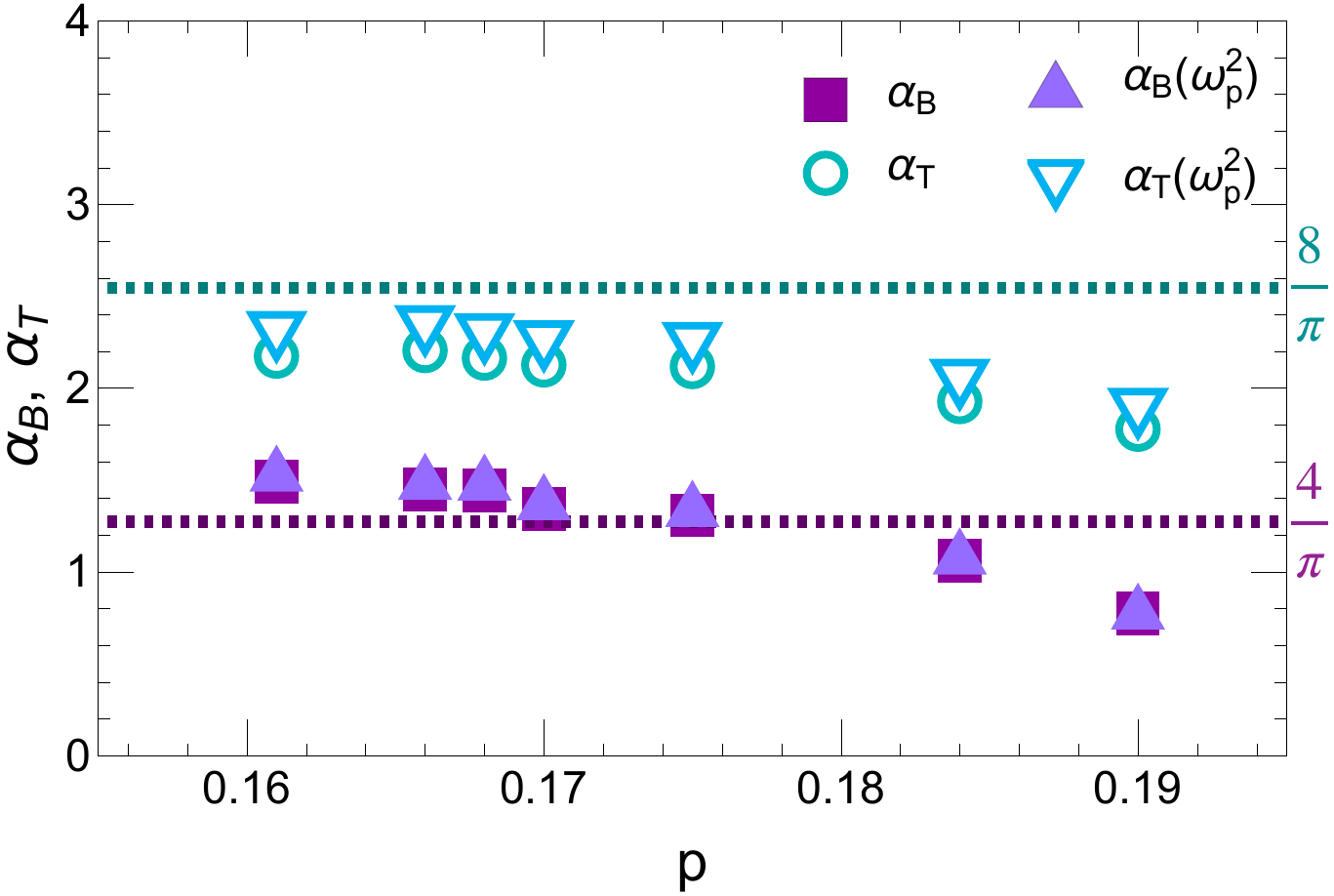}
    \end{center}
    \caption{The linear-in-temperature (empty symbols) and linear-in-field (filled symbols) Planckian coefficients versus doping for LSCO obtained from magneto--resistivity data reproduced from ~\cite{Beobinger-2018-Science-LSCO}. Empty circles and filled squares correspond to the Planckian coefficients $\alpha_T$ and $\alpha_B$ obtained from the estimation of effective mass $m^*$ from specific heat measurements as detailed above [\protect\autoref{fig:fit_mstar}]; while empty and filled triangles refer to  Planckian coefficients obtained by measurements of the plasma frequency $\omega_p^2$ denoted by $\alpha_T (\omega_p^2)$ and $\alpha_B(\omega_p^2)$ Horizontal purple and turquoise dashed lines correspond to theoretical prediction $\alpha_T=8/\pi\simeq 2.54$ and $\alpha_B=4/\pi\simeq 1.27$, respectively.
    For the experimental average value over doping for the Planckian coefficients we found $\langle\alpha_T^{exp}\rangle\simeq2.07$, and $\langle\alpha_B^{exp}\rangle\simeq1.27$ using effective mass estimations, and $\langle\alpha_T^{exp}(\omega_p^2)\rangle\simeq2.17$  and $\langle\alpha_B^{exp}(\omega_p^2)\rangle\simeq1.33$ from plasma frequency estimations, leading to a $5\%$ discrepancy between the two methods.}
    \label{fig:alpha_B_T_wp}
\end{figure}

\section{Scaling of AC resistivity}\label{AppE}

To extend our $B=T=0$ result for $\Sigma_{\xi}^{\prime\prime}(\omega,B=T=0)$, we now apply the conformal transformation assumming the system is in the quantum critical scaling regime. 

We start from the equation below for the spectral representation of correlation function \cite{georges-multichannel-kondo-PRB},
\begin{align}
    G_{\psi}(\tau)=-\int_{-\infty}^{\infty}\frac{e^{-\tau \varepsilon}}{1+e^{-\beta\varepsilon}}A_{\psi}(\varepsilon)d\varepsilon,
\label{eq:seeback-spectral}
\end{align}
where $0\leq\tau\le\beta$ and $A_\psi (\varepsilon) = (-1/\pi) G^{\prime\prime}_\psi( \omega+i0^+)$ denotes the spectral function for the $\psi$ field. When a system has conformal invariance, any fermionic correlation function, such as the self-energy, in the imaginary-time domain can be expressed as \cite{Georges-2021-PRR-seeback}
\begin{align}
    \Sigma(\tau)\propto e{}^{\alpha(\tau/\beta-1/2)}\left[\frac{\pi/\beta}{\sin\left(\pi\tau/\beta\right)}\right]^{1+\nu},
    \label{eq:seeback-self-energy-1}
\end{align}
where $\alpha$ and $\nu$ are constants. Here, $\alpha$ is a measure of particle-hole asymmetry. Its spectral representation is given by
\begin{align}
     & e^{\alpha(\tau/\beta-1/2)}\left[\frac{\pi}{\sin\left(\pi\tau/\beta\right)}\right]^{1+\nu} \nn
     =& C_{\alpha,\nu}\int_{-\infty}^{\infty}dx\frac{e^{-x\tau/\beta}}{1+e^{-x}}g_{\alpha,\nu}(x) 
     \label{eq:seeback-self-energy-2}
\end{align}
with
\begin{align} \label{eq:seeback-g} \notag
    g_{\alpha,\nu}(x) & =\left|\Gamma\left(\frac{1+\nu}{2}+i\frac{x+\alpha}{2\pi}\right)\right|^2 \\ \notag
    & \qquad \quad  \times \frac{\cosh(x/2)}{\cosh(\alpha/2)\Gamma[(1+\nu)/2]^{2}}, \\ 
     C_{\alpha,\nu} & =\frac{(2\pi)^{\nu}\cosh(\alpha/2)\Gamma\left[\frac{1+\nu}{2}\right]^{2}}{\pi\Gamma[1+\nu]}.
\end{align}

Following Eqs.~(\ref{eq:seeback-spectral})-(\ref{eq:seeback-g}), we will generalize $\Sigma_\xi$ that we obtain in the zero-temperature limit in the previous section  to the finite-temperature one. We will further derive its $\omega/T$ scaling behavior once the finite-temperature generalization  is acquired. 
Applying Eq. (\ref{eq:seeback-spectral}), we first evaluate the (imaginary) time dependence of $ \Sigma_{\xi}$ which shows a linear-in-frequency dependence in the frequency domain, $\Sigma_\xi \propto |\omega|$. It can be shown that $\Sigma_{\xi}$ exhibits a $\tau^{-2}$ dependence:
\begin{align}
    \Sigma_{\xi}(\tau)	= & \frac{1}{\pi}\int_{-\infty}^{\infty}\frac{|\varepsilon|e^{-\tau\varepsilon}}{1+e^{-\beta\varepsilon}}d\varepsilon \nn 
	= & \frac{1}{\pi}\left[\int_{0}^{\infty}|\varepsilon|e^{-\tau\varepsilon}+\int_{-\infty}^{0}|\varepsilon|e^{(\beta-\tau)\varepsilon}\right]d\varepsilon \nn
	=& \frac{1}{\pi}\left[\frac{1}{\tau^{2}}\int_{0}^{\infty}|x|e^{-x}dx+\frac{1}{(\beta-\tau)^{2}}\int_{-\infty}^{0}|x|e^{x}dx\right] \nn
	= & \frac{1}{\pi \tau^{2}}.
\end{align}
It implies that this situation corresponds to the case of $\alpha = 0$ (particle-hole symmetric) and $\nu=1$ (Planckian) as discussed in Refs.~\cite{georges-multichannel-kondo-PRB,Georges-2021-PRR-seeback}. The second term of the above equation vanishes when $T \to 0$.
Once $\alpha=0$ and $\nu = 1$  are decided, we can  generalize $\Sigma_{\xi}$ to the finite-temperature region by conformal transformation governed by Eqs. (\ref{eq:seeback-spectral})-(\ref{eq:seeback-g}), leading to the following expression for $\Sigma_{\xi}(\omega,T)$,
\begin{align}
    \Sigma_{\xi}^{\prime\prime}(\omega,T)=\lambda_{0}\beta^{-1}g_{0,1}(x)={\lambda_{0}\frac{\omega}{2}\coth\left(\frac{\omega}{4T}\right)}
    \label{eq:Sigma_xi-img-generalscaling}
\end{align}
with $\lambda_0$ being an unknown constant. $\lambda_0$ can be determined from its zero-temperature limit, where $\Sigma_{\xi}^{\prime\prime}(\omega,T=0) = -(2/\pi) |\omega| $. We thus find $\lambda_0 = -4/\pi$, leading to 
\begin{align}
     \Sigma_{\xi}^{\prime\prime}(\omega,T)=\lambda_{0}\beta^{-1}g_{0,1}(x)={-\frac{4}{\pi}\left[\frac{\omega}{2}\coth\left(\frac{\omega}{4T}\right)\right]}.
     \label{eq:Sigma_xi-img}
\end{align}
Using the relation of $\Sigma_c = (1/2)\Sigma_\xi$, the  scattering rate for the conduction $c$ electron can be obtained by $\hbar/\tau_c = -2\Sigma_c^{\prime\prime} (\omega)$ (here we restore $\hbar$ and $k_B$), ${\frac{\hbar}{\tau_c} =\frac{4}{\pi}\left[\frac{\hbar\omega}{2}\coth\left(\frac{\hbar \omega}{4k_B T}\right)\right]}$. It shows the following frequency-to-temperature scaling  behavior
\begin{align}
     \frac{\hbar/\tau_c}{k_B T} ={\frac{4}{\pi}\left[\frac{x}{2}\coth\left(\frac{x}{4}\right)\right]},
\end{align}
where $x\equiv \hbar \omega/k_B T$. 

Note that the $\omega/T$-scaling function in scattering rate at finite frequency and temperature is not uniquely determined by the zero-temperature conduction electron self-energy $\Sigma_{\xi}^{\prime\prime}(\omega,T=0)$ via conformal mapping by Eqs.~(\ref{eq:seeback-spectral})-(\ref{eq:seeback-g}). We find that in general the scaling function $\frac{\hbar/\tau}{k_BT} = \frac{2x}{\pi} \coth\frac{x}{2p}$ for $p \ge 1$ reduces to the same zero-temperature scattering rate $1/\tau(\omega,T=0) = -2\Sigma_{\xi}^{\prime\prime}(\omega,T=0) = \frac{2}{\pi} \omega$ in the limit of 
$\frac{k_BT}{\hbar\omega} \to 0$ 
(see also Fig.~S1 in Ref.~\cite{YYC-arXiv-2025}).

\bibliographystyle{apsrev4-2}
\bibliography{SBtJ.bib}

\end{document}